\documentclass[english,12pt]{iopart}
\usepackage[OT1]{fontenc}
\usepackage[latin1]{inputenc}
\usepackage{graphicx}

\usepackage{iopams}
\usepackage{amssymb}

\usepackage{babel}

\newcommand{\dd}{\mathrm{d}}

 5

\begin{document}

\noindent{\it \small Preprint: UTTG-14-06}\\

\title{Probing inflation with CMB polarization : weak lensing effect on
the covariance of CMB spectra}

\author{J Rocher$^{1,2}$, K Benabed$^{2}$ and F R Bouchet$^{2}$}

\address{$^{1}$ \emph{Theory Group, University of Texas at Austin}, 1
University Station, C1608, Austin, TX 78712, USA,}

\address{$^{2}$ \emph{Institut d'Astrophysique de Paris}, 98bis
boulevard Arago, 75014 Paris, France.}

\eads{\mailto{rocher@physics.utexas.edu}, \mailto{benabed@iap.fr}, \mailto{bouchet@iap.fr}}

\begin{abstract}
CMB anisotropies are modified by the weak lensing effect of
intervening large scale structures on the photon path from the
last scattering surface to the observer. This has to be accounted
for when observational data of sensitive experiments are used to
constrain cosmological models. A common approximation to analyze
the CMB angular power spectra is to include only the Gaussian part
of the lensing correction and to ignore the non-gaussian terms in
the error covariance matrix of the spectra. In order to
investigate the validity of this approximation, we computed these
non-Gaussian terms by using a perturbative expansion method. We
present a graphical method to write down any N-point correlation
functions at any order in lensing. We use a pedagogical approach to
demonstrate that neglecting non-gaussian terms is an accurate
approximation for all polarizations but B, and it will remain so
even for the analysis of very sensitive post-Planck experiments.
For the B polarization, non-gaussian contributions up to order 4
must be taken into account.
\end{abstract}

\pacs{98.80.-k,98.70.Vc,98.62.Sb,98.65.Dx}

\maketitle

\section{Introduction}

Measurements of the temperature anisotropies and polarization of the
Cosmic Microwave Background (CMB) provide very valuable limits and
constraints on our models of the early universe \cite{2006astro.ph..3449S}.
This will be even more so with the increased precisions of measurements
that future CMB experiments
promise\footnote{\texttt{http://www.rssd.esa.int/index.php?project=Planck}}
\cite{2006astro.ph..4101B}. However, while the hopes of high
precision cosmology become true, second order effects that were
formerly neglected or ignored have now to be taken into account. Among
them, one of the most important, in particular at small angular scale, is the
gravitational shear effect of the large scale structure, also
commonly called the weak lensing effect \cite{Bartelmann:1999yn}.
This effect generates B modes \cite{1998PhRvD..58b3003Z}, which are of
great interest for
testing theories of the early universe. Indeed, scalar primordial
perturbations do not generate B modes, whose detection could therefore
be considered as the ``smoking gun'' of the primordial gravitational
wave background generically predicted by inflation
theories \cite{Kinney:1998md}. In addition to determining the energy
scale of inflation, a detection of the primordial B-mode would also
test other aspects of the early universe physics, for example the
presence of cosmic strings
\cite{2006astro.ph..4143S,2006astro.ph..4141P}. From that point of
view, the lensing effect is an annoying foreground limiting the
detection capability of primordial B modes. It is however quite
interesting in its own right,
since it allows reconstructing the (projected) matter power spectrum
as well as extracting information on various aspects of
high energy physics such that neutrino masses
\cite{Lesgourgues:2005yv,Lesgourgues:2006nd}.

The gravitational deviations of photons along their paths as they
cross the potential wells of the large scale structures has been
studied in great details in the context of CMB. The gravitational
shear shifts power between scales and creates mode couplings in
the CMB power. It also deviates the distribution of temperature
and polarization anisotropies from the Gaussian statistic
\cite{Lewis:2006fu}. The specific and predictable signature of the
lensing effect has been used to propose various ways of detecting
and reconstructing its contribution to the temperature and
polarization of the CMB \cite{2003PhRvD..68h3002H,
 2003PhRvD..67h3002O, 2001PhRvD..63d3501B, 2000ApJ...540...14V,
 1999PhRvD..59l3507Z}.

Until recently however, few studies had been devoted to a detailed
assessment of the impact of weak lensing on the cosmological
parameter estimations from CMB data. Indeed, analyzes had in
general been concerned with including lensing corrections to the
power spectra of CMB temperature and polarization anisotropies,
but they neglected the non-gaussian corrections. Recently A. Lewis
showed numerically that this approximation is valid at the
sensitivity level of the Planck mission \cite{Lewis:2005tp}. This
paper presents an analysis which demonstrates that the lensing
induced non-gaussian terms of the error bars on the CMB
anisotropies power spectrum can be neglected as compared to the
Gaussian sample variance terms except for the B-mode power
spectrum. While developed independently, this works follows the
line of the recently published \cite{2006PhRvD..74l3002S, 2006astro.ph..7494L} and
fully confirm their relevant results. Given the availability of
the numerical results of these references, we rather focus here on
the physical origin of the various terms by using toy models which
allow understanding the final results.

Therefore, the aim of this article is in part to understand in detail
the origin of this result and above all to see whether this
approximation will remain warranted in the post-Planck era, or to
determine at which precision it might break down. To do so, we compute
analytically the lens effect using a perturbative expansion in terms
of the magnitude of the deflection field and give numerical results
for the dominant order.

\section{Weak lensing of the Cosmic Microwave Background anisotropies}
\label{sec:WLonCMB}

CMB photons, as they emerge from the last scattering surface, are
subject to the weak gravitational lensing effect. Measurements of
the temperature anisotropies, and polarization patterns are
perturbed by the cumulative effects of the large scale structure
gravitational wells from $z\sim1000$. The net result is that our
measurement of temperature or polarization of the CMB in a
direction $\mathbf{n}$ in fact provides an information on photons
emerging from the last scattering surface in the direction
$\mathbf{n}+\mathbf{\xi}$, $\mathbf{\xi}$ being the deflection
induced by the gravitational shear
\begin{eqnarray}
\widetilde{T}(\mathbf{n})  =  T(\mathbf{n}+\mathbf{\xi})~,\\
\widetilde{Q}(\mathbf{n})  =  Q(\mathbf{n}+\mathbf{\xi})~,\\
\widetilde{U}(\mathbf{n})  =  U(\mathbf{n}+\mathbf{\xi})~.
\end{eqnarray}
In this paper, we denote by $\widetilde{A}$ the lensed appearance of
any field $A$. The deflection
$\mathbf{\xi}$ is the cumulative effect of all deflections
produced by each gravitational well crossed during the propagation
of the photon. Since those deflection are small compared to the
size of the CMB anisotropies, it is sufficient to compute this
cumulative effect on the unperturbed path of the photon. This
approximation is often referred to as the Born approximation. By doing
so, we ignore higher order corrections to the weak lensing effect,
such as extra non gaussian correction of the lensing field due to the
lens-lens coupling and apparition of a curl component in the spin-2
shear field. Both have been shown to be negligible in the CMB
context \cite{2006JCAP...03..007S}.

In the Born approximation, the deflection field reduces
to \cite{Lewis:2006fu}
\begin{equation}
\mathbf{\xi}(\mathbf{n})=-2 \int_0^{\chi_{\rm cmb}}\dd \chi \,
\frac{f_K(\chi_{\rm cmb}-\chi)}{f_K(\chi_{\rm cmb})f_K(\chi)}
\mathbf{\nabla}\Psi(\chi\mathbf{n};\eta_0-\chi)~,
\end{equation}
where $f_K(\chi)$ is the comoving angular diameter distance and
$\Psi$ the gravitational potential, linked to the density perturbation
through the Poisson equation \cite{Bartelmann:1999yn}.

The deflection $\mathbf{\xi}$ being small, one can evaluate
accurately the effect on the CMB field by using a perturbative
expansion,
\begin{equation}
\widetilde{X}(\mathbf{n})=
X(\mathbf{n})+\xi_{i}\partial^{i}X(\mathbf{n})+
\frac{1}{2}\xi_{i}\xi_{j}\partial^{i}\partial^{j}X(\mathbf{n})+
\mathcal{O}(\xi^{3})~, \label{eq:expansion:X}
\end{equation}
where $X$ is either the temperature anisotropy field $T$ or
the polarization components $Q$ or
$U$~\cite{2001PhRvD..63d3501B,2000PhRvD..62d3007H}.

In the following, we will be interested in the effect of the
gravitational shear on the properties of the power spectra of
the CMB. Since our goal is only to evaluate the contribution of the
non-gaussian corrections due to weak lensing, and since
these corrections arise mostly at small scale, it is sufficient to
work in the flat space approximation and use a decomposition in
Fourier modes rather than in spherical harmonics $Y_m^l$.
Let us introduce the lensing potential
$\phi(\mathbf{n})$ such that $\mathbf{\xi}=\nabla
\phi(\mathbf{n})$ and its Fourier transform
\begin{equation}
\phi(\mathbf{l})=\int \mathrm{d}\mathbf{n}\, \phi(\mathbf{n})
e^{-i\mathbf{l\cdot n}}~.
\end{equation}
(\ref{eq:expansion:X}) can then be rewritten as
\begin{equation}
\widetilde{X}_\mathbf{l}=\widetilde{X}_\mathbf{l}^{(0)}+
\widetilde{X}_\mathbf{l}^{(1)}+\widetilde{X}_\mathbf{l}^{(2)}+
\mathcal{O}(\xi^{3})~.\label{eq:expansion:Fourier}
\end{equation}
When $X$ is a scalar quantity (namely for temperature), the first and
second order terms now read \cite{2000PhRvD..62d3007H}
\begin{equation}
\widetilde{X_{\mathbf{l}}}^{(1)}=
-\int\frac{\mathrm{d}^{2}\mathbf{l_{1}}}{(2\pi)^{2}}\,
X(\mathbf{l_{1}})\,\phi(\mathbf{l}-\mathbf{l_{1}})\,
[\mathbf{l_{1}}\cdot(\mathbf{l}-\mathbf{l_{1}})]~,
\label{eq:L:ordre1}
\end{equation}
\begin{equation}\label{eq:L:ordre2}
\widetilde{X_{\mathbf{l}}}^{(2)}=\frac{1}{2}\int
\frac{d^{2}\mathbf{l_{1}}d^{2}\mathbf{l_{2}}}{(2\pi)^{4}}\,
X(\mathbf{l_{1}})\,\phi(\mathbf{l_{2}})\phi(\mathbf{l}
-\mathbf{l_{1}}-\mathbf{l_{2}})(\mathbf{l_{1}}\cdot\mathbf{l_{2}})[\mathbf{l_{1}}
\cdot(\mathbf{l}-\mathbf{l_{1}}-\mathbf{l_{2}})]~.
\end{equation}

Finally, it is convenient to use another description of the
polarization field than the usual $Q$ and $U$ Stokes variables.
Indeed, these directly observable quantities have awkward
geometrical properties. They transform under rotation as the
components of a spin-2 vector. One then introduces a decomposition
in terms of a gradient and curl component, named $E$ and $B$ in
analogy with the electromagnetic field decomposition, which have
simpler properties under angular transformations. $E$ is a scalar
field and $B$ a pseudo scalar one (meaning that the $B$
polarization changes sign under a parity transformations). This
decomposition has the advantage of offering an important test of
the primordial cosmology model. The polarization sky pattern
essentially maps the local quadrupolar temperature anisotropies on
the last scattering surface and at the linear order, the scalar
metric perturbations can only produce an $E$ polarization
\cite{2002ARA&A..40..171H}. The tensorial contribution from
gravitational waves is therefore the unique source of primordial
$B$ polarization, which is concentrated at large scales. Since the
tensorial contribution yields only a weak contribution to the
temperature and E-type anisotropies, which is therefore very hard
to detect or differentiate from the scalar part, the $B$ modes
therefore offer a potentially unique opportunity to detect a
background of primordial gravitational waves, at least on the
largest scales. The lensing induced $B$ polarization may
nevertheless hide the primordial contribution and is in any case
dominant at small scales \cite{Lewis:2006fu,1998PhRvD..58b3003Z}.

The $E$ and $B$ decomposition in the flat sky approximation is
easily computed from the Stokes variables through \cite{2000PhRvD..62d3007H}
\begin{equation}
{}_{\pm}X(\mathbf{n})=Q(\mathbf{n})+iU(\mathbf{n})~.
\end{equation}
Since these ${}_{\pm}X$ are spin-2 quantities, their Fourier coefficients
contain an additional factor $\exp[\pm 2i(\varphi_{l_1}-\varphi_l)]$. We
recover the E and B Fourier coefficients with
\begin{equation}
{}_{\pm}X(\mathbf{l})=E(\mathbf{l})\pm i B(\mathbf{l})~,
\end{equation}
with which we obtain that the Eqs.
(\ref{eq:L:ordre1}-\ref{eq:L:ordre2}) translate into
\begin{eqnarray}\label{eq:E:ordre1}
\fl \widetilde{E_{\mathbf{l}}}^{(1)}= -\int\frac{d^{2}\mathbf{l}_{1}}{(2\pi)^{2}}
\left[E(\mathbf{l}_{1})\cos(2\varphi_{1})-B(\mathbf{l}_{1})\sin(2\varphi_{1})
\right]\nonumber\\
\times\phi(\mathbf{l}-\mathbf{l_{1}})\left[\mathbf{l_{1}}\cdot(\mathbf{l}
-\mathbf{l_{1}})\right]~,
\end{eqnarray}

\begin{eqnarray}\label{eq:B:ordre1}
\fl \widetilde{B_{\mathbf{l}}}^{(1)}= -\int\frac{d^{2}\mathbf{l}_{1}}{(2\pi)^{2}}
\left[B(\mathbf{l}_{1})\cos(2\varphi_{1})+E(\mathbf{l}_{1})\sin(2\varphi_{1})
\right]\nonumber\\
\times\phi(\mathbf{l}-\mathbf{l_{1}})\left[\mathbf{l_{1}}\cdot(\mathbf{l}
-\mathbf{l_{1}})\right]~,
\end{eqnarray}
for the first order and
\begin{eqnarray}\label{eq:E:ordre2}
\fl \widetilde{E_{\mathbf{l}}}^{(2)}=\frac{1}{2}\int\hspace{-.2cm}\int\frac{d^{2}
\mathbf{l}_{1}d^{2}\mathbf{l}_{2}}{(2\pi)^{4}}\left[E(\mathbf{l}_{1})
\cos(2\varphi_{1})-B(\mathbf{l}_{1})\sin(2\varphi_{1})\right]\nonumber\\
\times\phi(\mathbf{l_{2}})\phi(\mathbf{l}-\mathbf{l_{1}}-\mathbf{l_{2}})
(\mathbf{l_{1}}\cdot\mathbf{l_{2}})\left[\mathbf{l_{1}}\cdot(\mathbf{l}
-\mathbf{l_{1}}-\mathbf{l_{2}})\right]~,
\end{eqnarray}

\begin{eqnarray}\label{eq:B:ordre2}
\fl \widetilde{B_{\mathbf{l}}}^{(2)}=\frac{1}{2}\int\hspace{-.2cm}\int\frac{d^{2}
\mathbf{l}_{1}d^{2}\mathbf{l}_{2}}{(2\pi)^{4}}\left[B(\mathbf{l}_{1})
\cos(2\varphi_{1})+E(\mathbf{l}_{1})\sin(2\varphi_{1})\right] \nonumber\\
\times\phi(\mathbf{l_{2}})\phi(\mathbf{l}-\mathbf{l_{1}}-\mathbf{l_{2}})
(\mathbf{l_{1}}\cdot\mathbf{l_{2}})\left[\mathbf{l_{1}}\cdot
(\mathbf{l}-\mathbf{l_{1}}-\mathbf{l_{2}})\right]~,
\end{eqnarray}
for the second order in the perturbative development. We have used the
following definition for the angles,
\begin{equation}
\varphi_{1}\equiv \varphi_{l_1}-\varphi_l~.
\end{equation}

We can now turn to the computation of the power spectra of the CMB,
$C_{l}^{XY}$, defined by
\begin{equation}
\langle X_{\mathbf{l}}^{*}Y_{\mathbf{l}'}\rangle=(2\pi)^{2}\,
\delta(\mathbf{l}-\mathbf{l'})\, C_{l}^{XY}~,\label{eq:defspec}
\end{equation}
where $X$ and $Y$ can be either $T$, $E$ or $B$. Since no
confusion can arise, from now on we shall drop the tilde on the
lensed field in order to lighten notations. Putting
(\ref{eq:defspec}) and (\ref{eq:L:ordre1}-\ref{eq:L:ordre2})
in (\ref{eq:expansion:Fourier}), applying Wick theorem, and
truncating the results to second order in $\phi$ gives
\cite{2000PhRvD..62d3007H}
\begin{equation}\label{eq:ClT:lens:o2}
\widetilde{C_{l}^{TT}}=C_{l}^{TT}\left(1-\frac{l^{2}}{4\pi}\sigma_{0}^{2}\right)
+\frac{1}{(2\pi)^{2}}\int C_{l_{1}}^{TT}\, C_{|\mathbf{l}-\mathbf{l_{1}}|}^{\phi\phi}
[\mathbf{l_{1}}\cdot(\mathbf{l}-\mathbf{l_{1}})]^{2}\mathrm{d}^{2}\mathbf{l_{1}}~,
\end{equation}
where we have defined
\begin{equation}\label{eq:defR}
\sigma_{0}^{2}\equiv\int_{0}^{\infty}l_{1}^{3}\, C_{l_{1}}^{\phi\phi}\mathrm{d}l_{1}~.
\end{equation}

With the same computations, the polarization and temperature-polarization
cross spectra reads
\begin{eqnarray}\label{eq:ClE:lens:o2}
\fl \widetilde{C_{l}^{EE}}=C_{l}^{EE}\left(1-\frac{l^{2}}{4\pi}\sigma_{0}^{2}\right)+
\frac{1}{(2\pi)^{2}}\int\mathrm{d}^{2}\mathbf{l_{1}}\left[C_{l_{1}}^{EE}
\cos^{2}(2\varphi_{1})+C_{l_{1}}^{BB}\sin^{2}(2\varphi_{1})\right]\nonumber\\
\phantom{\left(1-\frac{l^{2}}{4\pi}\sigma_{0}^{2}\right)+\frac{1}{(2\pi)^{2}}\int\mathrm{d}^{2}\mathbf{l_{1}}}
\times C_{|\mathbf{l}-\mathbf{l_{1}}|}^{\phi\phi}[\mathbf{l_{1}}\cdot(\mathbf{l}
-\mathbf{l_{1}})]^{2}~,
\end{eqnarray}
\begin{eqnarray}\label{eq:ClB:lens:o2}
\fl \widetilde{C_{l}^{BB}}=C_{l}^{BB}\left(1-\frac{l^{2}}{4\pi}\sigma_{0}^{2}\right)+
\frac{1}{(2\pi)^{2}}\int\mathrm{d}^{2}\mathbf{l_{1}}\left[C_{l_{1}}^{BB}
\cos^{2}(2\varphi_{1})+C_{l_{1}}^{EE}\sin^{2}(2\varphi_{1})\right]\nonumber\\
\phantom{\left(1-\frac{l^{2}}{4\pi}\sigma_{0}^{2}\right)+\frac{1}{(2\pi)^{2}}\int\mathrm{d}^{2}\mathbf{l_{1}}}\times C_{|\mathbf{l}-\mathbf{l_{1}}|}^{\phi\phi}[\mathbf{l_{1}}\cdot(\mathbf{l}
-\mathbf{l_{1}})]^{2}~,
\end{eqnarray}
and
\begin{equation}\label{eq:ClTE:lens:o2}
\fl \widetilde{C_{l}^{TE}}=C_{l}^{TE}\left(1-\frac{l^{2}}{4\pi}\sigma_{0}^{2}\right)
+\frac{1}{(2\pi)^{2}}\int C_{l_{1}}^{TE}\cos(2\varphi_{1})\,
C_{|\mathbf{l}-\mathbf{l_{1}}|}^{\phi\phi}[\mathbf{l_{1}}\cdot(\mathbf{l}
-\mathbf{l_{1}})]^{2}\mathrm{d}^{2}\mathbf{l_{1}}~.
\end{equation}

\section{Covariance matrix of the power spectra}
\subsection{Gaussian assumption}\label{methodestd}

Lensing is quite well known to produce non-gaussian features
\cite{2001PhRvD..64h3005H,2000PhRvD..62f3510Z,1997A&A...324...15B}.
Still, in previous literature, the weak lensing effect has usually been
taken into account in the  partial way we recalled above. The
lensing corrections have been applied to the power spectrums only;
no effort have been made to reproduce the deviation from
gaussianity. In other words, the effect of lensing has been
reduced to a modification of the damping tails of the temperature
and $E$ power spectra, and the apparition of the small scale $B$
induced by lensing. In that case the power spectrum covariance
matrix is diagonal and reduces to the square of the $C_{\ell}$'s of
type $TT$, $TE$, $EE$ and $BB$ .

This approximation might turn out to be a valid one. This can be
the case if the deviation from Gaussian behavior induced by
lensing is small compared to the dominant Gaussian contribution,
at the level of accuracy of the planned experiments. A. Cooray has
argued qualitatively that this should be the case for temperature
anisotropies \cite{2002PhRvD..65f3512C}. A. Lewis has shown
numerically with some mock data that this approximation holds when
constraining the cosmological model at the level of precision of
the upcoming Planck experiment \cite{Lewis:2005tp}. The latter
approach has the great merit that it does not make assumptions on
the accuracy of the perturbative expansion used to compute the
weak lensing effect on the CMB. Moreover, one can easily
incorporate non-linear evolution of the matter density in the
simulation and produce very accurate predictions. A downside is
that such method has a relatively high computational cost and is
physically less transparent.

We propose here another approach to validate the simplification
described above. We evaluate analytically the impact of the
non-gaussian component of the covariance matrices of the power
spectrum. This provides us with a direct estimate of the extent to
which the diagonal covariance matrix approximation is a valid one.
To do so, we do not compute the full non-gaussian correction, and
restrict ourselves to the dominant order of the lens effect within
the perturbative framework introduced in the previous section.

In the simplest cases (e.g. full sky, noiseless case), one can
simply estimate the cross-correlation power spectrum of two fields
$X$ and $Y$, by
\begin{equation}
\hat{C}_{\ell}^{XY}\equiv\frac{1}{2V_{\ell}}\sum_{\mathbf{l}}
\left(X_{\mathbf{l}}Y_{\mathbf{l}}^{*}+X_{\mathbf{l}}^{*} Y_{\mathbf{l}}\right)
= \frac{1}{V_{\ell}}\sum_{\mathbf{l}} X_{\mathbf{l}}Y_{\mathbf{l}}^{*}~.
\end{equation}
since the $C_l$s are real. Here $X,Y=T, E, B$, $\left|\mathbf{l}\right|=\ell$
and $V_\ell$ is the volume of the sample of scales $\ell$ accessible by the
experiment. These estimators of the power spectra are unbiased (in the absence
of noise), and therefore $\left\langle \hat{C}_{\ell}^{XY}\right\rangle
=C_{\ell}^{XY}$.

The covariance matrix of the power spectra describes how two
estimators of the power spectra are correlated. It is given by
\begin{equation}\label{eq:covar:def}
\mathrm{Cov}(\ell,\ell^{'})_{XX-YY}=\left\langle
\hat{C}_{\ell}^{XX}\hat{C}_{\ell'}^{YY}\right\rangle -\left\langle
\hat{C}_{\ell}^{XX}\right\rangle \left\langle
\hat{C}_{\ell'}^{YY}\right\rangle~.
\end{equation}

In the case of a Gaussian random field $X$, and for a noise free experiment,
the Wick theorem gives
\begin{equation}
\mathrm{Cov}(\ell,\ell')_{XX-XX}=\frac{2}{V_{\ell}}\left[C_{\ell}^{XX}\right]^2
\delta(\ell-\ell')~,
\end{equation}
In the case where the field is sampled on the full sky, the volume
$V_{\ell}$ reduces to $2\ell+1$. Neglecting the exact shape of a
survey\footnote{Of course, this is only an approximation valid in the limit
of very small scales (large $\ell$) as compared to the size of the survey.
When this is not the case, different $\ell$'s can be correlated, see the
detailed work of \cite{Challinor:2004pr}.}, this function is often
approximated by $(2\ell+1)f_{\rm sky}$, the $f_{\rm sky}$ factor being the
ratio of the survey area to the full sky.

When one corrects for the experimental beam $B_{\ell}$, and
taking into account experimental noise, modeled by non-correlated
white noises, one reproduce the usual formulae for the power
spectrum covariances~\cite{Seljak:1996ti,Zaldarriaga:1996xe,Kamionkowski:1996zd}
\begin{equation}\label{eq:covar:diag:TEB}
\mathrm{Cov}(\ell)^{XX-XX}=\frac{2}{(2\ell+1)f_{\textrm{sky}}}
\left[C_{\ell}^{XX}+\left(w_{x}B_{\ell} \right)^{-1}\right]^{2}~,
\end{equation}
where $X=T,E,B$ and $w^{-1}_{T/P}$ denote the power spectra of non-polarized and
polarized noise respectively. The diagonal term for $TE$ is
\begin{eqnarray}\label{eq:covar:diag:TE}
\fl \mathrm{Cov}(\ell)^{TE-TE}=\frac{2}{(2\ell+1)f_{\textrm{sky}}}
\left\{
\left(C_{\ell}^{TE}\right)^{2}+\left[C_{\ell}^{TT}+\left(w_{T}B_{\ell}\right)^{-1}
\right]\left[C_{\ell}^{EE}+\left(w_{P}B_{\ell}\right)^{-1}\right]\right\}~.
\end{eqnarray}
Finally, there are three non-diagonal terms which are non zero
\begin{equation}
\mathrm{Cov}(\ell)^{TT-EE}=\frac{2}{(2\ell+1)f_{\textrm{sky}}}{C_{\ell}^{TE}}^{2}~,
\label{eq:covar:nodiag:TTEE}
\end{equation}
\begin{equation}
\mathrm{Cov}(\ell)^{TT-TE}=\frac{2}{(2\ell+1)f_{\textrm{sky}}}C_{\ell}^{TE}\left[C_{\ell}^{TT}+
\left(w_{T}B_{\ell}\right)^{-1}\right]~,\label{eq:covar:nodiag:TTTE}\end{equation}
and
\begin{equation}
\mathrm{Cov}(\ell)^{EE-TE}=\frac{2}{(2\ell+1)f_{\textrm{sky}}}C_{\ell}^{TE}\left[C_{\ell}^{EE}+
\left(w_{T}B_{\ell}\right)^{-1}\right]. \label{eq:covar:nodiag:EETE}
\end{equation}

\subsection{In presence of lensing}
\subsubsection{Temperature}
If we take lensing into account, the computation above is no more
valid. Indeed, we can no longer assume that the temperature and
polarization obey to Gaussian distribution. The first term in the
r.h.s. of (\ref{eq:covar:def}) will translate into a four
point moments that does not reduce to the usual Gaussian case
through Wick expansion \cite{2001PhRvD..64h3005H,1997A&A...324...15B}.

Of course, we don't have to deal with all four point functions,
but only with the reduced quantity
\begin{equation}
\sum_{\mathbf{l},\mathbf{l'}}\left\langle
X_{\mathbf{l}}Y_{\mathbf{l}}^{*}U_{\mathbf{l}'}V_{\mathbf{l}'}^{*}\right\rangle
-\left\langle X_{\mathbf{l}}Y_{\mathbf{l}}^{*}\right\rangle \left\langle
U_{\mathbf{l}'}V_{\mathbf{l}'}^{*}\right\rangle~.
\end{equation}
We use the same perturbation approach than in the previous
sections, and reduce our computations to second order in lensing
for all polarizations but $BB-BB$. This latter term is a
particular case, for which the expansion need to be done until
order four as we will develop later. We do not expect higher order
terms to modify significantly our results.

To calculate at order 2 in lensing the covariance for all possible
polarization, we replace each $X$, $Y$, $U$ and $V$ in the
formula above by their second order lensed version for $XY$ and
$UV$ taking the values $TT$, $EE$, $BB$, $TE$. Then assuming
the unlensed temperature anisotropies and polarization are
Gaussian, one can apply Wick theorem to compute the covariance. To
simplify the computation, and since we expect the most important
contribution to arise at small scale, we will keep the flat sky
approximation, and assume that the volume of sample is the full
plane. The complete development is given in \ref{annex2}. We only
summarize the result here.

For all $XY=UV$ covariances, we obtain two different terms. One
being non-null only at $\ell=\ell'$ that we refer in the following
as the diagonal or Gaussian term. The other term, being non-zero
for $\ell\neq\ell'$ will be called non-diagonal or non-gaussian.
In the case of the $TT-TT$ covariance, those reads respectively
\begin{equation}\label{eqcorrel4Tdiag}
\fl \mathcal{D}_{\ell}^{TT-TT}=2\left(C_{l}^{TT}\right)^{2}
\left(1-\frac{l^{2}}{2\pi}\sigma_{0}^{2}\right)
 +\frac{4}{(2\pi)^{2}}C_{l}^{TT}\,\int\dd^{2}\mathbf{l_{1}}\,
C_{l_{1}}^{TT}C_{|\mathbf{l}-\mathbf{l_{1}}|}^{\phi\phi}
[\mathbf{l_{1}}\cdot(\mathbf{l}-\mathbf{l_{1}})]^{2}~,
\end{equation}
\begin{equation}\label{eqcorrel4Tnd}
\fl  \mathcal{N}_{l,l'}^{TT-TT}=2\int\frac{\dd\varphi'}{2\pi}
C_{|\mathbf{l}-\mathbf{l'}|}^{\phi\phi}\Bigl\{ C_{l'}^{TT}
\left[\mathbf{l'}\cdot\left(\mathbf{l}-\mathbf{l'}\right)\right]
 +C_{l}^{TT}\left[\mathbf{l}\cdot\left(\mathbf{l'}-\mathbf{l}\right)
\right]\Bigr\}^{2}~.
\end{equation}

We recognize in (\ref{eqcorrel4Tdiag}) something very similar
to the classical result. The diagonal part is simply the square of
the lensed power spectrum, (\ref{eq:ClT:lens:o2}), truncated
to second order in lensing. This is the approximation commonly
used, where the power spectrum covariance with lensing are simply
computed by replacing the power spectra by their lensing
counterparts in
(\ref{eq:covar:diag:TEB}-\ref{eq:covar:nodiag:EETE}). This
approximation works as if the temperature anisotropies and
polarization remained Gaussian, ignoring the non-diagonal
contribution, (\ref{eqcorrel4Tnd}), that describes the
apparition of non-gaussian features.

\subsubsection{Effects of experimental limitations}
We have ignored in (\ref{eqcorrel4Tdiag}-\ref{eqcorrel4Tnd})
the experimental limitations of the considered experiment, such as
the sensitivity, resolution or sky coverage. Since we are
interested in the comparison between the diagonal and non-diagonal
parts of the covariance matrix, these experimental limitations
shouldn't affect the present work. Let us show how these
quantities will enter the results given in the present work. The
consequence of the introduction of the sensitivity and resolution
is that the observed fluctuations are now given by
\begin{equation}\label{Tobs}
X^{\rm obs}_\mathbf{l}=\widetilde{X}_\mathbf{l}B^{1/2}_l + n(l)~,
\end{equation}
where $X=\{T,E,B\}$. We have assumed a Gaussian, uncorrelated noise
$n(l)$ with power spectrum denoted
\begin{equation}\label{noisespectrum}
\langle n(l)n^*(l)\rangle \equiv \delta(\mathbf{l}-\mathbf{l'})w_l^{-1}
\end{equation}
and the beam transform $B_l$ is related to the FWHM of the beam $\theta_{\rm b}$ through
\begin{equation}
B_l=\exp\left[-l(l+1)\theta_{\rm b}^2/8\ln 2 \right]~.
\end{equation}
As a consequence,
\begin{equation}
\langle X^{\rm obs}_\mathbf{l}(X^{\rm obs}_\mathbf{l'})^* \rangle =
\delta(\mathbf{l}-\mathbf{l'})\left(\widetilde{C}_l B_l + w_l^{-1}(l)\right)~,
\end{equation}
and the estimator of the \emph{lensed} spectrum reads
\begin{equation}\label{defestimator}
\hat{\widetilde{C}}_l \equiv B^{-1}_l\left(\int \frac{\dd \varphi_l}{2\pi}
X^{\rm obs}_\mathbf{l}(X^{\rm obs}_\mathbf{l})^*-w_l^{-1}\right)~.
\end{equation}
In all the paper we will assume that the temperature noise
$n_T(l)$ is uncorrelated with the signal (and with the polarized
noise). Using the definition of the covariance matrix
\begin{equation}
\mathrm{Cov}(l,l')\equiv \langle \hat{\widetilde{C}}_l
\hat{\widetilde{C}}_{l'}\rangle - \langle \hat{\widetilde{C}}_l
\rangle\langle \hat{\widetilde{C}}_{l'}\rangle
\end{equation}
and (\ref{defestimator}), even in presence of the noise, the
covariance matrix reads
\begin{eqnarray}
\fl \mathrm{Cov}(l,l')=\Biggl(\int\frac{\dd \varphi_l}{2\pi}
\frac{\dd \varphi_{l'}}{2\pi}
\left\langle X^{\rm obs}_\mathbf{l}(X^{\rm obs}_\mathbf{l})^*
X^{\rm obs}_\mathbf{l'}(X^{\rm obs}_\mathbf{l'})^* \right\rangle \nonumber\\
\phantom{\Biggl(\int\frac{\dd \varphi_l}{2\pi}\frac{\dd \varphi_{l'}}{2\pi}}
- \left\langle X^{\rm obs}_\mathbf{l}(X^{\rm obs}_\mathbf{l})^* \right\rangle
\left\langle X^{\rm obs}_\mathbf{l'}(X^{\rm obs}_\mathbf{l'})^* \right\rangle\Biggr)B^{-1}_l B^{-1}_{l'}~.
\end{eqnarray}
We can re-introduce (\ref{Tobs}), and using the fact that the
noise is Gaussian and uncorrelated, we show that
\begin{equation}\label{Eq:covarfinalnoise}
\mathrm{Cov}(l,l')=\int\frac{\dd \varphi_l}{2\pi}
\frac{\dd \varphi_{l'}}{2\pi}
\left\langle \widetilde{X}_\mathbf{l}\widetilde{X}_\mathbf{l}^*
\widetilde{X}_\mathbf{l'}\widetilde{X}_\mathbf{l'}^* \right\rangle_{\rm c}
+4\widetilde{C}_l\, w_l^{-1}\,B^{-1}_l +2w_l^{-2}\,B^{-2}_l~,
\end{equation}
where
\begin{eqnarray}
\left\langle \widetilde{X}_\mathbf{l}\widetilde{X}_\mathbf{l}^*\widetilde{X}_\mathbf{l'}
\widetilde{X}_\mathbf{l'}^* \right\rangle_{\rm c}=\left\langle \widetilde{X}_\mathbf{l}
\widetilde{X}_\mathbf{l}^*\widetilde{X}_\mathbf{l'}\widetilde{X}_\mathbf{l'}^* \right\rangle
- \left\langle \widetilde{X}_\mathbf{l}\widetilde{X}_\mathbf{l}^*\right\rangle
\left\langle \widetilde{X}_\mathbf{l'}\widetilde{X}_\mathbf{l'}^*\right\rangle~.
\end{eqnarray}
Note that since the lensed anisotropies are not Gaussian, we
cannot use the Wick theorem to expand the first term of
(\ref{Eq:covarfinalnoise}). Rigorously, it has two
contributions one that is diagonal $\mathcal{D}_{\ell}$ and the
non-diagonal term $\mathcal{N}_{l,l'}$, given by
(\ref{eqcorrel4Tdiag}-\ref{eqcorrel4Tnd}). As a conclusion,
we obtain
\begin{eqnarray}\label{Eq:covarfinalnoise2}
\fl \mathrm{Cov}(l,l')=\delta(l-l')\Bigl[\mathcal{D}_{\ell}
+4\widetilde{C}_l\, \left(w_l B_l\right)^{-1} +2\left(w_l B_l\right)^{-2}\Bigr]+\mathcal{N}_{l,l'}~,
\nonumber \\
\simeq 2\delta(l-l')\left[\widetilde{C}_l+ \left(w_l B_l\right)^{-1}\right]^2 +\mathcal{N}_{l,l'}~.
\end{eqnarray}
This generalizes (\ref{eq:covar:diag:TEB}). Note that the
previous formulae are easily transposed to other polarization
terms of the covariance by introducing a cross-correlated spectrum
estimator $\hat{\widetilde{C}}_l^{TE}$. We obtain generalizations
of (\ref{eq:covar:diag:TEB}-\ref{eq:covar:nodiag:EETE}), each
time introducing the quantities $\mathcal{D}_{\ell}$ and
$\mathcal{N}_{l,l'}$. In the rest of the paper these quantities
are calculated and compared for every polarization terms of the
covariance matrix.

\subsubsection{Numerical evaluation of temperature}
To evaluate and understand better the structure of the covariance
matrix, instead of showing comparisons between numerical
results for (\ref{eqcorrel4Tdiag}-\ref{eqcorrel4Tnd}) for a
few sets of cosmological parameters, we rather show results for some
simple approximations of the power spectrum, with increasing
complexity.

\begin{figure}[hhh]
\begin{centering}
\includegraphics[scale=0.5]{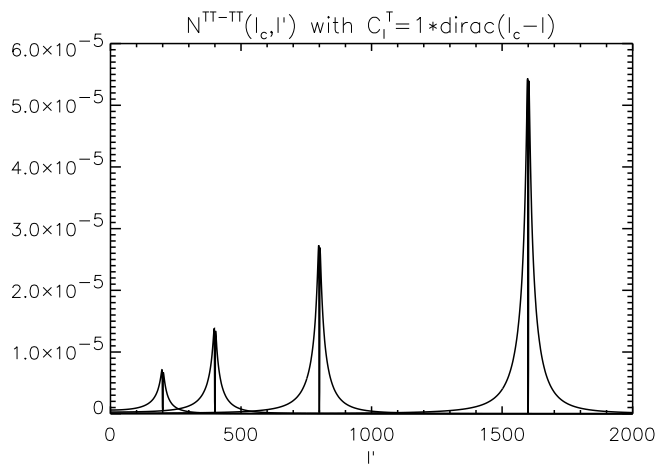}
\par\end{centering}
\caption{Non-diagonal ($l\neq l'$) contribution to the covariance
matrix $\mathcal{N}_{l_{c},l'}^{TT-TT}$ for a Dirac temperature
power spectrum $C_{l}^{TT}=\delta(l_{c}-l)$ and for various values
of $l_{c}$. From left to right, $l_{c}=200$, $400$, $800$, and
$1600$.}\label{Fig:CovarDiracT}
\end{figure}

As a first step, figure~\ref{Fig:CovarDiracT} shows the comparison
when $C_{\ell}=\delta(\ell-\ell_{c})$, and $\ell_{c}=(200,\ 400,\
800,\ 1600)$. This illustrates the effect of lensing on the
covariance matrix for a single mode. The non-diagonal terms are at
least a factor $10^{-4}$ below the diagonal part. We see that the
lensing effect spreads the covariance matrix around the $\ell_{c}$
mode in a symmetric way; the amplitude of the effect grows with
$\ell_{c}$, the coupling between modes being more important at
small scales.

\begin{figure}[hhh]
\begin{centering}\includegraphics[scale=0.5]{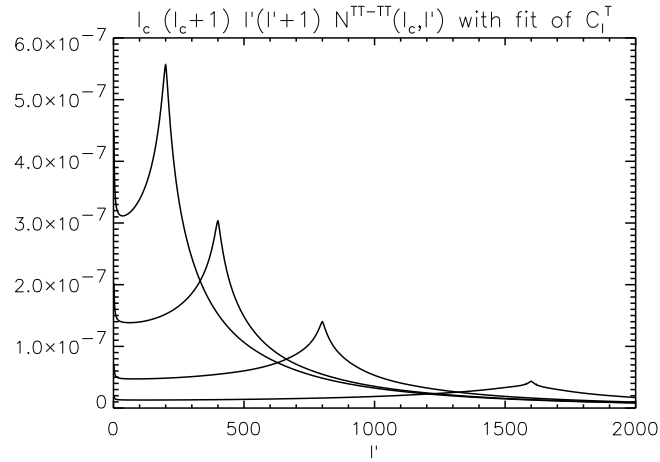}
\par\end{centering}
\caption{Non-diagonal ($l\neq l'$) contribution to the covariance
matrix $\mathcal{N}_{l_{c},l'}^{TT-TT}$ for a temperature power
spectrum approximated by
$C_{l}^{TT}=l^{-2}\,\exp\left(-\frac{l^{2}}{2l_{c}^{2}}\right)$ and
for various values of $l_{c}$. From left to right, $l_{c}=200$,
$400$, $800$, and $1600$.}
\label{Fig:CovarFitT}
\end{figure}

Of course, the temperature power spectra is more complicated than
a Dirac function. In fact, ignoring the acoustic peaks, it is
rather well described by
$C_{\ell}^{TT}\sim\ell^{-2}\,\exp\left(-\frac{\ell^{2}}{2\ell_{c}^{2}}\right)$.
We show in figure~\ref{Fig:CovarFitT} the non-diagonal term for this
approximation. The figure shows
$l(l+1)l^\prime(l^\prime+1)\mathcal{N}_{l l^\prime}$ in order to
see further than the dominant variation due to the $\ell^{-2}$
behavior of the power spectra, as is usually done with the
$C_{\ell}$. The figure exhibits the same features than the
previous one. Namely it spreads over a broad range of $\ell$, but
with a very small amplitude. This spread is more important at
large scales (small $\ell$) than at small scales, which seems to
contradict the idea that weak lensing is essentially a small scale
effect. Of course, this is due to the exponential suppression of
the small scales in the power spectrum. We showed above that the
effect is essentially symmetrical.

Full results for a concordance model power spectrum are showed in
figure~\ref{Fig:Covar}, page \pageref{Fig:Covar}, first column. It
is also illustrated on figure \ref{Fig:CovarRealT} below.
They display the same general features we demonstrated on our two
simple examples. The structure of acoustic peaks (and especially
the first one) complicates somewhat the results. However, the fact
that the non-diagonal contribution is far below the diagonal part
remains true.

\begin{figure}[hhh]
\begin{centering}
\includegraphics[scale=0.5]{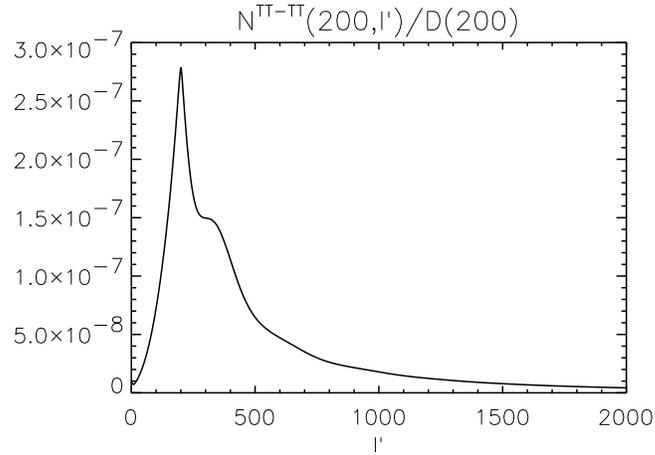}
\par\end{centering}
\caption{Non-diagonal contribution to the covariance
matrix $\mathcal{N}_{200,l'}^{TT-TT}$, normalized by the diagonal
value $\mathcal{D}_{200}^{TT-TT}$, and for the temperature power
spectrum of the concordance model. The cosmological parameters are those
measured by WMAP 3 \cite{2006astro.ph..3449S}.}
\label{Fig:CovarRealT}
\end{figure}

\subsubsection{Polarization terms}
\begin{figure}[hhh]
\begin{centering}\includegraphics[scale=0.5]{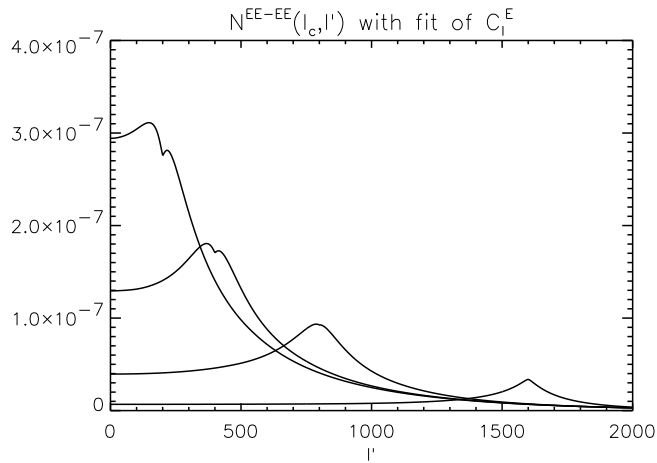}
\par \end{centering}
\caption{Non-diagonal ($l\neq l'$) contribution to the covariance
matrix $\mathcal{N}_{l_{c},l'}^{EE-EE}$ for a E polarization power
spectrum approximated by
$C_{l}^{EE}=\exp\left(-\frac{l^{2}}{2l_{c}^{2}}\right)$ and for
various values of $l_{c}$. From left to right, $l_{c}=200$, $400$,
$800$, and $1600$. }
\label{Fig:CovarFitE}
\end{figure}

We can perform a similar analysis for the $E$-type polarization.
The diagonal and non-diagonal terms read (see \ref{annex2} for a
complete development)
\begin{eqnarray}\label{eq:correl4Ediag}
\fl \mathcal{D}_{\ell}^{EE-EE}=2\left(C_{l}^{EE}\right)^{2}\left(1
-\frac{l^{2}}{2\pi}\sigma_{0}^{2}\right)
 +\frac{4}{(2\pi)^{2}}C_{l}^{EE}\int\dd^{2}\mathbf{l_{1}}C_{|\mathbf{l}
-\mathbf{l_{1}}|}^{\phi\phi}\Bigl[C_{l_{1}}^{EE}\cos^{2}(2\varphi_{1})\nonumber\\
 \phantom{\frac{4}{(2\pi)^{2}}C_{l}^{EE}\int}+C_{l_{1}}^{BB}\sin^{2}
(2\varphi_{1})\Bigr][\mathbf{l_{1}}\cdot(\mathbf{l}-\mathbf{l_{1}})]^{2}\,~,
\end{eqnarray}

\begin{eqnarray}\label{eq:correl4End}
\fl \mathcal{N}_{l,l'}^{EE-EE}= 2\int\frac{\dd\varphi'}{2\pi}
\, C_{|\mathbf{l}-\mathbf{l'}|}^{\phi\phi}\cos^{2}(2\varphi')\Bigl\{
C_{l'}^{EE}\left[\mathbf{l'}\cdot\left(\mathbf{l}-\mathbf{l'}\right)\right]
 +C_{l}^{EE}\left[\mathbf{l}\cdot\left(\mathbf{l'}-\mathbf{l}\right)\right]
\Bigr\}^{2}~.
\end{eqnarray}
Before giving the full numerical result for a concordance model
power spectrum, we again demonstrate the features of the
non-diagonal term on the simplified model
$C_{\ell}^{EE}=\exp\left(-\frac{\ell^{2}}{2\ell_{c}^{2}}\right)$.
figure~\ref{Fig:CovarFitE} shows the corresponding result.

\begin{figure}[hhh]
\begin{centering}
\includegraphics[scale=0.5]{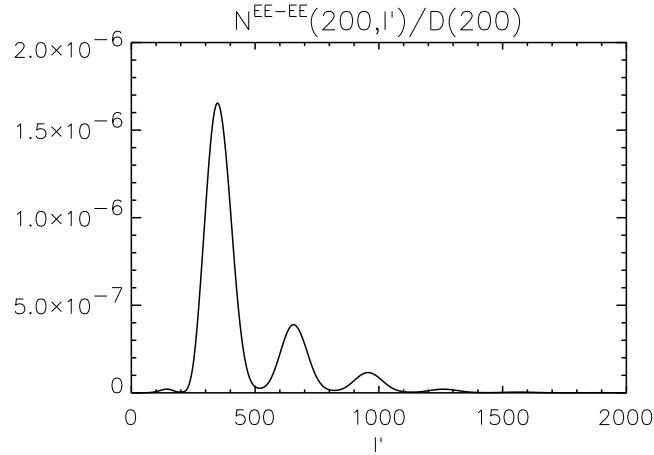}
\par\end{centering}
\caption{Non-diagonal contribution to the covariance
matrix $\mathcal{N}_{200,l'}^{EE-EE}$, normalized by the diagonal
value $\mathcal{D}_{200}^{EE-EE}$, and for the power spectrum of
the concordance model. The cosmological parameters are those
measured by WMAP 3 \cite{2006astro.ph..3449S}.}
\label{Fig:CovarRealE}
\end{figure}

The results for the full concordance model are represented in
figure~\ref{Fig:Covar}, second column. The main idea is
illustrated in figure \ref{Fig:CovarRealE}, where one can
see the effect of the oscillations of the E spectrum and
the sub-dominant amplitude of the non-diagonal contribution.

The terms for the B polarization at second order in lensing are
obtained by replacing E by B and B by E in
(\ref{eq:correl4Ediag}-\ref{eq:correl4End}). Surprisingly,
there seems to be no non-diagonal contributions from the $E$
polarization to the covariance matrix of the $B$ polarization. In
fact, this is an artifact of the truncation of the perturbative
development, which  explains why $\mathcal{N}_{l,l'}^{BB-BB}$ is
so small compared to the Gaussian part (see figure~\ref{Fig:Covar},
third column). Indeed, contributions from the $E$ mode to the $B$
covariance cannot appear at second order (in lensing), they appear
only at fourth order. Moreover, if we restrict ourselves to order
2 for the diagonal parts, there are no contribution proportional
to $(C_l^{EE})^2$, which may be dominant.

It is at first sight surprising that, while doing a perturbative
expansion, a 4th order term may end up being larger than a 2nd
order term. The problem comes from the fact that this development
is in fact done on the $Q$ and $U$ Stokes parameters, where
second order terms are indeed greater than 4th order ones.
However, we are combining these $Q$ and $U$ terms to form the $E$
and $B$ fields in a way that enhances the 4th order terms relatively
to the 2nd order ones in $B$. The most trivial of those fourth
order terms is given by the square of the second order lensed $B$
power spectrum which contributes to the diagonal part of the
covariance
\begin{eqnarray}\label{eq:correl4Bo4diag}
\fl \mathcal{D}_{\ell}^{BB-BB}=\left(C_{l}^{BB}\right)^{2}\left(2
-\frac{l^{2}}{\pi}\sigma_{0}^{2}\right)\nonumber\\
 +\frac{4}{(2\pi)^{2}}C_{l}^{BB}\int\dd^{2}\mathbf{l_{1}}C_{|\mathbf{l}
-\mathbf{l_{1}}|}^{\phi\phi}\Bigl[C_{l_{1}}^{BB}\cos^{2}(2\varphi_{1}) +C_{l_{1}}^{EE}\sin^{2}
(2\varphi_{1})\Bigr][\mathbf{l_{1}}\cdot(\mathbf{l}-\mathbf{l_{1}})]^{2}\nonumber\\
 +\frac{2}{(2\pi)^{4}} \left( \int\dd^{2}\mathbf{l_{1}}C_{|\mathbf{l}
-\mathbf{l_{1}}|}^{\phi\phi}\Bigl[C_{l_{1}}^{BB}\cos^{2}(2\varphi_{1})
+C_{l_{1}}^{EE}\sin^{2}
(2\varphi_{1})\Bigr][\mathbf{l_{1}}\cdot(\mathbf{l}-\mathbf{l_{1}})]^{2}\right)^2
\,~.
\end{eqnarray}
This term, in addition to be the most trivial one, is also the
dominant 4th order diagonal contribution. Indeed it is a
configuration ``1+1+1+1'', and involves therefore a term
containing only E modes (contrary to configurations ``2+2+0+0'',
``3+1+0+0'', or ``4+0+0+0''). Note that, as detailed in
\ref{annex2}, a configuration is called ``1+1+1+1''  when
the four-point correlation function involves four fields at order
1 in lensing, see e.g. the configurations represented in
figure~\ref{Fig:graphTlens4}.

We may think that the new diagonal term dominates the diagonal
part of the covariance matrix, as lower order terms depend on the
$B$ power spectra which is much smaller than the $E$ modes.
However, on the other hand, this term is of order four in lensing
and this effect suppresses the term: it finally represents a
correction of order $0.01\%$ to the order 2.

Let us now turn to the order four non-diagonal contributions. If
we assume, temporarily, that the B modes are negligible compared
to E modes \emph{at the same order in lensing}, we can see that
the dominant non-diagonal 4th order terms involve four first order
E modes from $\widetilde{B}^{(1)}$. All possible terms of the form
``1+1+1+1'' are represented graphically in figure~\ref{Fig:graphTlens4}.
\begin{equation}
\mathcal{N}_{l,l'~(4)}^{BB-BB}=\frac{2}{(2\pi)^2}(A+B+C)~,
\end{equation}
where the three terms are given by
\begin{eqnarray}\label{eq:correl4Bo4term1}
\fl A=\int\frac{\dd\varphi'}{2\pi}\int \dd^2 \mathbf{l_1}
\,\left(C_{l_1}^{EE}\right)^2 C_{|\mathbf{l}-\mathbf{l_1}|}^{\phi\phi}
C_{|\mathbf{l'}-\mathbf{l_1}|}^{\phi\phi}\sin^{2}(2\varphi_1)\sin^{2}(2\varphi'_1)\nonumber\\
\times \left[\mathbf{l_1}\cdot\left(\mathbf{l}-\mathbf{l_1}\right)\right]^2
 \left[\mathbf{l_1}\cdot\left(\mathbf{l'}-\mathbf{l_1}\right)\right]^2~,
\end{eqnarray}
\begin{eqnarray}\label{eq:correl4Bo4term2}
\fl B=\int\frac{\dd\varphi'}{2\pi}\int \dd^2 \mathbf{l_1}\dd^2 \mathbf{l_2}
\,C_{l_1}^{EE}C_{l_2}^{EE} C_{|\mathbf{l}-\mathbf{l_1}|}^{\phi\phi}
C_{|\mathbf{l}-\mathbf{l_2}|}^{\phi\phi}
\sin(2\varphi_1)\sin(2\varphi'_1)\sin(2\varphi_2)\sin(2\varphi'_2)\nonumber\\
\times \left[\mathbf{l_1}\cdot\left(\mathbf{l}-\mathbf{l_1}\right)\right]
\left[\mathbf{l_1}\cdot\left(\mathbf{l}-\mathbf{l_2}\right)\right]
\left[\mathbf{l_2}\cdot\left(\mathbf{l}-\mathbf{l_2}\right)\right]
\left[\mathbf{l_2}\cdot\left(\mathbf{l}-\mathbf{l_1}\right)\right]
\delta(\mathbf{l}+\mathbf{l'}+\mathbf{l_1}-\mathbf{l_2})
\end{eqnarray}
and
\begin{eqnarray}\label{eq:correl4Bo4term3}
\fl C=\int \frac{\dd\varphi'}{2\pi}\int \dd^2 \mathbf{l_1}\dd^2 \mathbf{l_3}
\,C_{l_1}^{EE}C_{l_3}^{EE} \left(C_{|\mathbf{l}-\mathbf{l_1}|}^{\phi\phi}\right)^2
\sin^{2}(2\varphi_1)\sin^{2}(2\varphi'_3)\nonumber\\
\times \left[\mathbf{l_1}\cdot\left(\mathbf{l}-\mathbf{l_1}\right)\right]^2
 \left[\mathbf{l_3}\cdot\left(\mathbf{l}-\mathbf{l_1}\right)\right]^2
\delta(\mathbf{l}+\mathbf{l'}-\mathbf{l_1}-\mathbf{l_3})~.
\end{eqnarray}
In these formulae, $\delta$ is the Dirac distribution and we have used the
following definitions for the angles
\begin{equation}
\varphi_{i}\equiv \varphi_{l_i}-\varphi_l~, \quad \varphi'_{i}\equiv \varphi_{l_i}-\varphi_{l'}~.
\end{equation}
Our results agree with those of \cite{2004PhRvD..70d3002S,Smith:2005ue}.

The TE-TE contributions read,
\begin{eqnarray}
\fl \mathcal{D}_{\ell}^{TE-TE}=\left(C_{l}^{TE}\right)^{2}\left(1-\frac{l^{2}}{2\pi}
  \sigma_{0}^{2}
\right)+C_{l}^{TT}C_{l}^{EE}\left(1-\frac{l^{2}}{2\pi}\sigma_{0}^{2}\right)\nonumber\\
+\frac{1}{(2\pi)^{2}}C_{l}^{EE}\,\int\dd^{2}\mathbf{l_{1}}\,
C_{l_{1}}^{TT}
C_{|\mathbf{l}-\mathbf{l_{1}}|}^{\phi\phi}[\mathbf{l_{1}}\cdot(\mathbf{l}-
\mathbf{l_{1}})]^{2}\nonumber\\
+\frac{1}{(2\pi)^{2}}C_{l}^{TT}\,\int\dd^{2}\mathbf{l_{1}}[\mathbf{l_{1}}\cdot
(\mathbf{l} - \mathbf{l_{1}})]^{2}\,\left[C_{l_{1}}^{EE}\cos^{2}(2\varphi_{1})+C_{l_{1}}^{BB}
\sin^{2}( 2\varphi_{1})\right]C_{|\mathbf{l}-\mathbf{l_{1}}|}^{\phi\phi}\,\nonumber\\
 +\frac{2}{(2\pi)^{2}}C_{l}^{TE}\,\int\dd^{2}\mathbf{l_{1}}\,
 C_{l_{1}}^{TE}
 \cos(2\varphi_{1})C_{|\mathbf{l}-\mathbf{l_{1}}|}^{\phi\phi}[\mathbf{l_{1}} \cdot
(\mathbf{l} - \mathbf{l_{1}})]^{2}~,
\end{eqnarray}
for the diagonal part whereas the non-diagonal part is given by
\begin{eqnarray}
\fl \mathcal{N}_{l,l'}^{TE-TE}=\int\frac{\dd\varphi'}{2\pi}\biggl\{
\left[\left(C_{l'}^{TE}\right)^{2}+C_{l'}^{EE}C_{l'}^{TT}\right]\,\left[\mathbf{l'}
\cdot\left(\mathbf{l}-\mathbf{l'}\right)\right]^{2}\nonumber\\
 +\left[\left(C_{l}^{TE}\right)^{2}+C_{l}^{EE}C_{l}^{TT}\right]\,\left[\mathbf{l}
\cdot\left(\mathbf{l}-\mathbf{l'}\right)\right]^{2}\biggr\}
C_{|\mathbf{l}-\mathbf{l'}|}^{\phi\phi}\cos(2\varphi')\nonumber\\
 +\left[\mathbf{l'}\cdot\left(\mathbf{l}-\mathbf{l'}\right)\right]
\left[\mathbf{l}\cdot\left(\mathbf{l}-\mathbf{l'}\right)\right]\Bigl\{
C_{l}^{TE}C_{l'}^{TE}\left[1+\cos^{2}(2\varphi')\right]\nonumber\\
 \phantom{+2\left[\mathbf{l'}\cdot\left(\mathbf{l}-\mathbf{l'}\right)\right]}
+\left[C_{l}^{EE}C_{l'}^{TT}+C_{l'}^{EE}C_{l}^{TT}\right]\cos(2\varphi')\,\Bigr\}
C_{|\mathbf{l}-\mathbf{l'}|}^{\phi\phi}~.
\end{eqnarray}

As already mentioned, until order 2 in lensing, the covariance
terms involving polarization TB and EB are null. At this order,
there are also no cross-correlations involving BB polarization.
Thus there are no other terms in the diagonal.

We are left with six off-diagonal ($UV\neq XY$) terms. Their
analytical expressions are given in \ref{annex}. We would like
to point out that among these 6 off-diagonal terms three of them
represent corrections to Gaussian terms $\mathcal{D}_{\ell}^{UV-XY}$. But the
three others involve polarization of the form $BB-XY$ with $XY\in \{TT,
EE, TE\}$, polarizations for which the Gaussian terms were null. This
introduces another lensing modification to the
covariance matrix.\\

\subsubsection{Numerical evaluation of the corrections}
Let's now turn to the numerical calculation of the non-gaussian
terms ($\mathcal{N}_{l,l'}^{XY-UV}$) compared when possible to the Gaussian ones
($\mathcal{D}_{\ell}^{XY-UV}$). This is represented in figure~\ref{Fig:Covar},
\ref{Fig:Covar2}, and \ref{Fig:Covar4}. These numerical calculations mainly
extend to all polarizations and to different values of multipoles the results
represented for TT-TT and EE-EE in figures \ref{Fig:CovarRealT} and
\ref{Fig:CovarRealE}. On the figures \ref{Fig:Covar} and \ref{Fig:Covar2}, we
can see that the non-gaussian contributions are completely sub-dominant compared
to the Gaussian contributions, namely of the order of
$10^{-2}\% $ or lower. At the end of \ref{annex}, in table \ref{tableD},
are given the absolute values of the Gaussian contribution to the covariance
matrix, in order to calculate the absolute values of the non-gaussian
corrections represented on figure~\ref{Fig:Covar}, and \ref{Fig:Covar2}.
The term BBBB also receives important non-gaussian contributions from the
order 4 in lensing. It has been shown in \cite{2004PhRvD..70d3002S, Smith:2005ue}
that these corrections are negligible at low multipoles but become
important at higher multipoles ($\ell\gtrsim 800$). Despite the fact that
these terms are of order four in lensing, the fact that they involve the E
spectrum and not the B spectrum dominates and make these terms dominant.
This is confirmed by the semi-analytical approach of A. Lewis~\cite{Lewis:2005tp}.
The lensing also introduce new correlations in the covariance matrix. These
terms are represented on figure~\ref{Fig:Covar4}. The amplitude of these new
terms is clearly sub-dominant compared to any Gaussian term in the matrix.

Our results fully agree with those of
\cite{2006PhRvD..74l3002S, 2006astro.ph..7494L}, although they both
assumed that the primordial B modes are vanishing. However, in our
calculations, we found out that this approximation is not recommended in
the sense that this arbitrarily sets the second order non-diagonal
contributions to the BB-BB covariance to zero. For low multipoles,
these terms represent the leading corrections to the Gaussian assumption.
On the other hand, one should turn to \cite{2006PhRvD..74l3002S, 2006astro.ph..7494L} do
see how the non-gaussian corrections propagate to the errors on cosmological
parameters, and the consequences on the observation strategies.

\section*{Conclusions}

CMB anisotropies and polarization data are a powerful tool to
constrain the cosmological model. Indeed, their statistical
properties can be computed and compared to the actual data. In the
minimal case, when weak lensing is neglected, the theory predicts
that, at dominant order, the anisotropies and polarization must
obey a Gaussian distribution, thus allowing the well known data
compression that reduces the experimental data to a set of power
spectra. This is why most article put a strong emphasis on the
evaluation of the power spectra of the CMB, taking into account or
not secondary effects, and reduce all experimental results to a
set of $C_{\ell}$'s. This can turn out to be a poor approximation;
by doing so, one would ignore any deviation from the Gaussian
behavior that can arise from those secondary effects.

We have computed analytically and numerically the Gaussian part of
the covariance matrix as well as the non-gaussian contributions
due to lensing. This last contribution was usually assumed to be
negligible. We prove that this assumption is justified, for all
polarizations except BB-BB and independently of the sensitivity of the
experiment considered. The error made is always completely
sub-dominant, of the order of $0.01\%$ or lower. The covariance
matrix can thus be computed using the Gaussian assumption, as
described in sec. \ref{methodestd} : the covariance matrix can be
computed by assuming that lensed $C_{l}$'s are Gaussian and by
using the unlensed
formulae~(\ref{eq:covar:diag:TEB}-\ref{eq:covar:nodiag:EETE}), but
for the replacement of the $C_{l}$'s by the lensed spectra
$\widetilde{C}_{l}$'s. We can see with
\ref{Eq:covarfinalnoise2} that our conclusions remain valid
independently of the considered experiment, even the most
sensitive one. These results confirm the recently published
\cite{2006PhRvD..74l3002S, 2006astro.ph..7494L} where the lensing effect on
covariance has been studied as well.

The case of BB-BB polarization requires a more extended expansion in
lensing, until order four, in order to take into account all dominant
effects in non-gaussian corrections. Indeed, at order four, terms where
only E modes contribute to the covariance matrix can exist. They have
been found \cite{2004PhRvD..70d3002S} (see also \cite{Smith:2005ue})
numerically dominant over Gaussian terms only when considering
multipoles higher than $\ell \gtrsim 800$.

The weak lensing has another effect on the covariance matrix. It
introduces new correlations between BB and TT, EE, and
TE spectra. Their amplitude is strongly sub-dominant compared to any
Gaussian contribution to the covariance matrix.\\

\ack

It is a pleasure to thank S. Prunet, J. Lesgourgues, and L. Perotto
for stimulating discussions. The work of J.R. was partially supported
by the National Science Foundation under Grant No. PHY-0455649, and by
the University of Texas at San Antonio, Texas 78249, USA.


\appendix

\section{Graphical representation of different contributions to the
  four-point correlation function}

\label{annex2}

The covariance matrix is proportional to the connected part of the
four-point correlation function of the lensed anisotropy fields, defined
by\footnote{This definition follows the one used in previous works, for e.g.
\cite{Zaldarriaga:1996xe, 2006PhRvD..74l3002S, 2004PhRvD..70d3002S}.}
\begin{equation}
\left\langle \widetilde{X}_{l}\widetilde{Y}_{l}^{*}\widetilde{U}_{l'}
\widetilde{V}_{l'}^{*}\right\rangle_C\equiv
\left\langle \widetilde{X}_{l}\widetilde{Y}_{l}^{*}\widetilde{U}_{l'}
  \widetilde{V}_{l'}^{*}\right\rangle -\left\langle
  \widetilde{X}_{l}\widetilde{Y}_{l}^{*}\right\rangle \left\langle
  \widetilde{U}_{l'}\widetilde{V}_{l'}^{*}\right\rangle ~.
\label{eq:rappeldefcovar}
\end{equation}
We have chosen the same Fourier transform conventions as W. Hu
\cite{2000PhRvD..62d3007H,2001PhRvD..64h3005H},
in which,
\begin{equation}\label{defspectredeuxpoints}
\langle X_{l_1}Y_{l_2}^{*}\rangle=(2\pi)^{2}\delta(l_1-l_2)C_{l_1}^{XY}
=(2\pi)^{2}\delta(l_1-l_2)C_{l_2}^{XY}~,
\end{equation}
for $X,Y=\phi$, $T$, $E$, $B$ and the four-point correlation
function may be written in order to define a diagonal
$\mathcal{D}_{l}^{XY-UV}$ and a non-diagonal part
$\mathcal{N}_{l,l'}^{XY-UV}$ as
\begin{equation}\label{deffunccorrelcovar}
\left\langle \widetilde{X}_{l}\widetilde{Y}_{l}^{*}\widetilde{U}_{l'}
\widetilde{V}_{l'}^{*}\right\rangle_C =(2\pi)^{4}\delta(l-l')
\mathcal{D}_{l}^{XY-UV}  +(2\pi)^{2}\mathcal{N}_{l,l'}^{XY-UV}~.
\end{equation}

For simplicity, we will illustrate our method of calculation on
the $XY-UV=TT-TT$ polarization term in the covariance matrix which
will clarify the various contributions to the four-point
correlation function. The generalization to other polarization
terms is straightforward. The method we describe here can be
extended to all orders in lensing and all $n-$point correlation
functions.

\subsection{General method}
Listing and calculating all contributions can be done graphically
\cite{Bernardeau:2001qr}: the task is then simpler and more
efficient. For this, we can use the analogy with Feynman
graphs. We will focus on the four-point correlation function of the
temperature, up to order 2 in lensing. Let us represent by a cross
$\times$ a field $T(\mathbb{\ell})$ and by a circle
$\circ$ the lensing potential $\phi$. The term we calculate contains
two fields $T$ at the multipole $\mathbf{l}$ and two at the
multipole $\mathbf{l}'$. We must remember that in general, each
field $\widetilde{T}_{l}$ is the sum of lensed temperature field
lensed at each order
\begin{equation}
\widetilde{T}_{l}=T_{l}+\sum_{i=1}^{\infty}\widetilde{T}_{l}^{(i)}~,
\end{equation}
where the first $\widetilde{T}^{(i)}$ are given in
(\ref{eq:L:ordre1}-\ref{eq:B:ordre2}). This development can be
represented graphically as
\begin{equation}
\widetilde{T}_{l}=\times ~~+~~ \times \circ ~~+~~ \times \circ\circ ~~+~~ \cdots ~,
\end{equation}

As a consequence, to consider all contributions we can proceed order
by order in lensing. Once the order is set (say to $n$), all possible
combinations of the four fields $\widetilde{T}$ are written such that
the sum of the four orders is equal to $n$. All ways of correlating
them (represented by a solid line) give all the graphs contributing
to the order $n$. Considering that we calculate the covariance
matrix given in (\ref{eq:rappeldefcovar}), among the fields
$\widetilde{T}$, two must be at multipole $l$, two at multipole $l'$
and one field at each multipole carry a $*$ for the complex conjugate.
Then we need to use the the following rules (for temperature) derived
from (\ref{eq:L:ordre1}-\ref{eq:B:ordre2}),
\begin{eqnarray}\label{rulescorrelations}
\times = T ~,\nonumber\\
\times \circ = -\int\frac{\mathrm{d}^{2}\mathbf{l_{1}}}{(2\pi)^{2}}\,
T(\mathbf{l_{1}})\,\phi(\mathbf{l}-\mathbf{l_{1}})\,
[\mathbf{l_{1}}\cdot(\mathbf{l}-\mathbf{l_{1}})]~,\\
\times \circ\circ = \frac{1}{2}\int
\frac{d^{2}\mathbf{l_{1}}d^{2}\mathbf{l_{2}}}{(2\pi)^{4}}\,
T(\mathbf{l_{1}})\,\phi(\mathbf{l_{2}})\phi(\mathbf{l}
-\mathbf{l_{1}}-\mathbf{l_{2}})(\mathbf{l_{1}}\cdot\mathbf{l_{2}})[\mathbf{l_{1}}
\cdot(\mathbf{l}-\mathbf{l_{1}}-\mathbf{l_{2}})]\nonumber ~,
\end{eqnarray}
where $\sigma_{0}$ is given by (\ref{eq:defR}).
Finally, we use (\ref{defspectredeuxpoints}) to calculate the
contribution from the graph to the four point correlation function. Two
classes of terms will appear : the diagonal ones proportional to
$\delta(l-l')$ that contribute to $\mathcal{D}_{l}^{XY-UV}$ and the
others that contribute to $\mathcal{N}_{l,l'}^{XY-UV}$ [Note the factors
$(2\pi)$ due to the definition (\ref{deffunccorrelcovar})]. To calculate
the contribution to the covariance matrix, we need to introduce the
integration over the angles of $\mathbf{l}$ and $\mathbf{l'}$,
$$\int\frac{\dd \varphi}{2\pi}\int\frac{\dd \varphi'}{2\pi}~,$$
equivalent to the summation over $m$ and $m'$ in the full sky calculation.

\subsection{Temperature at order 0}
\begin{figure}[hhh]
\begin{centering}\includegraphics[scale=0.4]{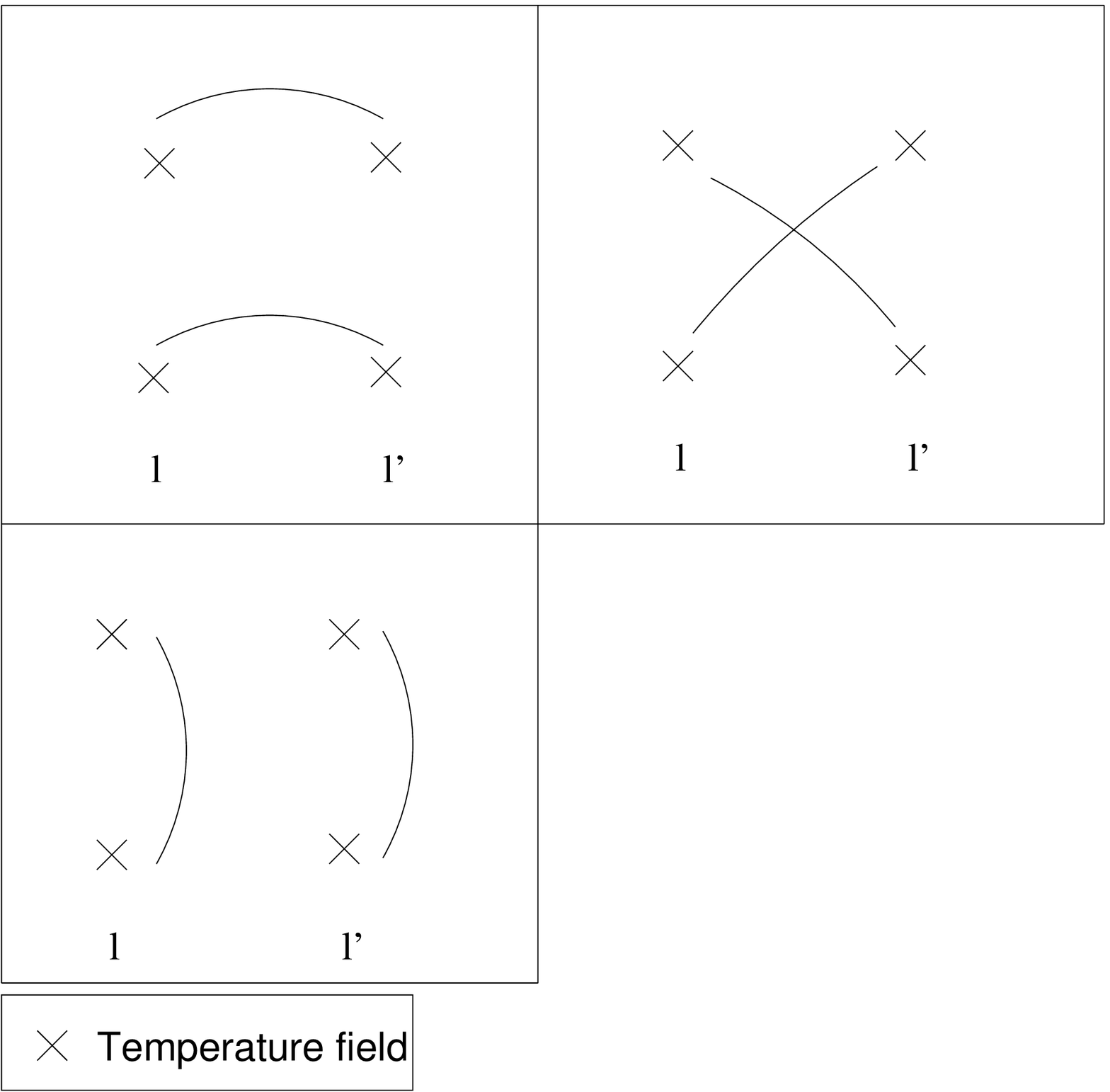}
\par\end{centering}
\caption{Configurations of correlations contributing to the order 0 of the
covariance matrix.}
\label{Fig:graphTnolens}
\end{figure}

At order 0, the graphs will not involve any $\circ$; they are given
in figure~\ref{Fig:graphTnolens}. We arbitrarily chose
to put fields at multipole $l$ in the left part of the graph and
fields at multipole $l'$ in the right part. We note that the two
graphs on the first row do connect the different multipoles: they
are called connected graphs. The graph on the second row is said
not connected and will not contribute to the covariance matrix.

The expression for a graph, for instance the first one of
figure~\ref{Fig:graphTnolens}, is simply the product of the two correlations
\begin{equation}
\langle T_{l}T_{l'}^{*}\rangle\langle T_{l}^{*}T_{l'}\rangle
=\left[(2\pi)^{2}\delta(l-l')C_{l}^{TT}\right]^{2}~.
\end{equation}
This graph will contribute to  $\mathcal{D}_{\ell}$. We can check that the
second graph of the first row gives the same contribution and thus
the total contribution to the covariance matrix of graphs of order
0 is twice the previous expression. This yields the very first
term in (\ref{eqcorrel4Tdiag}).

\subsection{Temperature at higher order}
The contributions of order one (as well as all other odd orders) are
vanishing since we are considering fields $T$ and $\phi$ that are
uncorrelated and of vanishing mean value
\begin{equation}
\langle T_{l}\phi_{l'}\rangle=\langle T_{l}\rangle=\langle\phi_{l}\rangle=0~.
\end{equation}
Thus, in graphs with only one $\circ$, the lensing field cannot
correlate to any other fields and one gets a term proportional to
$\langle\phi\rangle$ which is null. For the same reason, all
odd-point correlation functions of the CMB are null.

\begin{figure}[hhh]
\begin{centering}\includegraphics[scale=0.4]{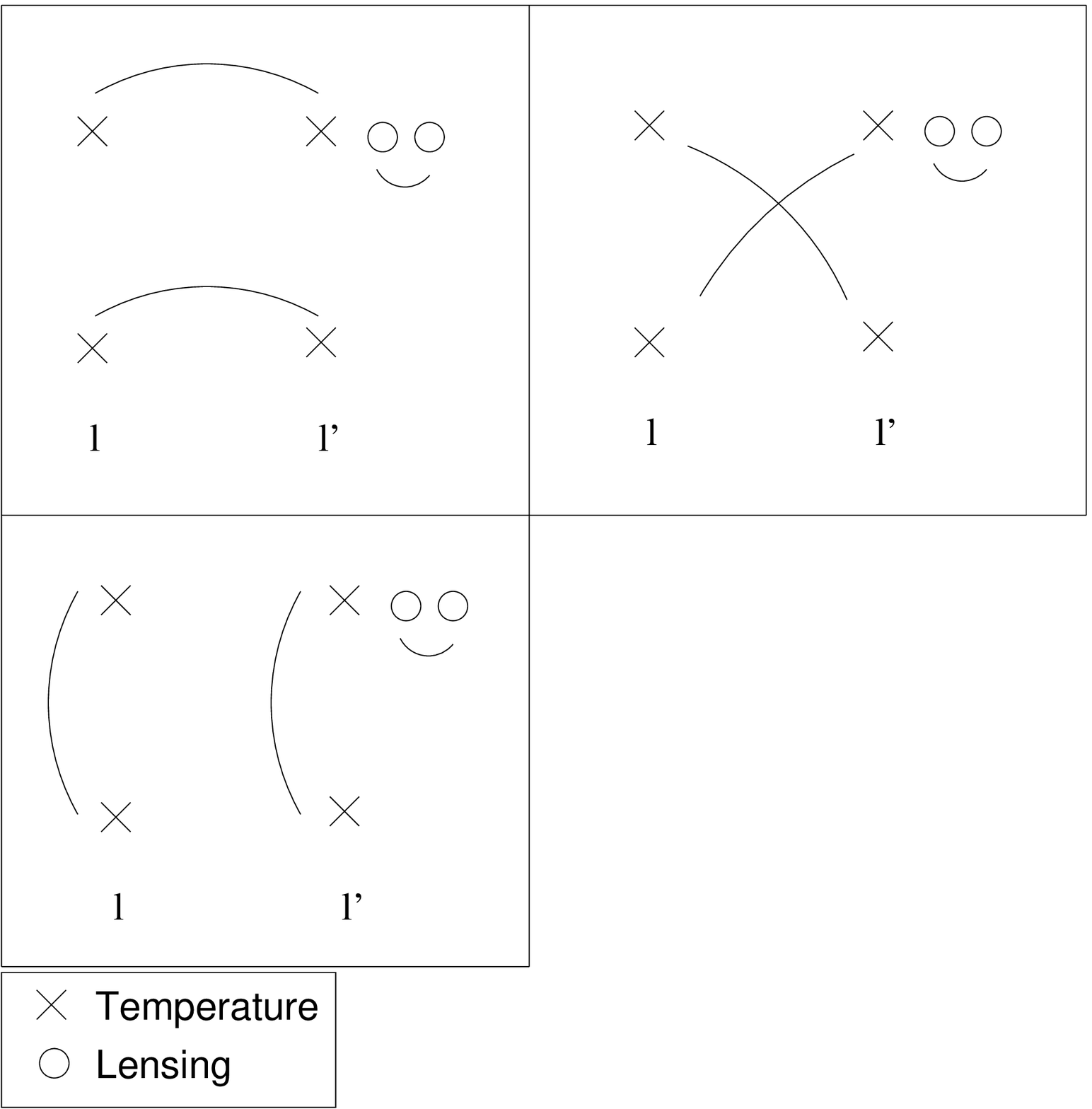}
\par\end{centering}
\caption{First set of configurations of correlations contributing
to the order 2 involving three unlensed temperature fields. More
precisely, they contribute to the order ``0+0+0+2''.}
\label{Fig:graphTlens1}
\end{figure}

Let us turn to the order 2 in lensing. Graphically, we need now to
add two $\circ$ in all possible configurations (each field
$\widetilde{T}$ can be expanded until order two). The easiest
possibilities are to consider three unlensed temperature fields
and one order 2 field. We will call this first class of graphs, the
order ``0+0+0+2''. Then, similarly to the figure~\ref{Fig:graphTnolens},
there are three ways to correlate these configurations since the
two lensing fields must correlate together. For a chosen lensed
field, the contributions are given in figure~\ref{Fig:graphTlens1}.
The nine other contributions to the same order are identical but
with permutations of the lensing fields.

As above, the configuration of the second row does not contribute
to the covariance matrix. To calculate a term, we simply read which
fields are correlated and use (\ref{rulescorrelations}). For example,
the first term reads
\begin{eqnarray}
\fl  \langle\widetilde{T}_{l}^{(2)}T_{l'}^{*}\rangle\langle T_{l}^{*}T_{l'}
\rangle=\left[(2\pi)^{2}\delta(l-l')C_{l}^{TT}\right]
\Biggl[\frac{1}{2}\frac{1}{(2\pi)^{4}}\int\dd^{2}\mathbf{l_{1}}
\dd^{2}\mathbf{l_{2}}\langle T_{l}^{*}T_{l_{1}}\rangle\langle\phi
(\mathbf{l_{2}})\phi(\mathbf{l'-l_{1}-l_{2}})\rangle\nonumber\\
  \times(\mathbf{l_{1}\cdot l_{2}})\left[\mathbf{l_{1}\cdot(l'-l_{1}-l_{2}})
\right]\Biggr]\\
  =-\frac{1}{2}(2\pi)^{2}\delta(l-l')\left(C_{l}^{TT}\right)^{2}
\frac{l^{2}}{4\pi}\sigma_{0}^{2}~,
\label{eq:term0011}
\end{eqnarray}
We can equivalently place the two lensing
fields on all four temperature fields, and we can check that the
second graph of first row of figure~\ref{Fig:graphTlens1} give the
same contribution. Thus, the total contribution to the covariance
matrix of graphs of order ``0+0+0+2'' is 8 times the previous
expression of (\ref{eq:term0011}). Finally, this term contributes
to $\mathcal{D}_{l}$ and gives the second (negative) term of
(\ref{eqcorrel4Tdiag}).\\

\begin{figure}[hhh]
\begin{centering}\includegraphics[scale=0.4]{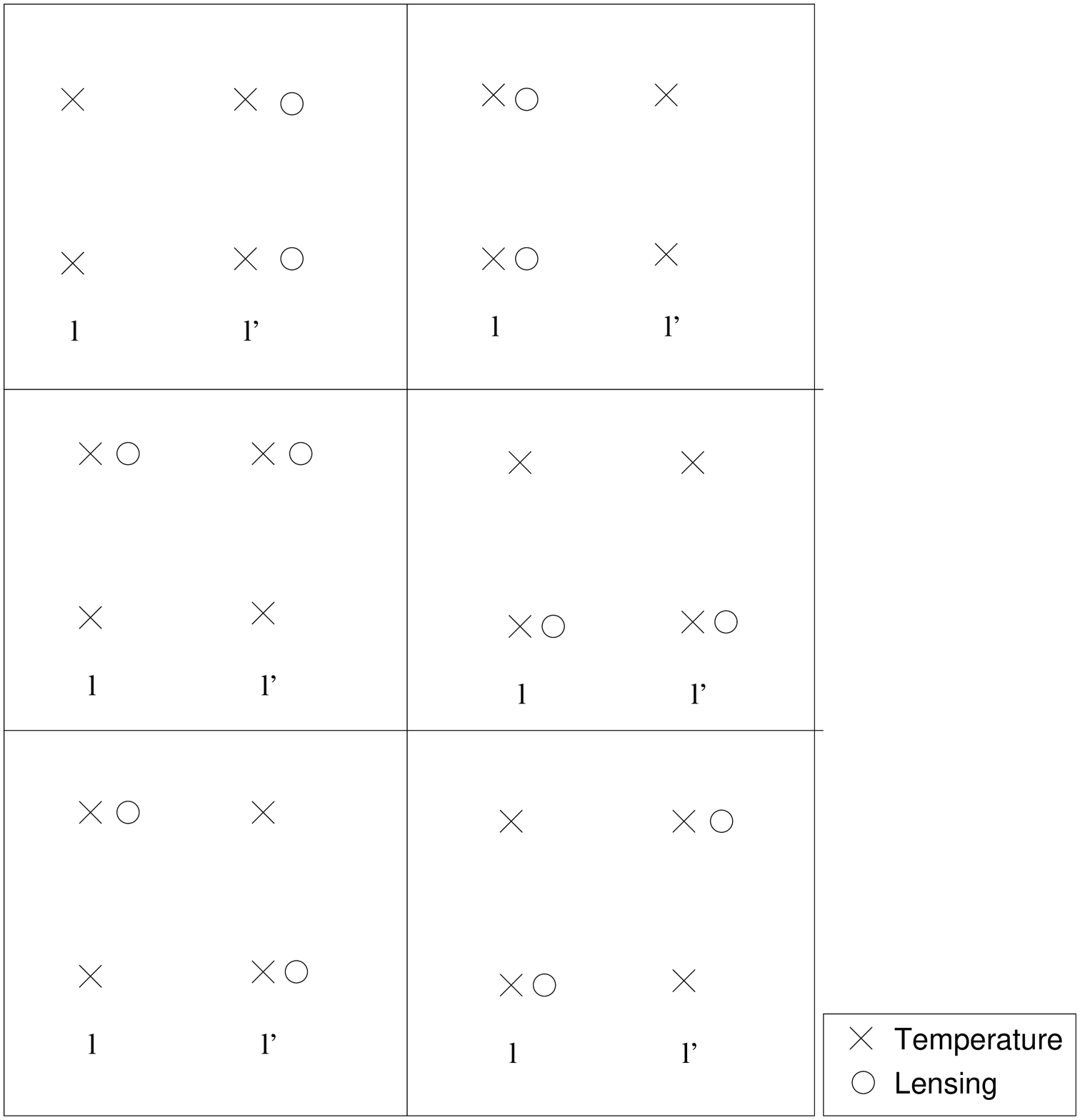}
\par\end{centering}
\caption{All possible configurations of temperature and lensing
fields that contribute to the order ``0+0+1+1''.}
\label{Fig:graphTlens2}
\end{figure}

Still restricting to the order 2 in lensing, we are now left with
all the possibilities involving two temperature fields of order 1
and two unlensed fields. These graphs will be called the order
``0+0+1+1''. To obtain all possible terms, we write graphs for all
possible way of placing the lensing fields $\circ$ : there are 6
possibilities, represented on figure~\ref{Fig:graphTlens2}.

Then for each graph, we write all possible ways of correlating the
fields: there are still three possible ways. In total, there are
18 graphs of order ``0+0+1+1'', two of them being graphs that do
not contribute to the covariance matrix (unconnected graphs). Four
of them are ``not topologically connected'' which means that they
correlate the two multipoles but can be separated into two
different two-point correlation functions: these graphs then
contribute to $\mathcal{D}_{l}^{TT-TT}$. They give the third and last term of
(\ref{eqcorrel4Tdiag}).

\begin{figure}[hhh]
\begin{centering}
\includegraphics[scale=0.4]{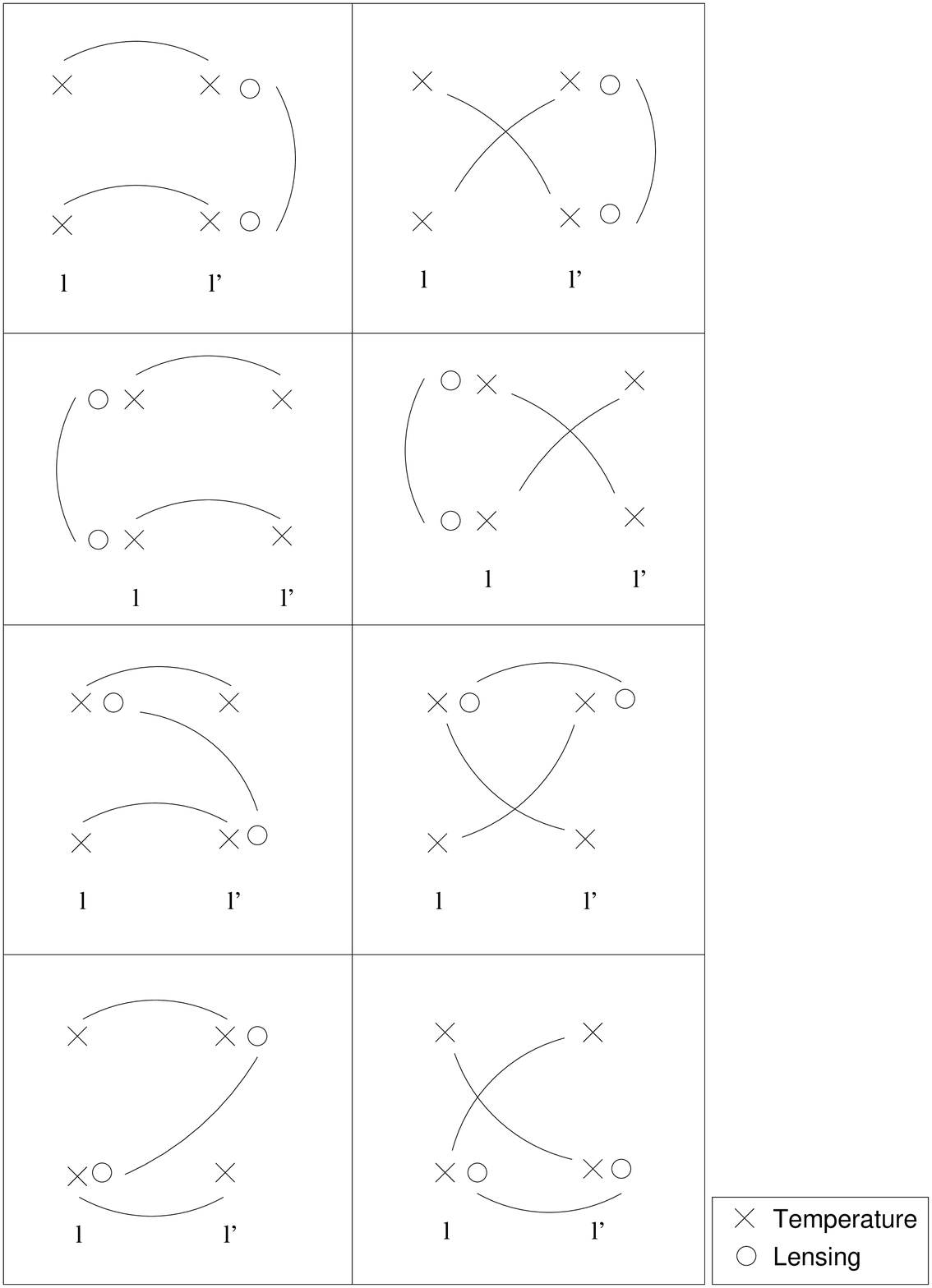}
\par\end{centering}
\caption{Non vanishing, ``topologically connected'',
configurations of correlations contributing to the order
``0+0+1+1''. These graphs are all the contributions to
$\mathcal{N}_{l,l'}^{TT-TT}$.}
\label{Fig:graphTlens3}
\end{figure}

We are now left with 12 graphs that are either ``connected'' or
``topologically connected''. These contributions will give new
graphs that \emph{cannot} be obtained via the product of lensed
$C_{l}$s. Four of them are vanishing. The eight other graphs are
represented in figure~\ref{Fig:graphTlens3}.

The two graphs of the first (resp. second) row give the
contributions to $\mathcal{N}_{l,l'}^{TT-TT}$ proportional to
$(C_{l}^{TT})^{2}$ (resp. $(C_{l'}^{TT})^{2}$) in
(\ref{eqcorrel4Tnd}). The last two rows give the contributions
proportional to $C_{l}^{TT}C_{l'}^{TT}$.

\subsection{Rules for the polarization}
Let us finish this section with the few explanation for the polarization
case. The method explained at the beginning of \ref{annex2} are still
valid except that the rules are slightly modified. If $+$ (resp. $-$)
denotes the primordial $E$ (resp. $B$) field, the rules for temperature
are replaced by
\begin{equation}
\widetilde{E}_{l}= +_E ~~+~~ +_E\circ ~~+~~ -_E\circ ~~+~~ +_E \circ\circ ~~+~~ -_E \circ\circ ~~+~~ \cdots ~,
\end{equation}
where for example $+_E\circ$ (resp. $-_E\circ$) denotes the E (resp. B)
modes contributing to $\widetilde{E}$. Once the graphs are written,
their contributions can be calculated using the following rules, derived
from (\ref{eq:E:ordre1}-\ref{eq:B:ordre2})
\begin{eqnarray}\label{rulescorrelationspolarE}
\fl +_E = E ~,\nonumber\\
\fl +_E \circ = -\int\frac{\mathrm{d}^{2}\mathbf{l_{1}}}{(2\pi)^{2}}\,
E(\mathbf{l_{1}})\cos(2\varphi_1)\,\phi(\mathbf{l}-\mathbf{l_{1}})\,
[\mathbf{l_{1}}\cdot(\mathbf{l}-\mathbf{l_{1}})]~,\nonumber\\
\fl -_E \circ = \int\frac{\mathrm{d}^{2}\mathbf{l_{1}}}{(2\pi)^{2}}\,
B(\mathbf{l_{1}})\sin(2\varphi_1)\,\phi(\mathbf{l}-\mathbf{l_{1}})\,
[\mathbf{l_{1}}\cdot(\mathbf{l}-\mathbf{l_{1}})]~,\\
\fl +_E \circ\circ = \frac{1}{2}\int
\frac{d^{2}\mathbf{l_{1}}d^{2}\mathbf{l_{2}}}{(2\pi)^{4}}\,
E(\mathbf{l_{1}})\cos(2\varphi_1)\,\phi(\mathbf{l_{2}})\phi(\mathbf{l}
-\mathbf{l_{1}}-\mathbf{l_{2}})(\mathbf{l_{1}}\cdot\mathbf{l_{2}})[\mathbf{l_{1}}
\cdot(\mathbf{l}-\mathbf{l_{1}}-\mathbf{l_{2}})] ~,\nonumber\\
\fl -_E \circ\circ = -\frac{1}{2}\int
\frac{d^{2}\mathbf{l_{1}}d^{2}\mathbf{l_{2}}}{(2\pi)^{4}}\,
B(\mathbf{l_{1}})\sin(2\varphi_1)\,\phi(\mathbf{l_{2}})\phi(\mathbf{l}
-\mathbf{l_{1}}-\mathbf{l_{2}})(\mathbf{l_{1}}\cdot\mathbf{l_{2}})[\mathbf{l_{1}}
\cdot(\mathbf{l}-\mathbf{l_{1}}-\mathbf{l_{2}})] ~.\nonumber
\end{eqnarray}
Using these rules, all the procedure is strictly identical to the temperature
case. For example, the graphs contributing to the polarized terms of the
covariance are given by the same diagrams given in figures
\ref{Fig:graphTnolens}, \ref{Fig:graphTlens1} and \ref{Fig:graphTlens3}. The only difference is
that each field contain now 2 terms from the E modes and the B modes. However
a lot of terms are vanishing due to the fact that
\begin{equation}
\langle T_{l_1}B_{l_2}^{*}\rangle=\langle E_{l_1}B_{l_2}^{*}\rangle = 0~.
\end{equation}
We can write the same expressions for the expansion of $\widetilde{B}_{l}$, by
replacing E by B and B by E
\begin{equation}
\widetilde{B}_{l}= -_B ~~+~~ -_B\circ ~~+~~ +_B\circ ~~+~~ -_B \circ\circ ~~+~~ +_B
\circ\circ ~~+~~ \cdots ~,
\end{equation}
At every order, clearly the main contribution will come from the the E
modes denoted $+_B \circ$, $+_B \circ\circ \dots $. This diagrammatic
representation is employed in section \ref{annexsec2} for the
calculation of contributions to the covariance BB-BB at order 4 in lensing.

\section{Full expressions for the covariance matrix}\label{annex}

\subsection{Cross-correlated terms of the covariance matrix at order two}

We give here the expressions of the off-diagonal terms of the covariance
matrix. Non-gaussian contributions $\mathcal{N}_{l,l'}^{UV-XY}$ are
numerically evaluated and represented on figure~\ref{Fig:Covar2}. They
are compared to the Gaussian part $\mathcal{D}_{l}^{UV-XY}$, which is usually
taken into account.

\begin{eqnarray}
\fl   \mathcal{D}_{\ell}^{TT-EE}=2\left(C_{l}^{TE}\right)^{2}\left(1-\frac{l^{2}}{2\pi}
 \sigma_{0}^{2}\right) \nonumber\\
  +\frac{4}{(2\pi)^{2}}C_{l}^{TE}\,\int\dd^{2}\mathbf{l_{1}}\,
 C_{l_{1}}^{TE}
 \cos(2\varphi_{1})C_{|\mathbf{l}-\mathbf{l_{1}}|}^{\phi\phi}[\mathbf{l_{1}}
 \cdot(\mathbf{l}-\mathbf{l_{1}})]^{2}~,
\end{eqnarray}

\begin{eqnarray}
\fl
  \mathcal{N}_{l,l'}^{TT-EE}=2\int\frac{\dd\varphi'}{2\pi}
  \biggl\{\left(C_{l'}^{TE}
  \right)^{2}\left[\mathbf{l'}\cdot\left(\mathbf{l}-
      \mathbf{l'}\right)\right]^{2}
 +\left(C_{l}^{TE}\right)^{2}\left[\mathbf{l}\cdot\left(\mathbf{l'}-\mathbf{l}
\right)\right]^{2}\cos^{2}(2\varphi')\nonumber\\
 +2C_{l}^{TE}C_{l'}^{TE}\cos(2\varphi')\left[\mathbf{l'}\cdot\left(\mathbf{l}-
 \mathbf{l'}\right)\right]\left[\mathbf{l}\cdot\left(\mathbf{l'}-\mathbf{l}
 \right) \right]\biggr\} C_{|\mathbf{l}-\mathbf{l'}|}^{\phi\phi}~.
\end{eqnarray}

\begin{eqnarray}
\fl   \mathcal{D}_{\ell}^{TT-TE}=2\, C_{l}^{TT}C_{l}^{TE}
\left(1-\frac{l^{2}}{2\pi}\sigma_{0}^{2}\right)
  +\frac{2}{(2\pi)^{2}}C_{l}^{TE}\,\int\dd^{2}\mathbf{l_{1}}\,
 C_{l_{1}}^{TT}C_{|\mathbf{l}-\mathbf{l_{1}}|}^{\phi\phi}[\mathbf{l_{1}}\cdot(\mathbf{l}-
 \mathbf{l_{1}})]^{2}\nonumber\\
  +\frac{2}{(2\pi)^{2}}C_{l}^{TT}\,\int\dd^{2}\mathbf{l_{1}}\,
 C_{l_{1}}^{TE}
 \cos(2\varphi_{1})C_{|\mathbf{l}-\mathbf{l_{1}}|}^{\phi\phi}[\mathbf{l_{1}}
 \cdot(\mathbf{l}-\mathbf{l_{1}})]^{2}~,
\end{eqnarray}

\begin{eqnarray}
 \fl
  \mathcal{N}_{l,l'}^{TT-TE}=2\int\frac{\dd\varphi'}{2\pi}
 \Bigl\{ C_{l'}^{TT}C_{l'}^{TE}\left[\mathbf{l'}\cdot\left(\mathbf{l}-
 \mathbf{l'}\right)\right]^{2}
 +C_{l}^{TT}C_{l}^{TE}\cos(2\varphi')\left[\mathbf{l}\cdot\left(\mathbf{l'}-
 \mathbf{l}\right)\right]^{2}\nonumber\\
 +\left[C_{l}^{TT}C_{l'}^{TE}+C_{l'}^{TT}C_{l}^{TE}\cos(2\varphi')\right]
\left[\mathbf{l'}\cdot\left(\mathbf{l}-\mathbf{l'}\right)\right]
\left[\mathbf{l}\cdot\left(\mathbf{l'}-\mathbf{l}\right)\right]\Bigr\}
C_{|\mathbf{l}-\mathbf{l'}|}^{\phi\phi}~.
\end{eqnarray}

The last polarization terms are given by
\begin{eqnarray}
\fl   \mathcal{D}_{\ell}^{EE-TE}=2\, C_{l}^{TE}C_{l}^{EE}
\left(1-\frac{l^{2}}{2\pi}\sigma_{0}^{2}\right)
 +\frac{2}{(2\pi)^{2}}C_{l}^{TE}\,\int\dd^{2}\mathbf{l_{1}}\,\Bigl[C_{l_{1}}^{EE}
\cos^{2}(2\varphi_{1})\nonumber\\
 \phantom{+\frac{2}{(2\pi)^{2}}C_{l}^{TE}}+C_{l_{1}}^{BB}\sin^{2}(2\varphi_{1})
\Bigr]C_{|\mathbf{l}-\mathbf{l_{1}}|}^{\phi\phi}[\mathbf{l_{1}}\cdot(\mathbf{l}-
\mathbf{l_{1}})]^{2}\nonumber\\
  +\frac{2}{(2\pi)^{2}}C_{l}^{EE}\,\int\dd^{2}\mathbf{l_{1}}\,
 C_{l_{1}}^{TE}
\cos(2\varphi_{1})C_{|\mathbf{l}-\mathbf{l_{1}}|}^{\phi\phi}[\mathbf{l_{1}}\cdot
(\mathbf{l}-\mathbf{l_{1}})]^{2}~,
\end{eqnarray}

\begin{eqnarray}
\fl  \mathcal{N}_{l,l'}^{EE-TE}=2\int\frac{\dd\varphi'}{2\pi}
\Bigl\{
C_{l'}^{EE}C_{l'}^{TE}\cos^{2}(2\varphi')\left[\mathbf{l'}\cdot
\left(\mathbf{l}-\mathbf{l'}\right)\right]^{2}\nonumber\\
 +C_{l}^{EE}C_{l}^{TE}\cos(2\varphi')\left[\mathbf{l}\cdot\left(\mathbf{l'}-
\mathbf{l}\right)\right]^{2}\nonumber\\
 +\left[C_{l}^{EE}C_{l'}^{TE}\cos^{2}(2\varphi')+C_{l'}^{EE}C_{l}^{TE}\cos(2\varphi')
\right]\nonumber\\
 \times\left[\mathbf{l'}\cdot\left(\mathbf{l}-\mathbf{l'}\right)\right]
\left[\mathbf{l}\cdot\left(\mathbf{l'}-\mathbf{l}\right)\right]\Bigr\}
C_{|\mathbf{l}-\mathbf{l'}|}^{\phi\phi}~.
\end{eqnarray}

There are finally terms that are vanishing if we restrict
ourselves to Gaussian terms only, involving cross terms with the B
polarized spectrum. They are given by
\begin{equation}
\mathcal{N}_{\mathbf{l},\mathbf{l'}}^{BB-TT}= 2\int\frac{\dd
\varphi'}{2\pi} C_{|\mathbf{l}-\mathbf{l'}|}^{\phi\phi}
\sin^2(2\varphi')C_{l '}^{TE}C_{l '}^{EE} \left[\mathbf{l'} \cdot
\left(\mathbf{l}-\mathbf{l'} \right)\right]^2~.
\end{equation}

\begin{eqnarray}
\fl \mathcal{N}_{l,l'}^{BB-EE}=2\int\frac{\dd \varphi'}{2\pi}
C_{|\mathbf{l}-\mathbf{l'}|}^{\phi\phi} \Bigl\{
C_{l '}^{BB} \sin(2\varphi') \left[\mathbf{l'} \cdot \left(\mathbf{l}-\mathbf{l'} \right)\right]
+ C_l^{EE} \sin(2\varphi') \left[
\mathbf{l}\cdot\left(\mathbf{l'}-\mathbf{l}\right)\right]\Bigr\}^2~.
\end{eqnarray}

\begin{eqnarray}
\fl \mathcal{N}_{\mathbf{l},\mathbf{l'}}^{BB-TE}= 2\int\frac{\dd
\varphi'}{2\pi} C_{|\mathbf{l}-\mathbf{l'}|}^{\phi\phi}
\sin^2(2\varphi')\Bigl\{ C_{l '}^{TE}C^{EE}_{l '}\left[\mathbf{l'}
\cdot \left(\mathbf{l}-\mathbf{l'} \right)\right]^2 \nonumber\\
+ C_{l '}^{TE}C^{BB}_{l } \left[
\mathbf{l}\cdot\left(\mathbf{l'}-\mathbf{l} \right)\right] \left[
\mathbf{l'}\cdot\left(\mathbf{l}-\mathbf{l'}
\right)\right]\Bigr\}\, ~.
\end{eqnarray}

\subsection{Dominant contributions to BB-BB at order four}\label{annexsec2}
\begin{figure}[hhh]
\begin{centering}
\includegraphics[scale=0.4]{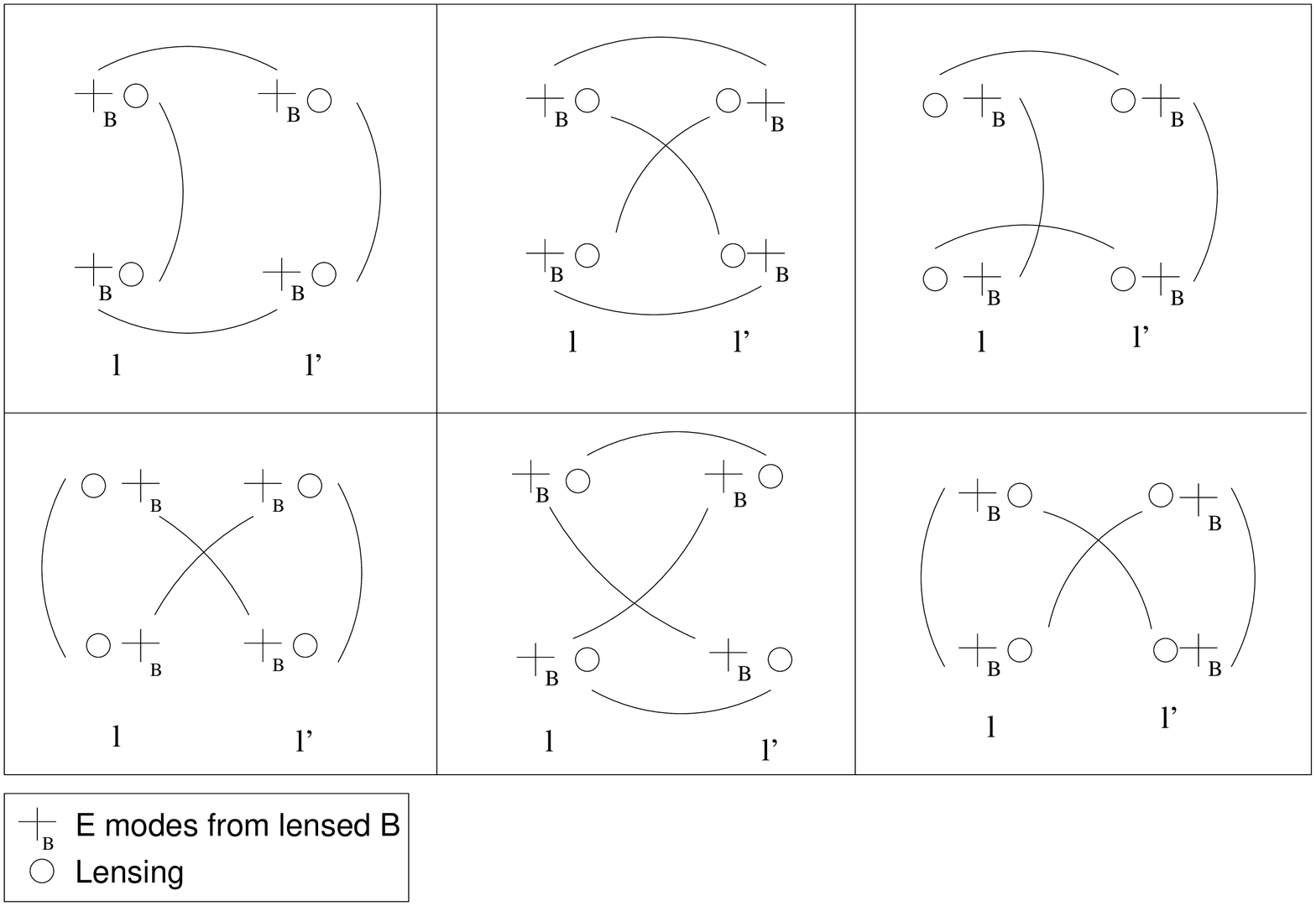}
\par\end{centering}
\caption{Non-diagonal ``topologically connected'' configurations
of correlations contributing to the covariance matrix, for the
polarization BB-BB at order ``1+1+1+1''. These graphs are the
dominant terms contributing to $\mathcal{N}_{\mathbf{l},\mathbf{l'}~(4)}
^{BB-BB}$ at order four in lensing.}
\label{Fig:graphTlens4}
\end{figure}

If we temporarily assume
that the B modes are negligible compared to the E modes \emph{at the same order in
lensing}, we can see that the dominant non-diagonal 4th order terms
involve four first order E modes from $\widetilde{B}^{(1)}$. As explained in
\ref{annex2}, they are denoted $+_B\circ$. All possible terms of the
form ``1+1+1+1'' are represented graphically in figure~\ref{Fig:graphTlens4}.
For each column, the two graphs have the same contribution,
respectively given by (\ref{eq:correl4Bo4term1}-\ref{eq:correl4Bo4term3}).

\subsection{Numerical calculation of the non-gaussian contributions}

We now turn to the numerical calculation of all these terms. They
are represented on the following figures~\ref{Fig:Covar},
\ref{Fig:Covar2}, and \ref{Fig:Covar4}. When possible
(figures~\ref{Fig:Covar}, and \ref{Fig:Covar2}) these non-gaussian
corrections are renormalized by the Gaussian value
$\mathcal{D}_\ell^{XY-UV}$ so that the amplitude of the graph
reflects the amplitude of the correction. These renormalization
factors are given in Table \ref{tableD}. This table can be used to
evaluate the absolute amplitude of non-gaussian corrections of
figures~\ref{Fig:Covar}, and \ref{Fig:Covar2} and can then be
compared to the corrections of figure~\ref{Fig:Covar4}.

\begin{table*}
\begin{center}
\begin{tabular}{l|c|c|c|c|c|c|c}
Multipole & $TT-TT$ & $EE-EE$ & $BB-BB$ & $TE-TE$ & $TT-TE$ & $TT-EE$ & $EE-TE$ \\ \hline
$\ell=200$ &  $2.6\times 10^7$ & $11$ & $7.2$ & $8.4\times 10^3$ & $5.2\times 10^4$ & $10^2$ & $34$ \\
$\ell=400$ &  $2.4\times 10^6$ & $9.2 \times 10^2$ & $1.4\times 10^2$ & $2.3\times 10^4$ & $2.0\times 10^4$ & $1.6 \times 10^2$ & $3.9\times 10^2$ \\
$\ell=800$ &  $5.5\times 10^6$ & $6.7\times 10^3$ & $4.9\times 10^3$ & $9.8\times 10^4$ & $1.3\times 10^5$ & $2.8\times 10^3$ & $4.4\times 10^3$ \\
$\ell=1600$ &  $4.3\times 10^5$ & $1.2\times 10^6$ & $1.2\times 10^6$ & $3.7\times 10^5$ & $1.1\times 10^4$ & $2.8\times 10^2$ & $1.9\times 10^4$ \\
\end{tabular}
\caption{Absolute values of the covariance matrix in the Gaussian
limit $\mathcal{D}_\ell^{XY-UV}$ with the pre-factor
$[\ell(\ell+1)/2\pi]^2$. The unit is $(\mu K)^4$.}\label{tableD}
\end{center}
\end{table*}

\begin{figure*}
\begin{centering}
\includegraphics[scale=0.21]{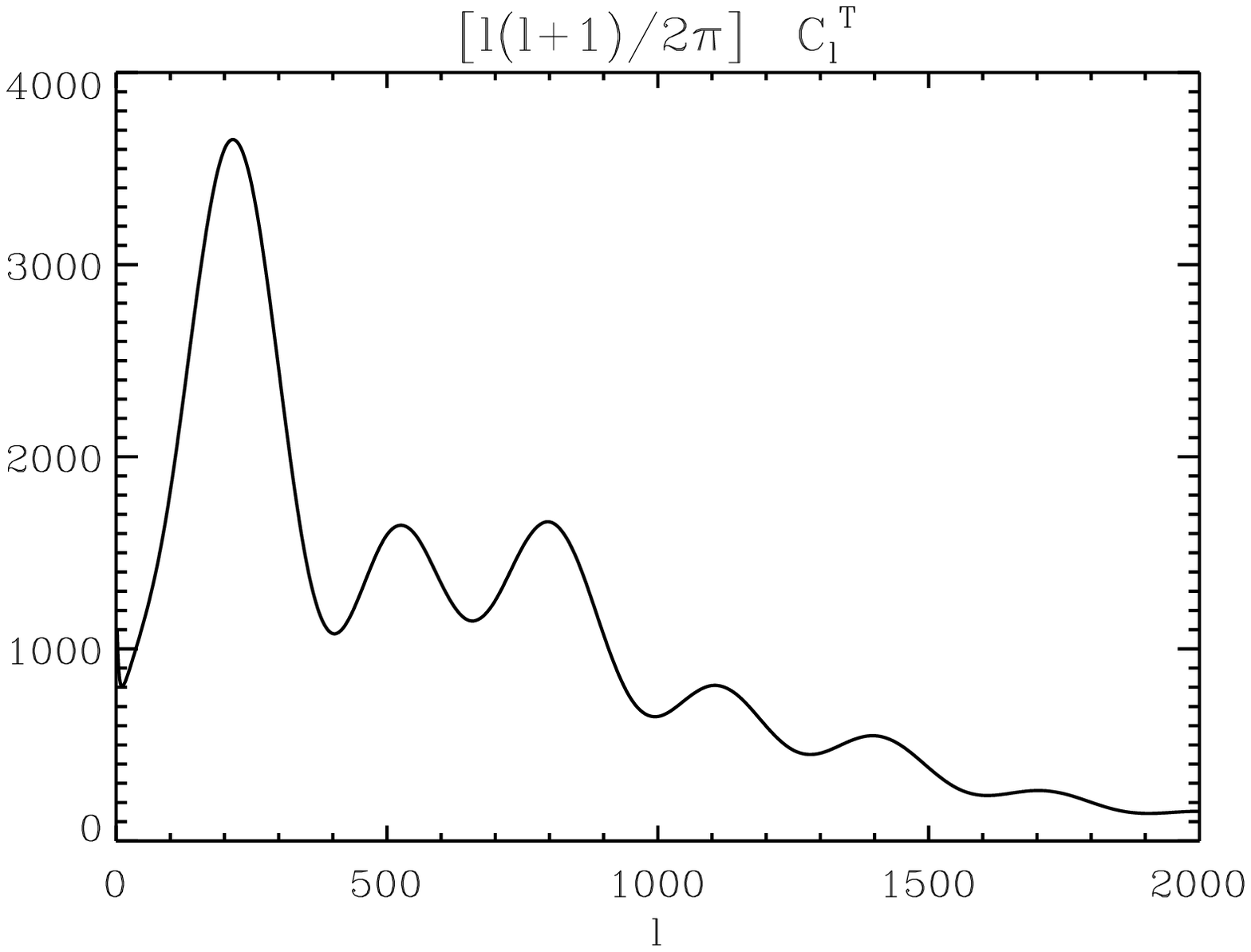}
\includegraphics[scale=0.21]{specInitE}
\includegraphics[scale=0.21]{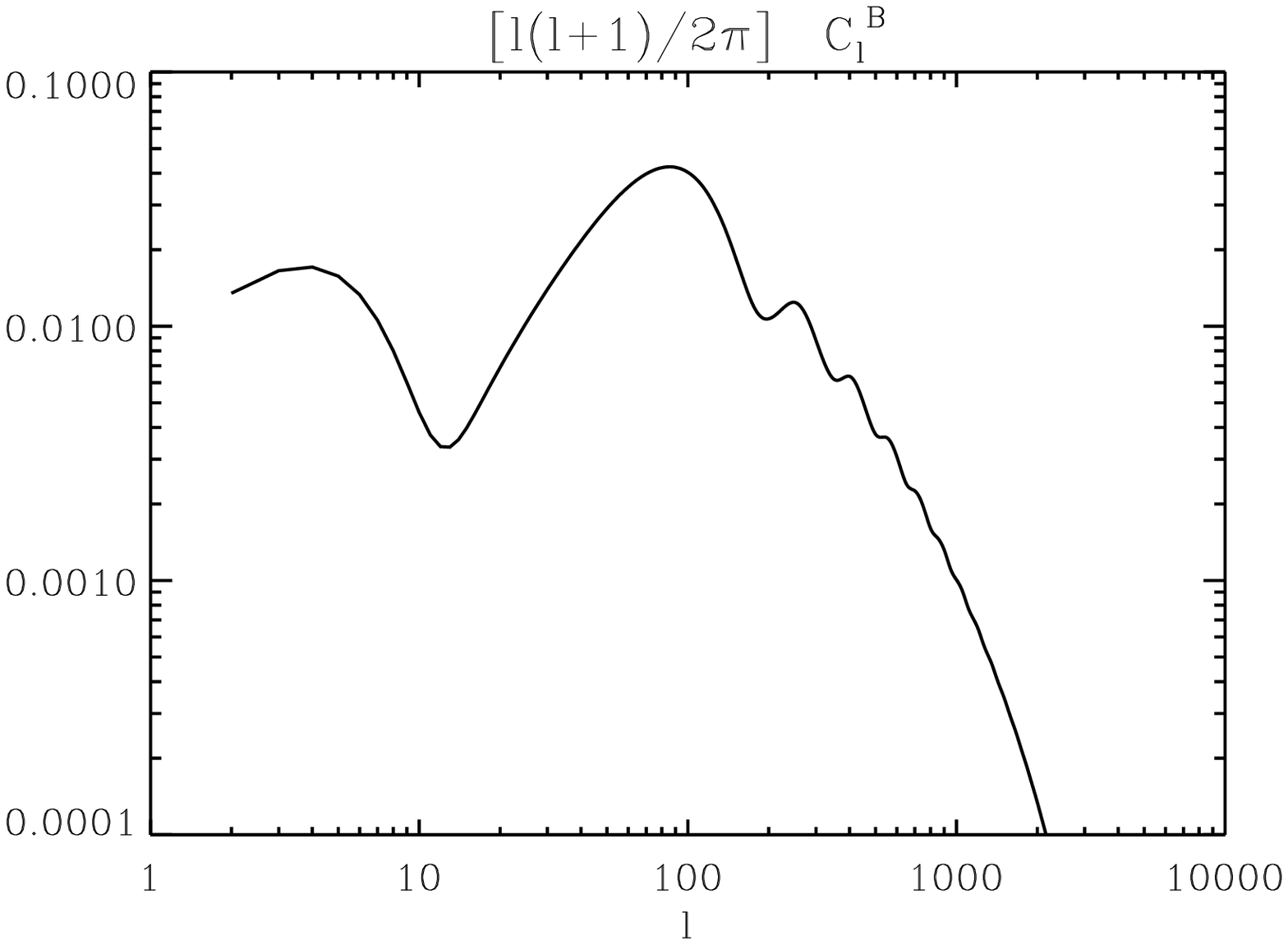}
\includegraphics[scale=0.21]{specInitTE}\par\end{centering}

\vspace*{1cm}

\includegraphics[scale=0.21]{contribnondiagT200}
\includegraphics[scale=0.21]{contribnondiagE200}
\includegraphics[scale=0.21]{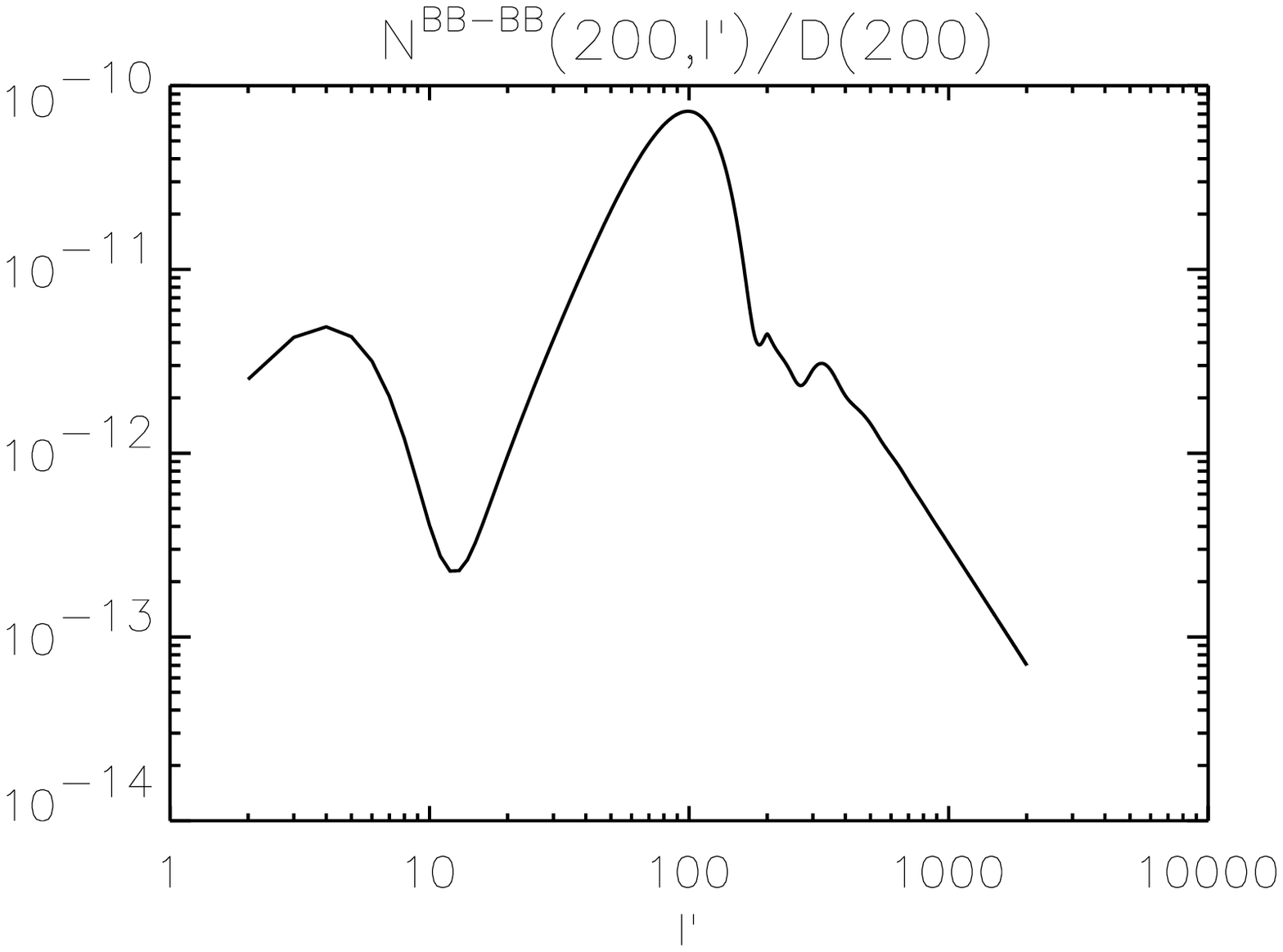}
\includegraphics[scale=0.21]{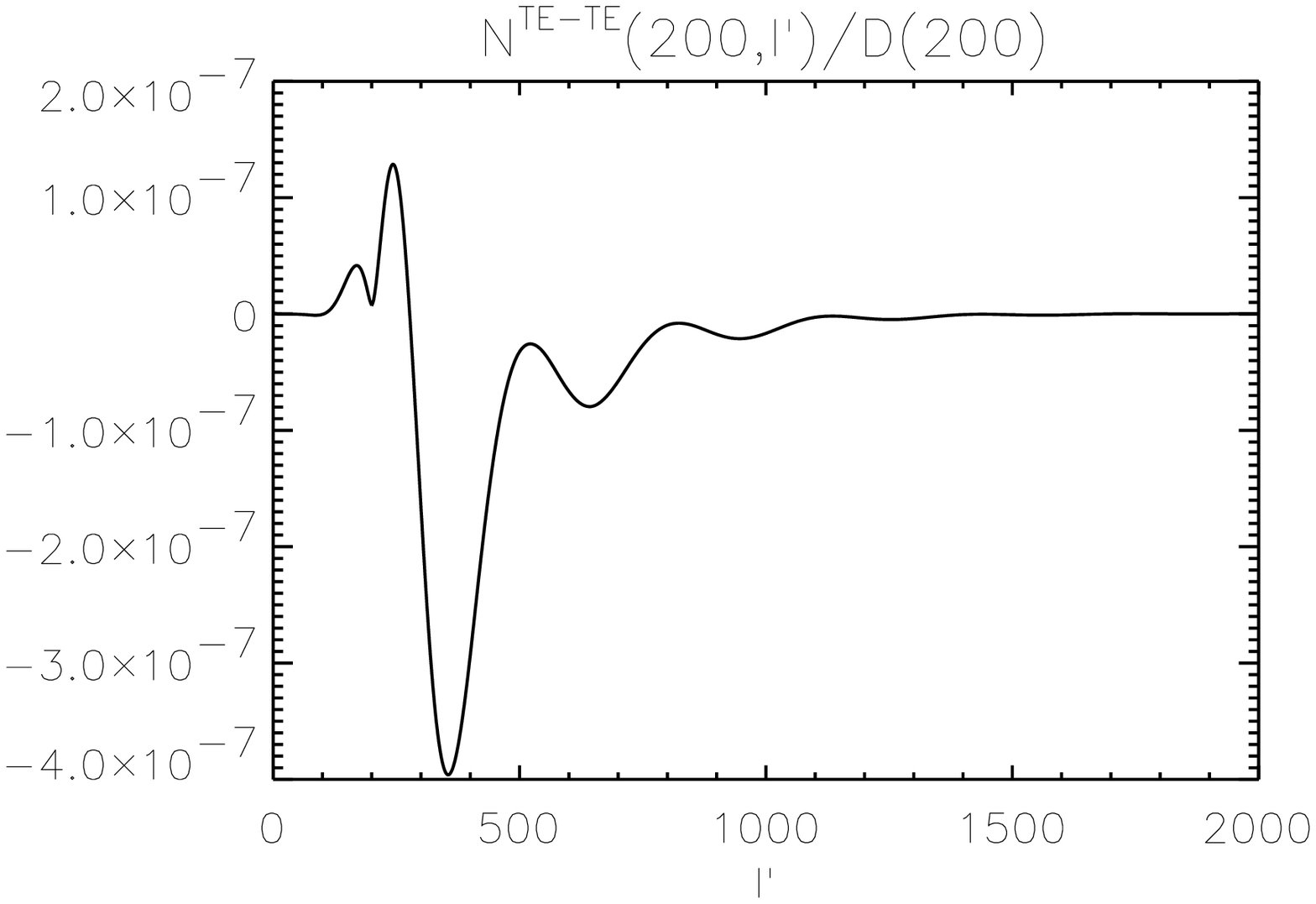}
\includegraphics[scale=0.21]{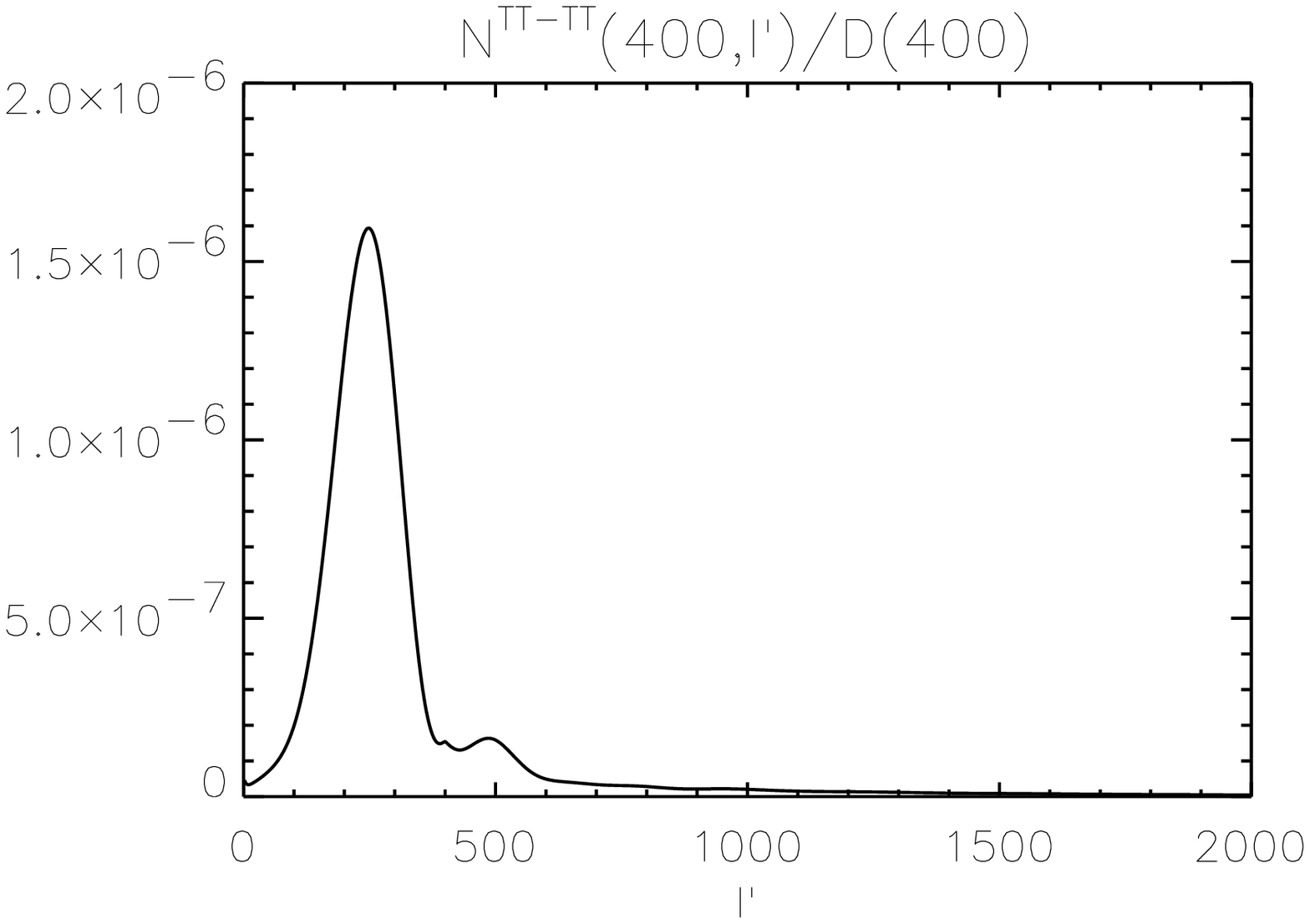}
\includegraphics[scale=0.21]{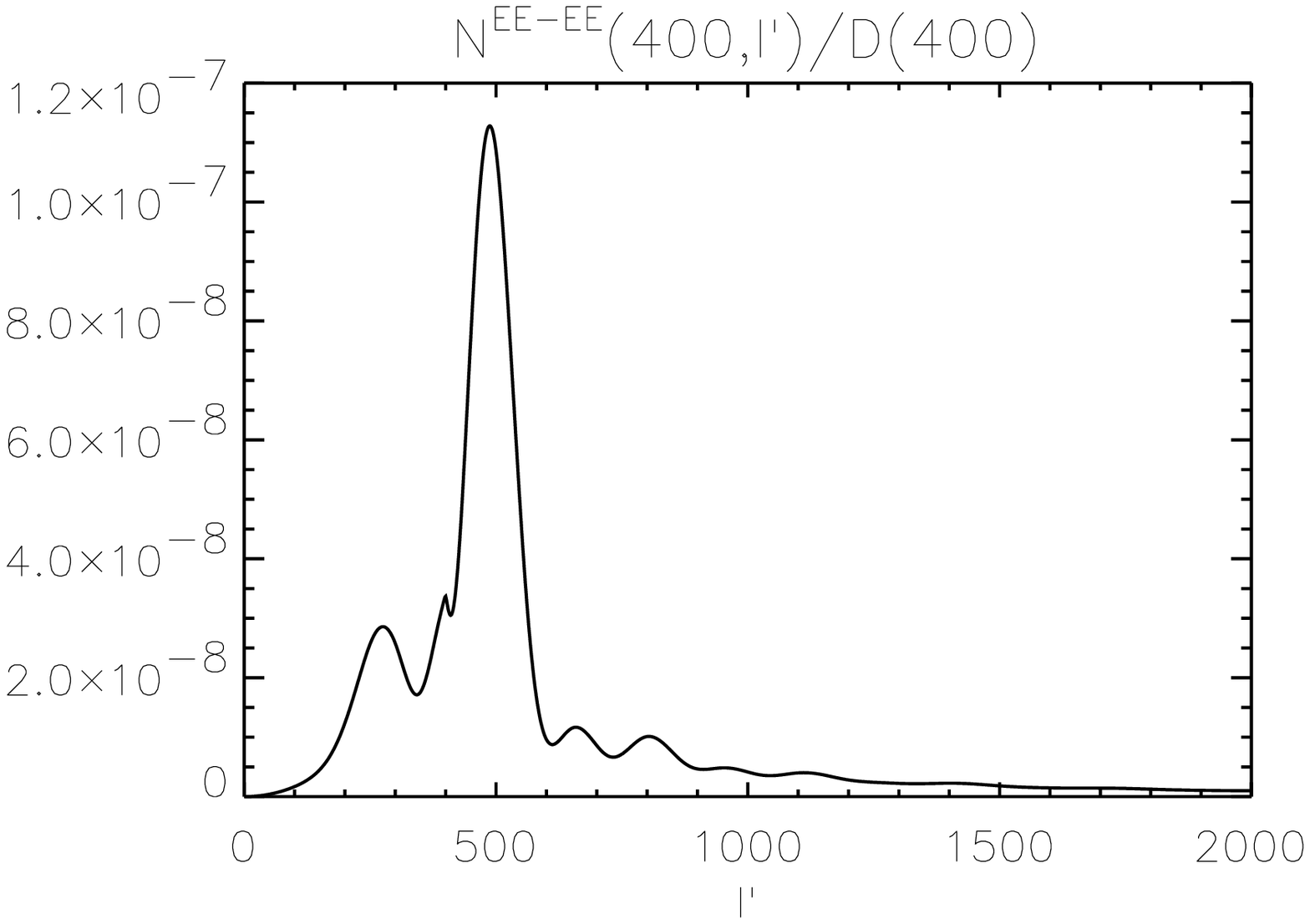}
\includegraphics[scale=0.21]{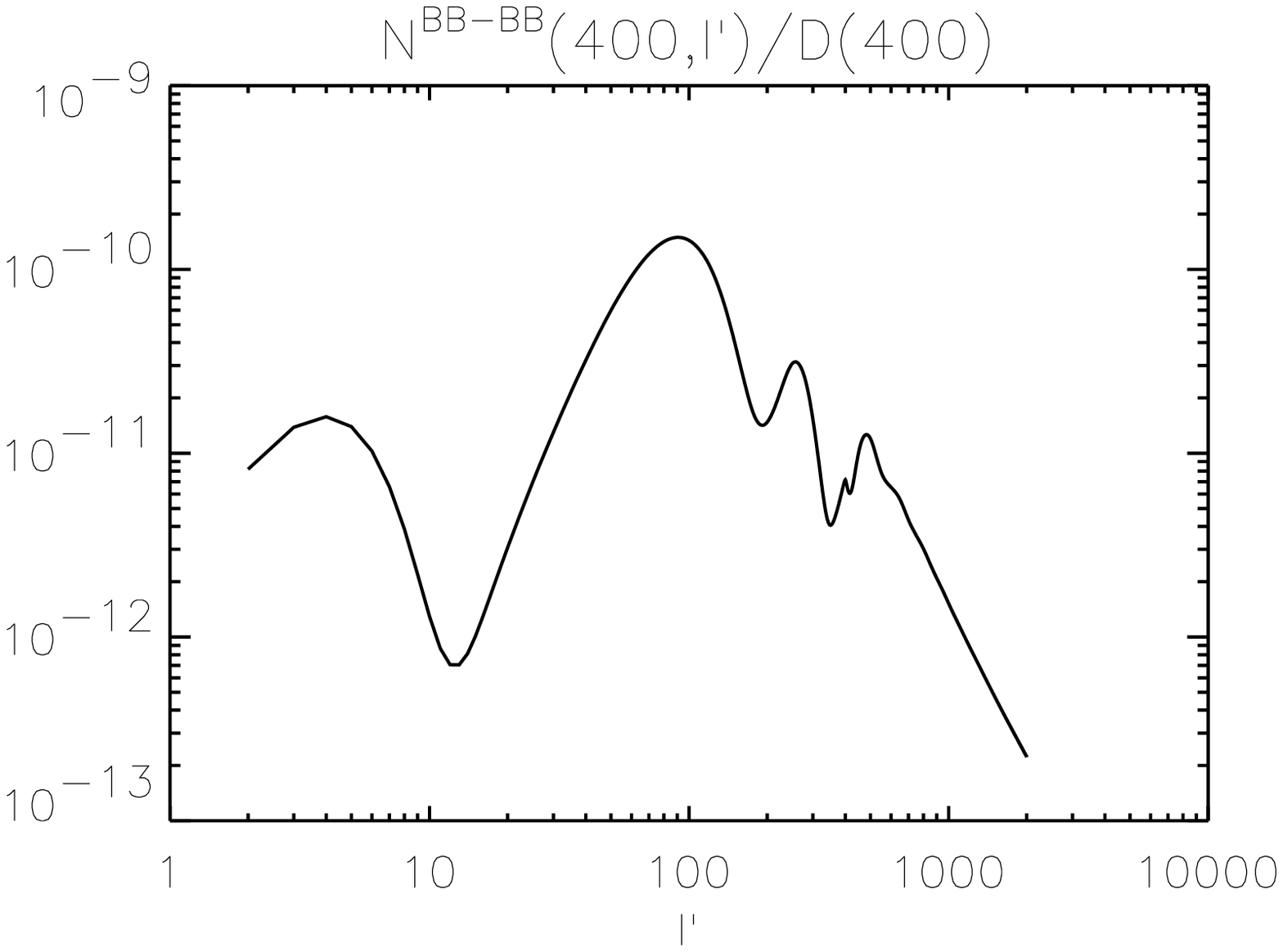}
\includegraphics[scale=0.21]{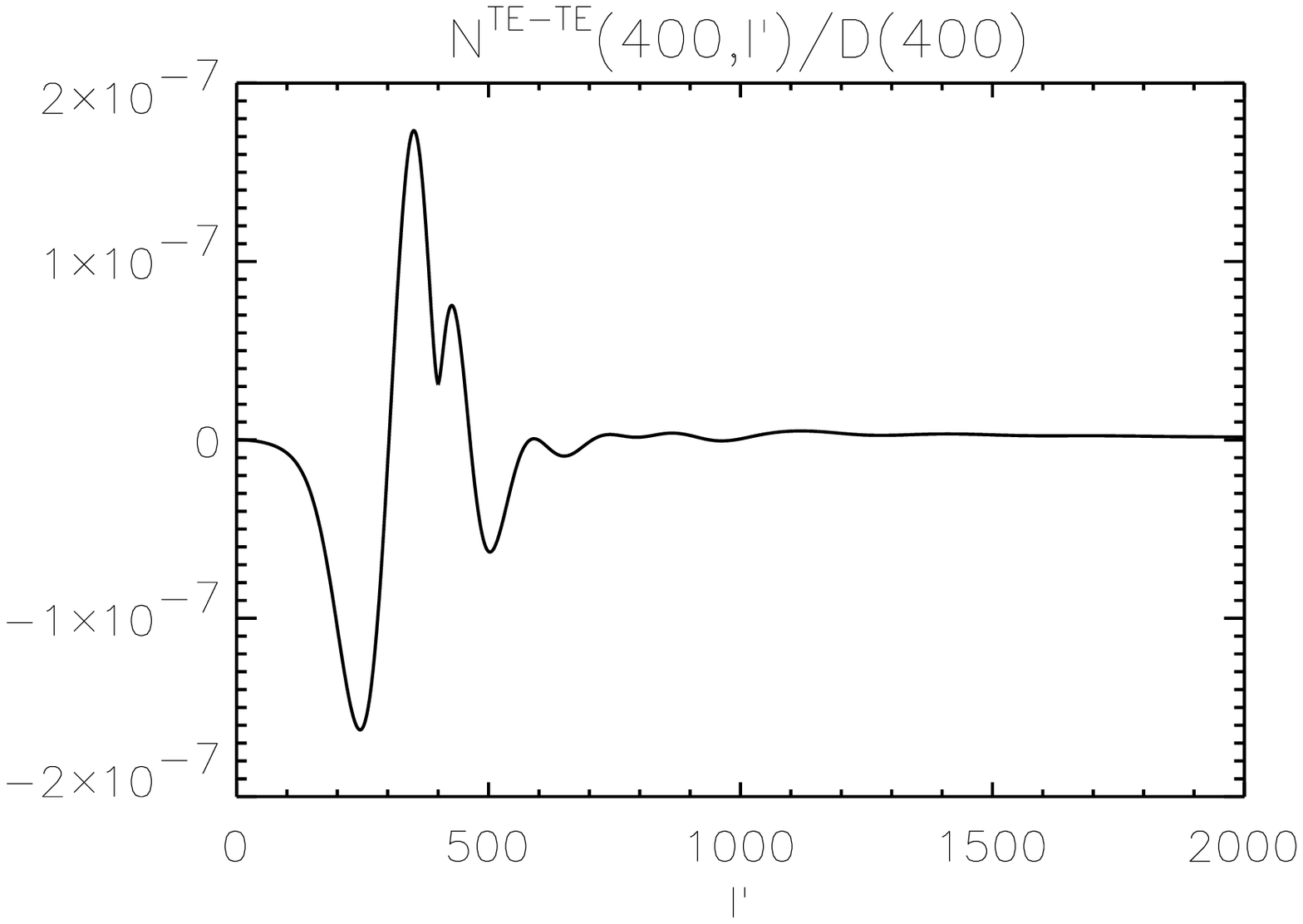}
\includegraphics[scale=0.21]{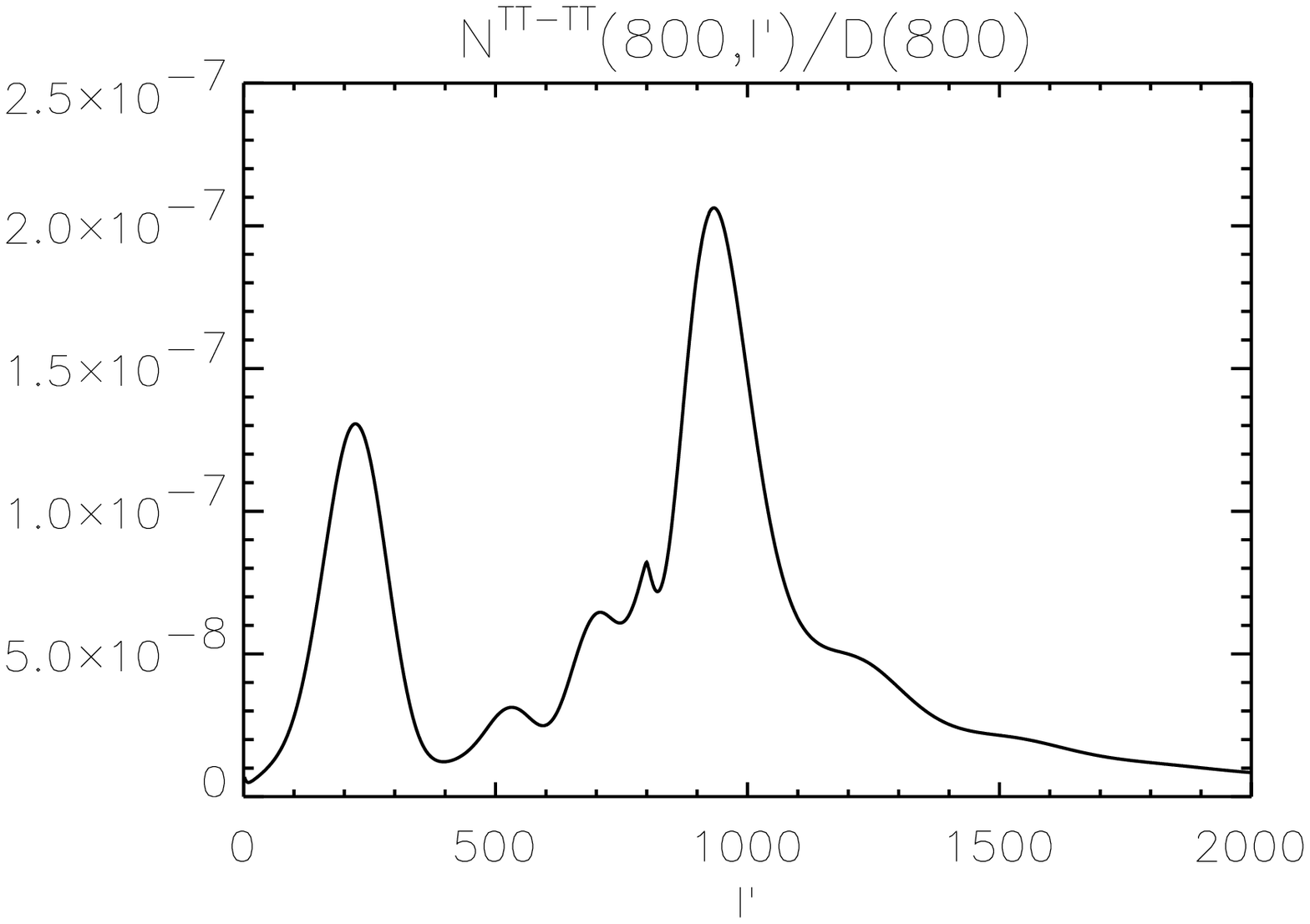}
\includegraphics[scale=0.21]{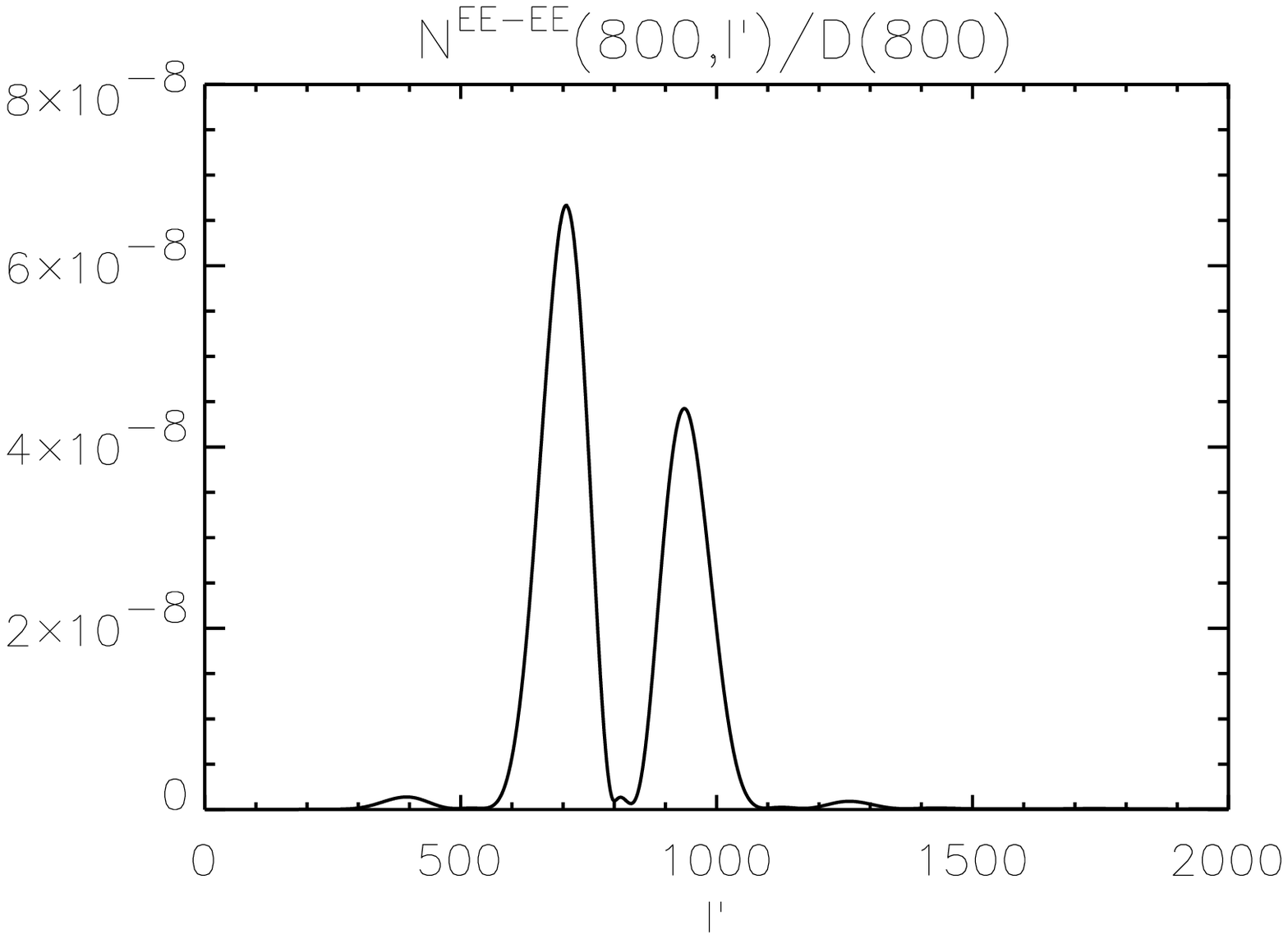}
\includegraphics[scale=0.21]{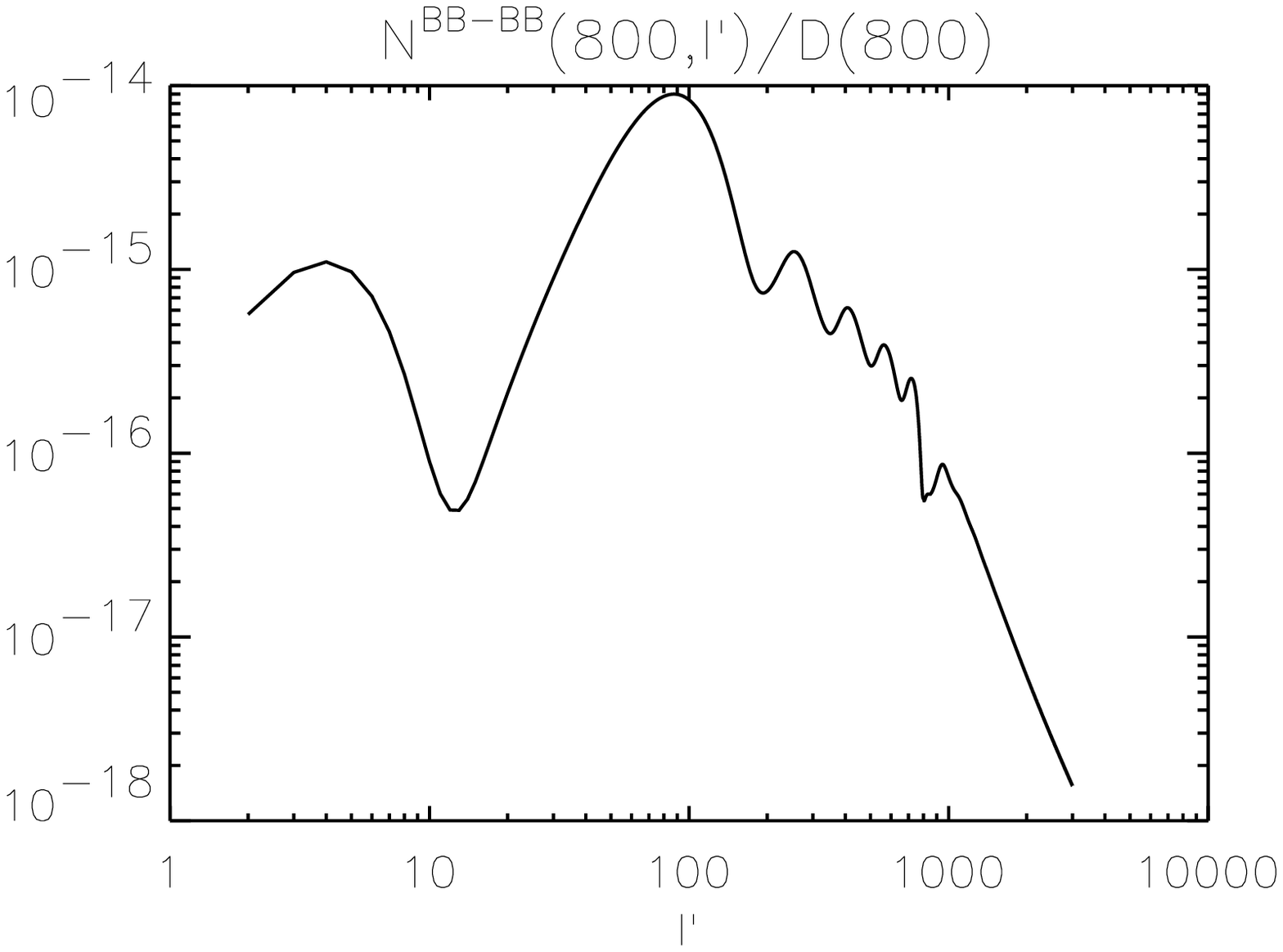}
\includegraphics[scale=0.21]{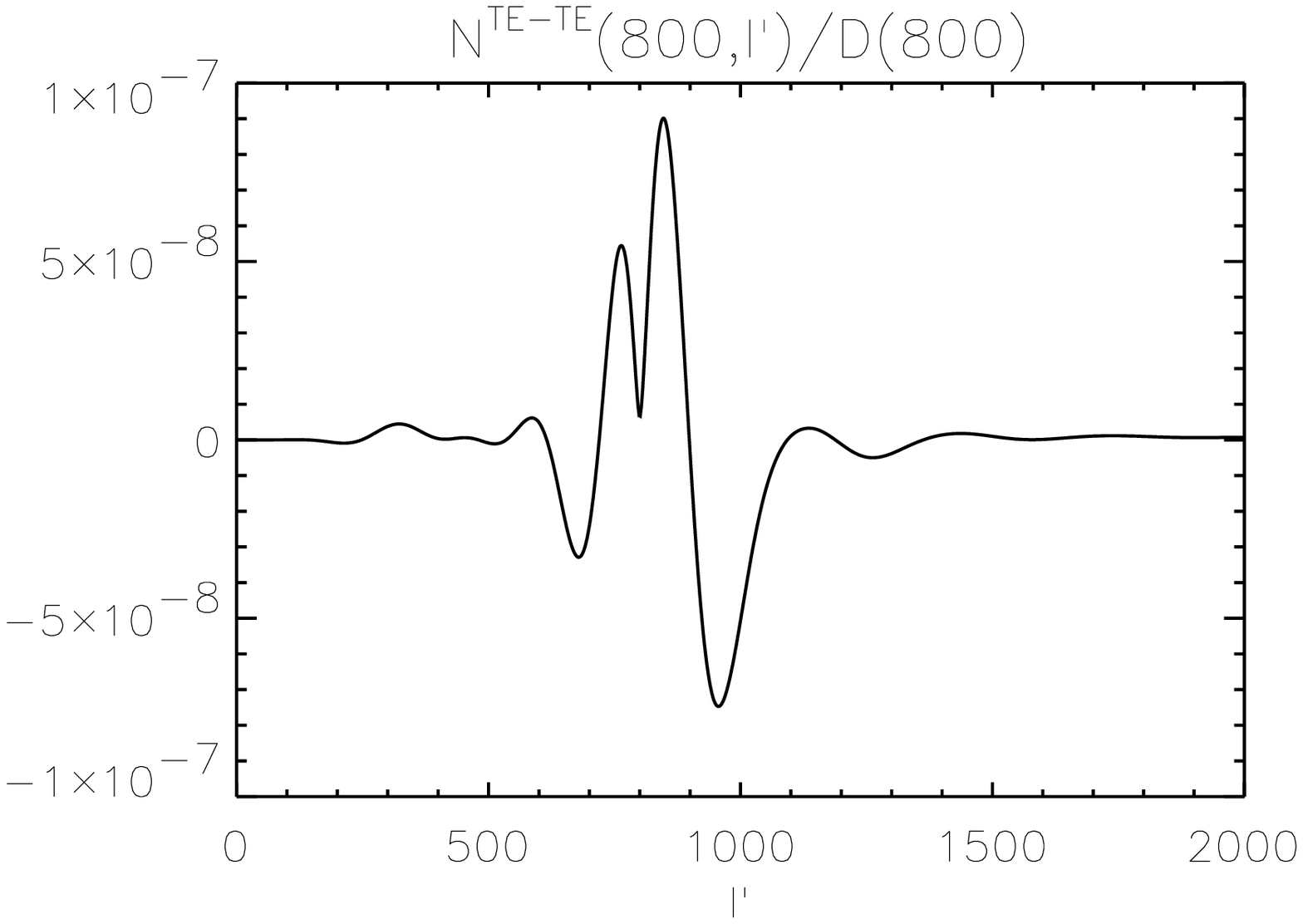}
\includegraphics[scale=0.21]{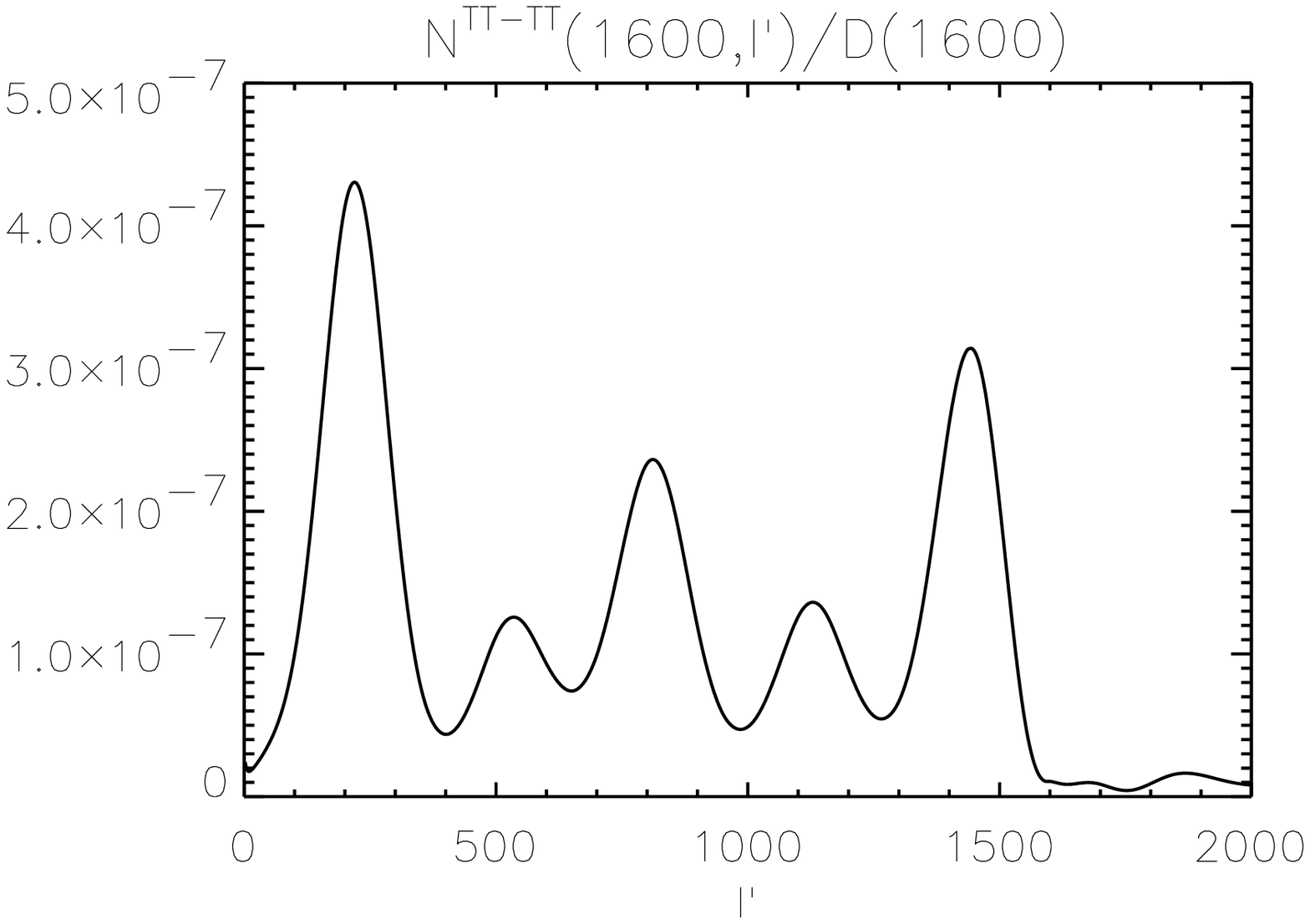}
\includegraphics[scale=0.21]{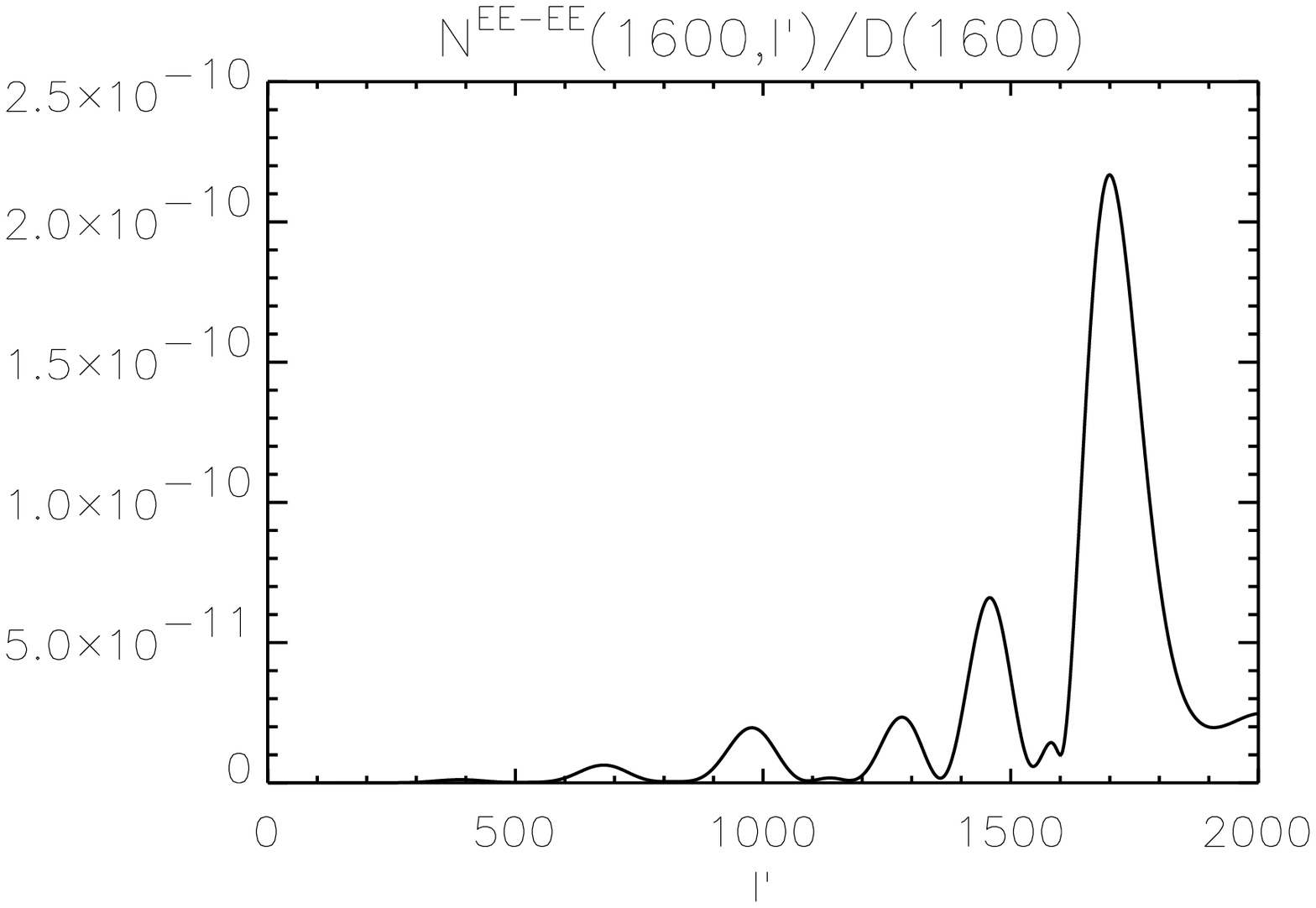}
\includegraphics[scale=0.21]{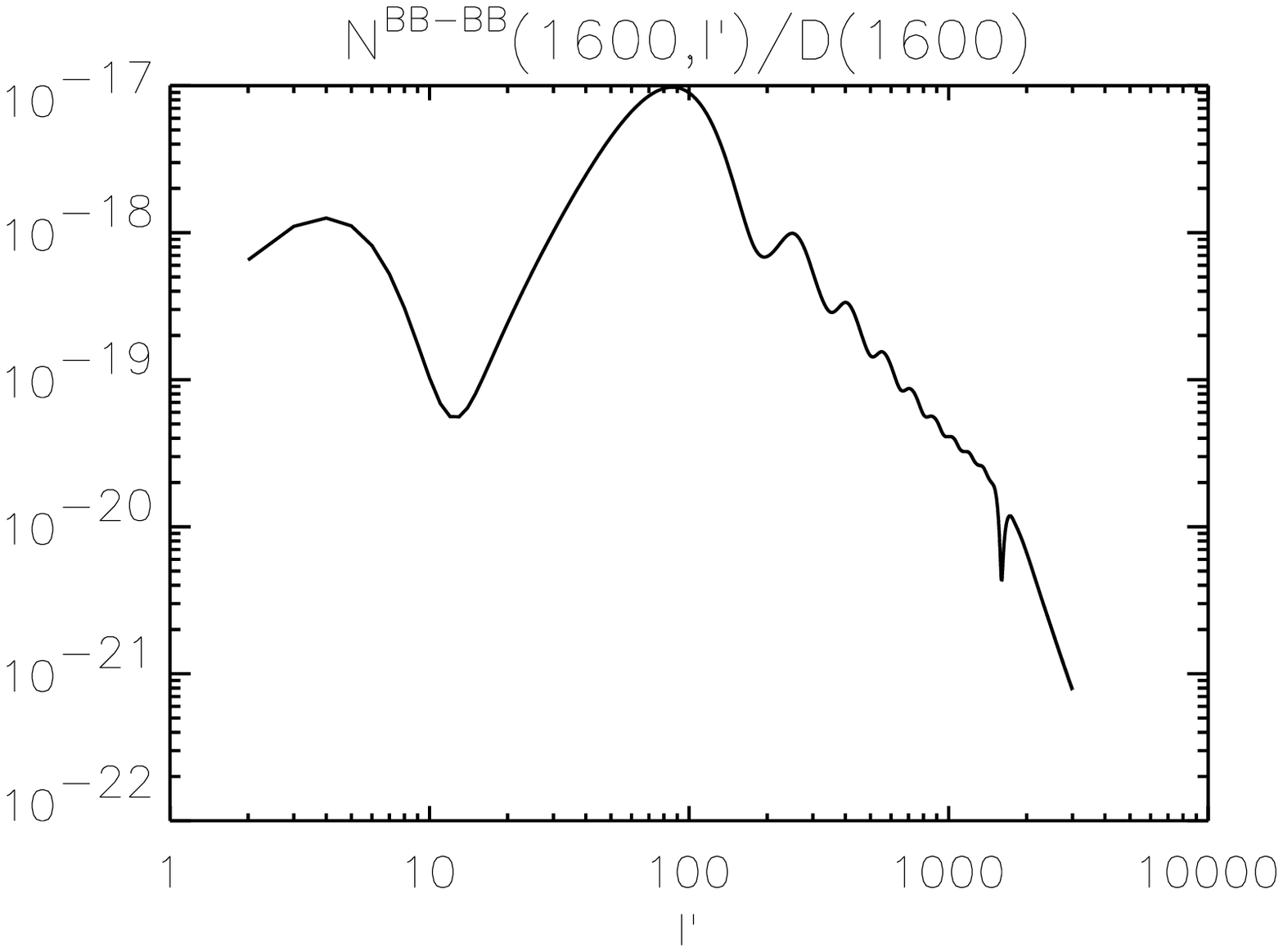}
\includegraphics[scale=0.21]{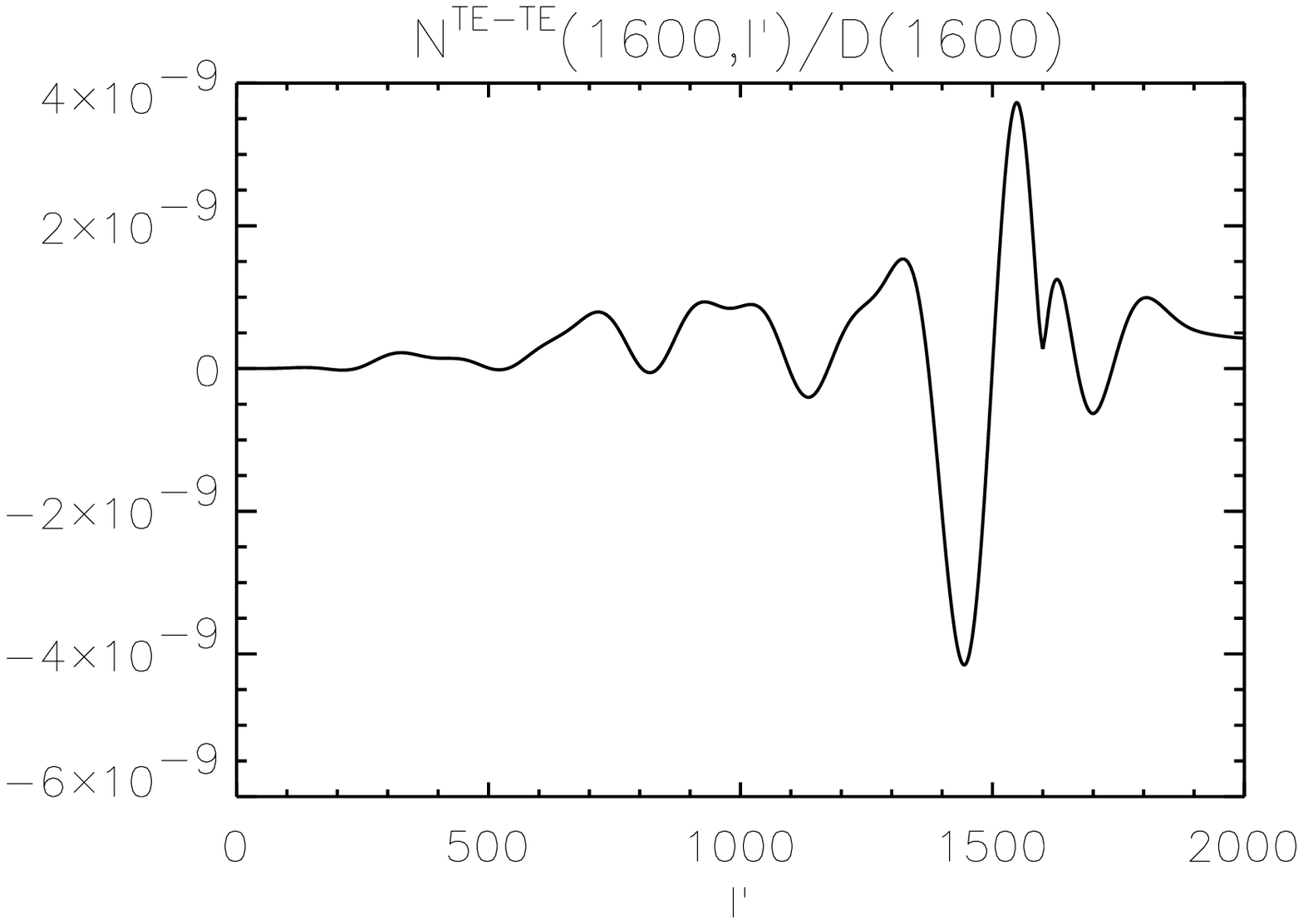}
\caption{On the first row, we recall the form of the TT, EE, BB,
and TE power spectrum in the absence of lensing. The cosmological
parameters are given by WMAP 3 \cite{2006astro.ph..3449S} and we
assume a tensor/scalar ratio $r=1$. Below is the non-diagonal ($l\neq l'$)
contribution to the covariance matrix $\mathcal{N}_{l,l'}^{UV-XY}$
for several polarization and various values of $l$. From left to
right, $UV-XY=TT-TT$, $EE-EE$, $BB-BB$, and $TE-TE$, and from top
to bottom, $l=200$, $400$, $800$, and $1600$. The amplitude is
normalized to by the diagonal value $\mathcal{D}_{\ell}^{UV-XY}$ to have the
relative contribution.}

\label{Fig:Covar}
\end{figure*}

\begin{figure*}
\begin{centering}\includegraphics[scale=0.3]{specInitTE}\par
\end{centering}

\vspace*{1cm}
\includegraphics[scale=0.3]{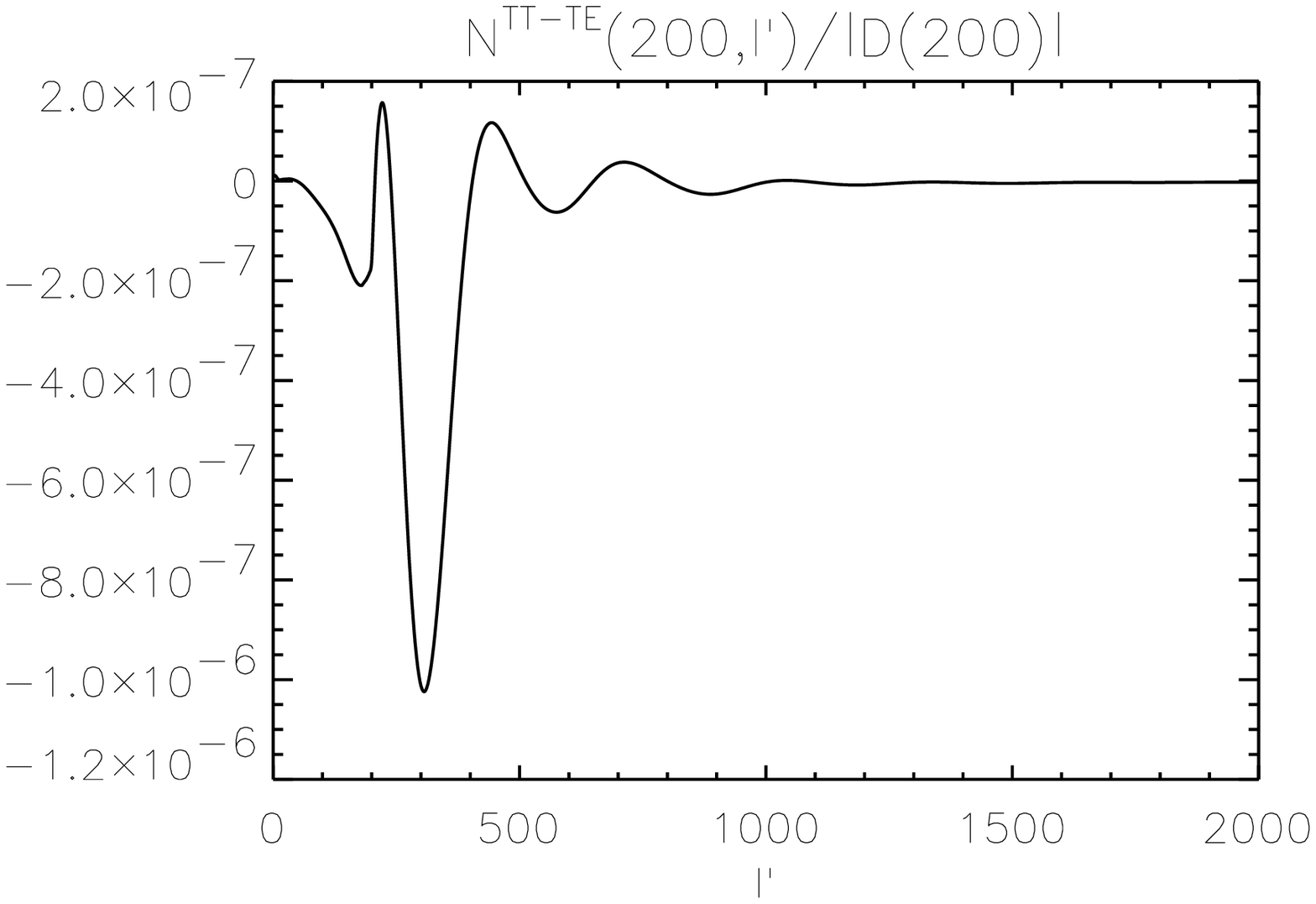}
\includegraphics[scale=0.3]{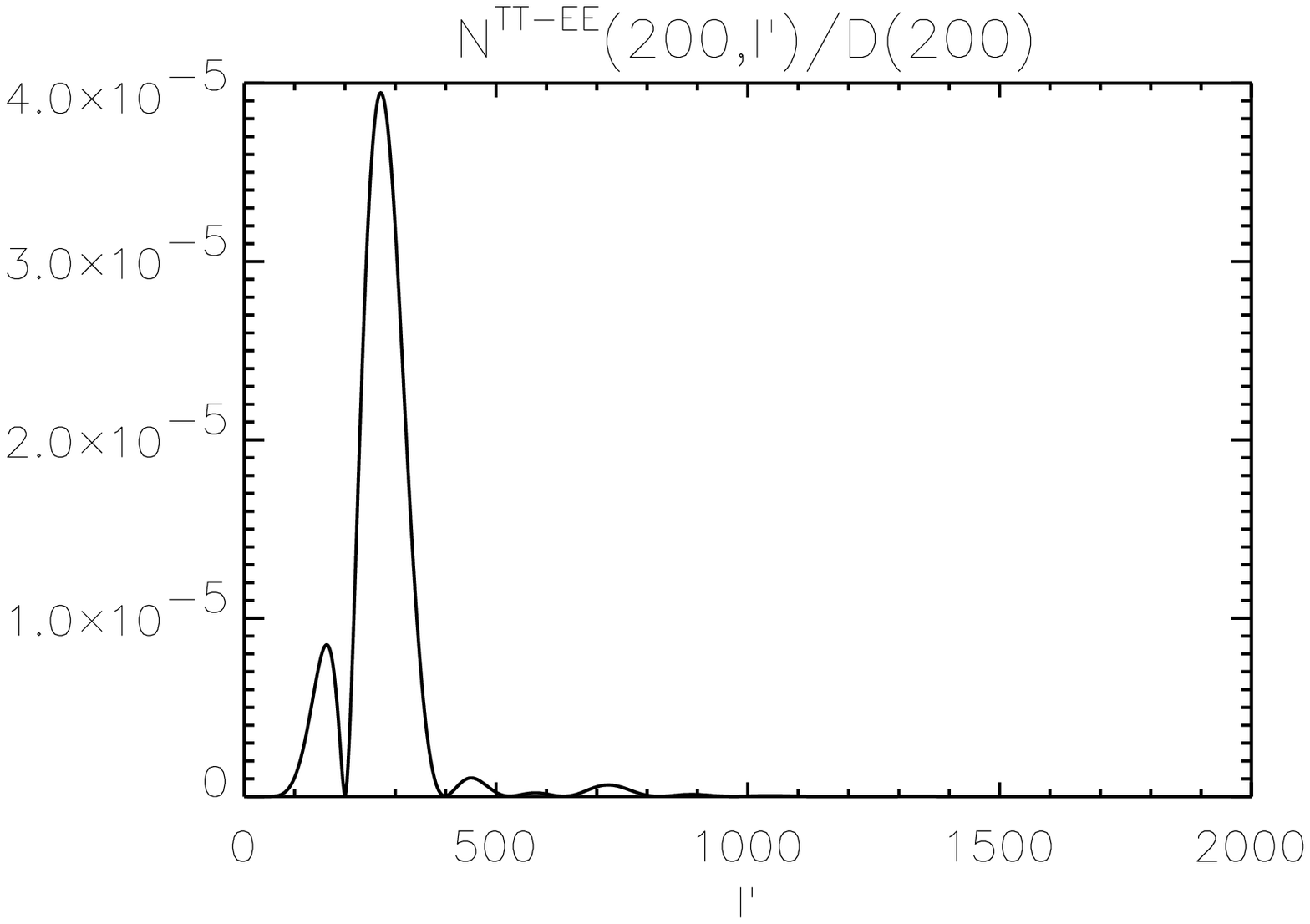}
\includegraphics[scale=0.3]{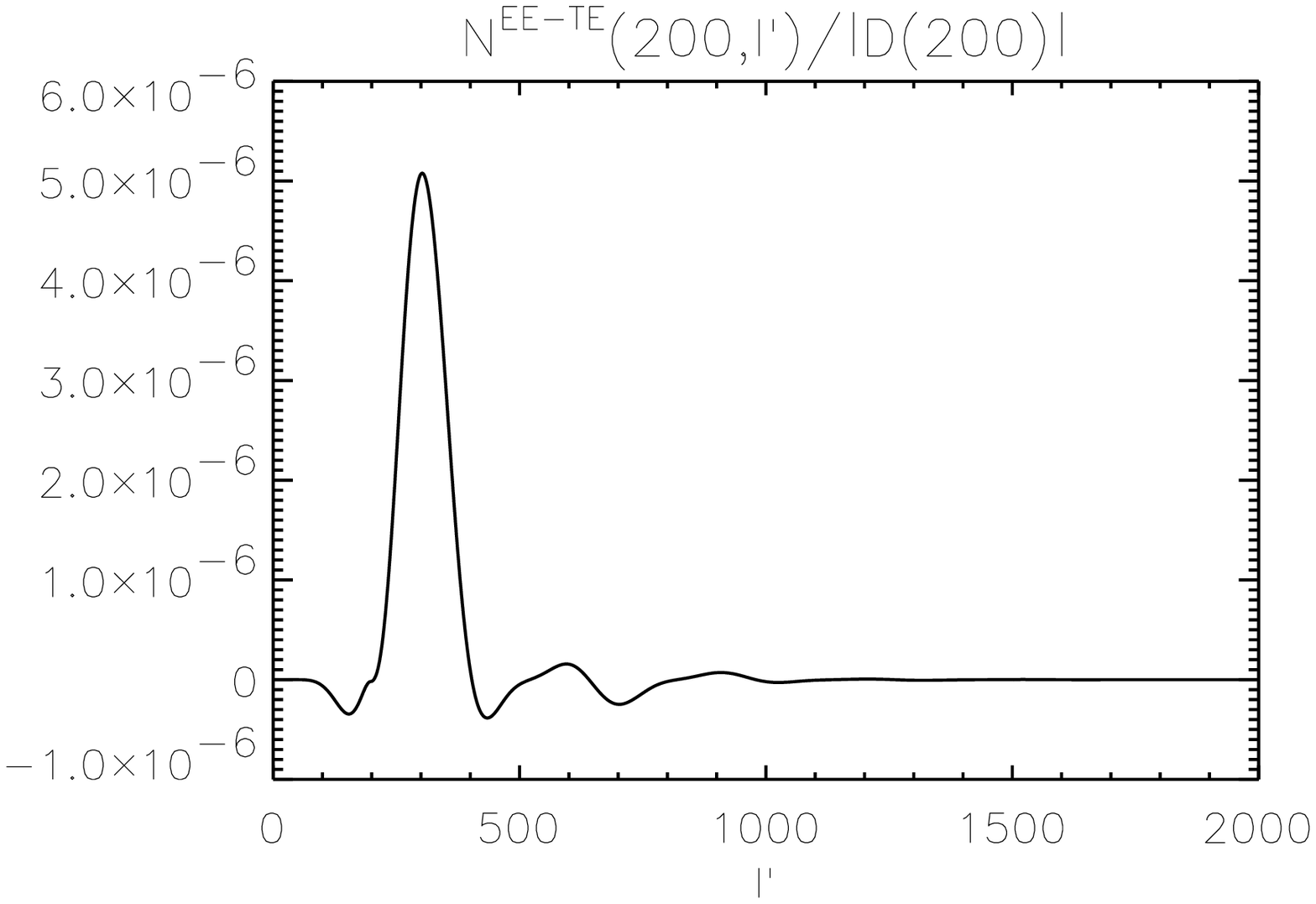}
\includegraphics[scale=0.3]{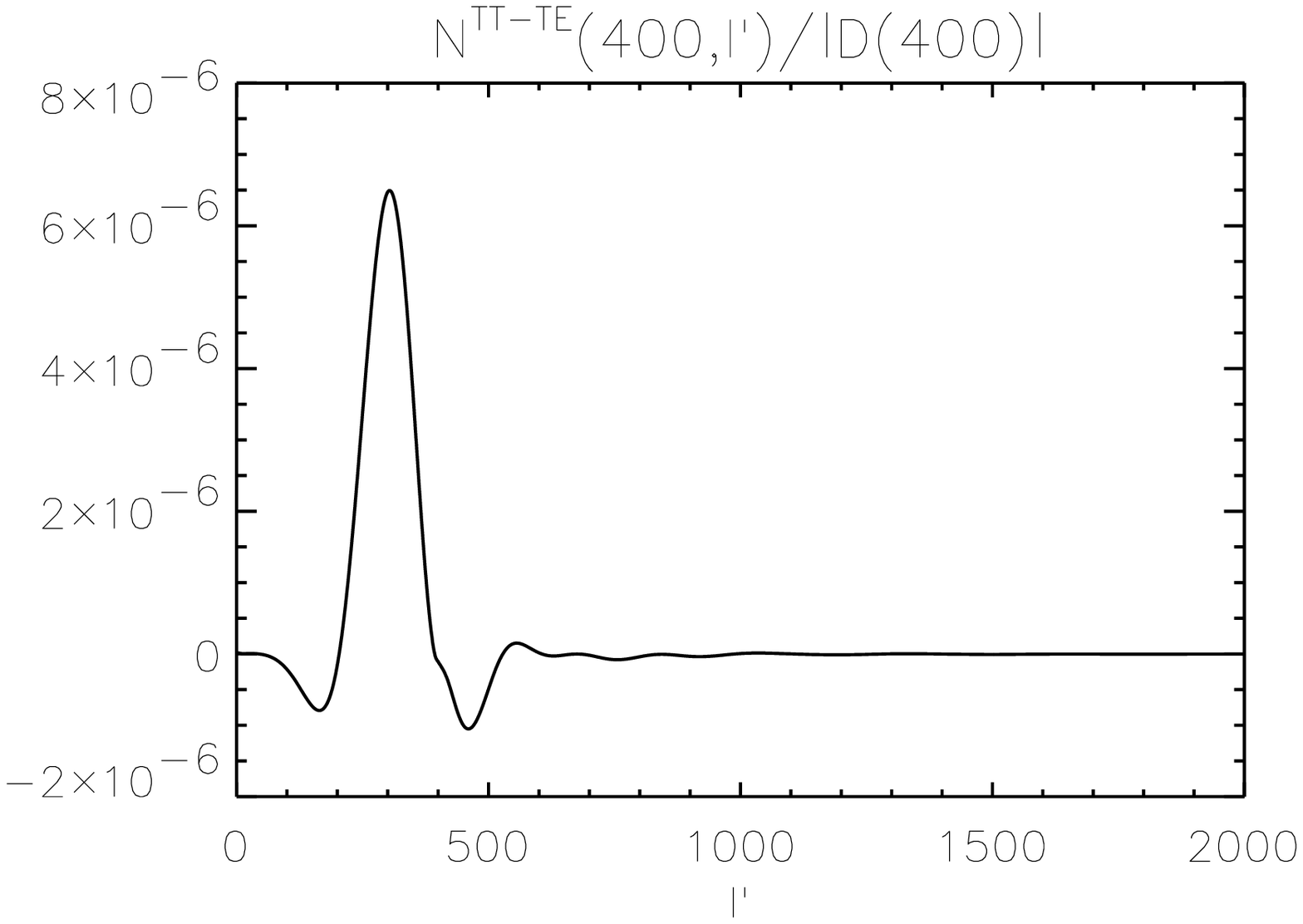}
\includegraphics[scale=0.3]{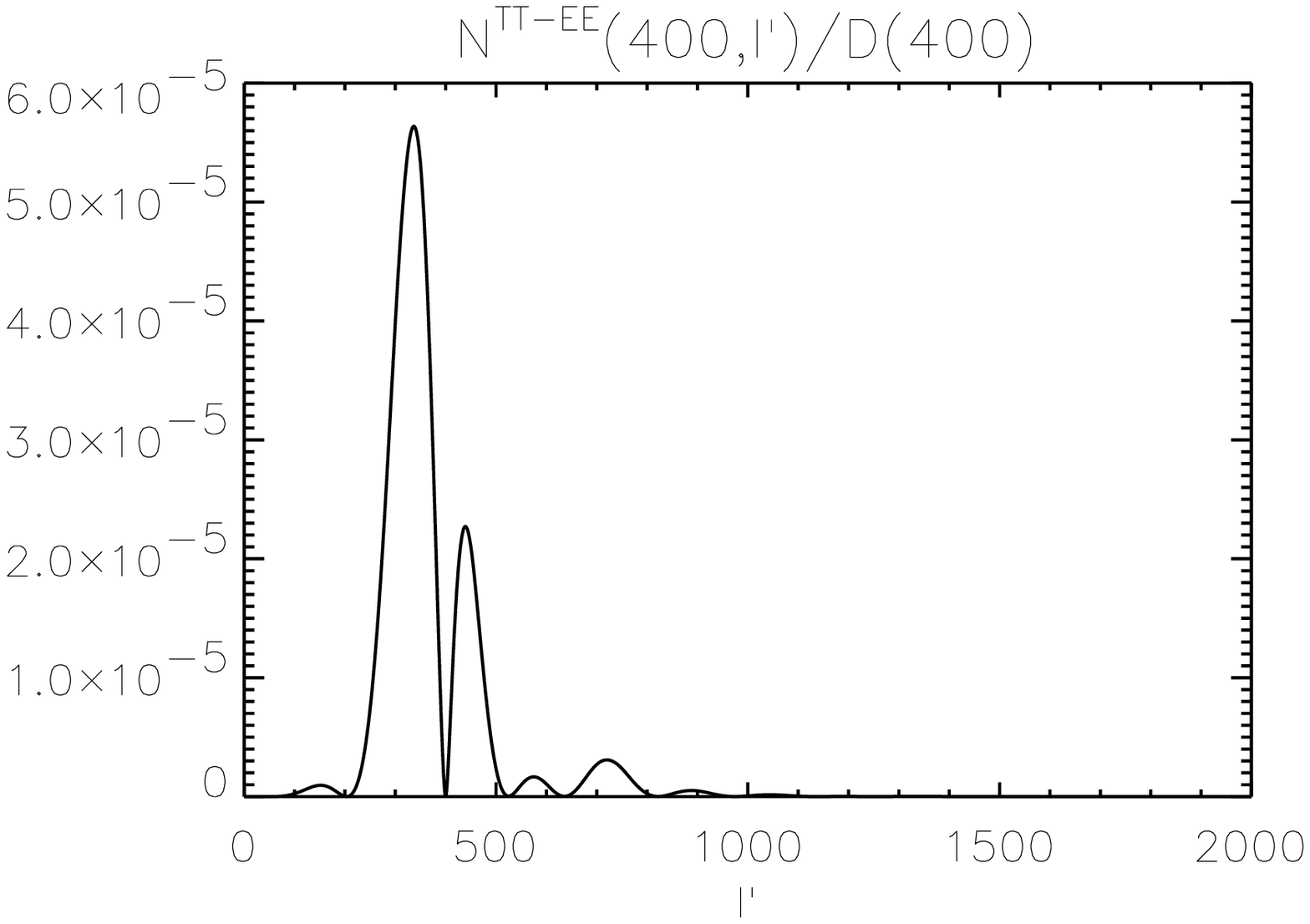}
\includegraphics[scale=0.3]{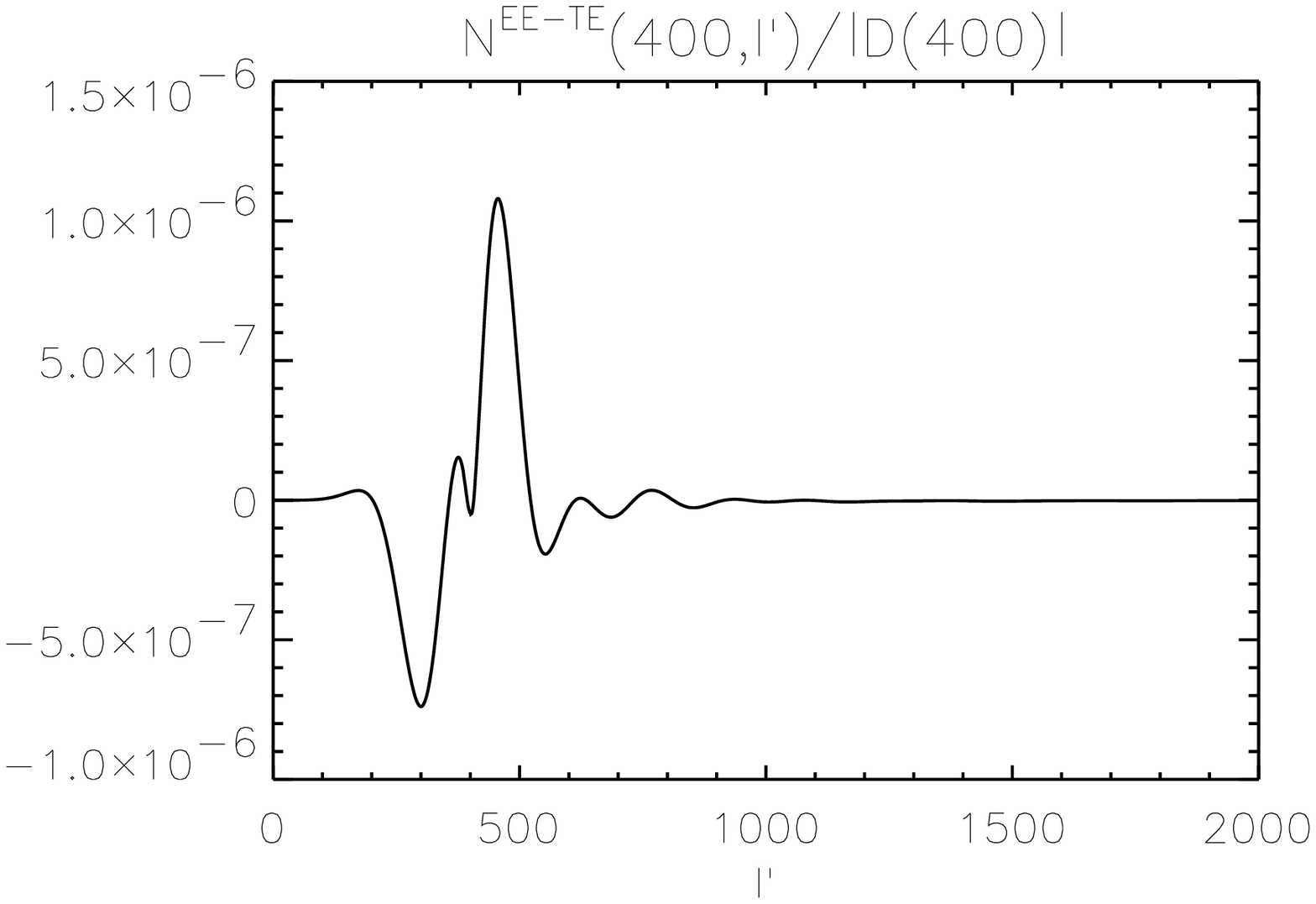}
\includegraphics[scale=0.3]{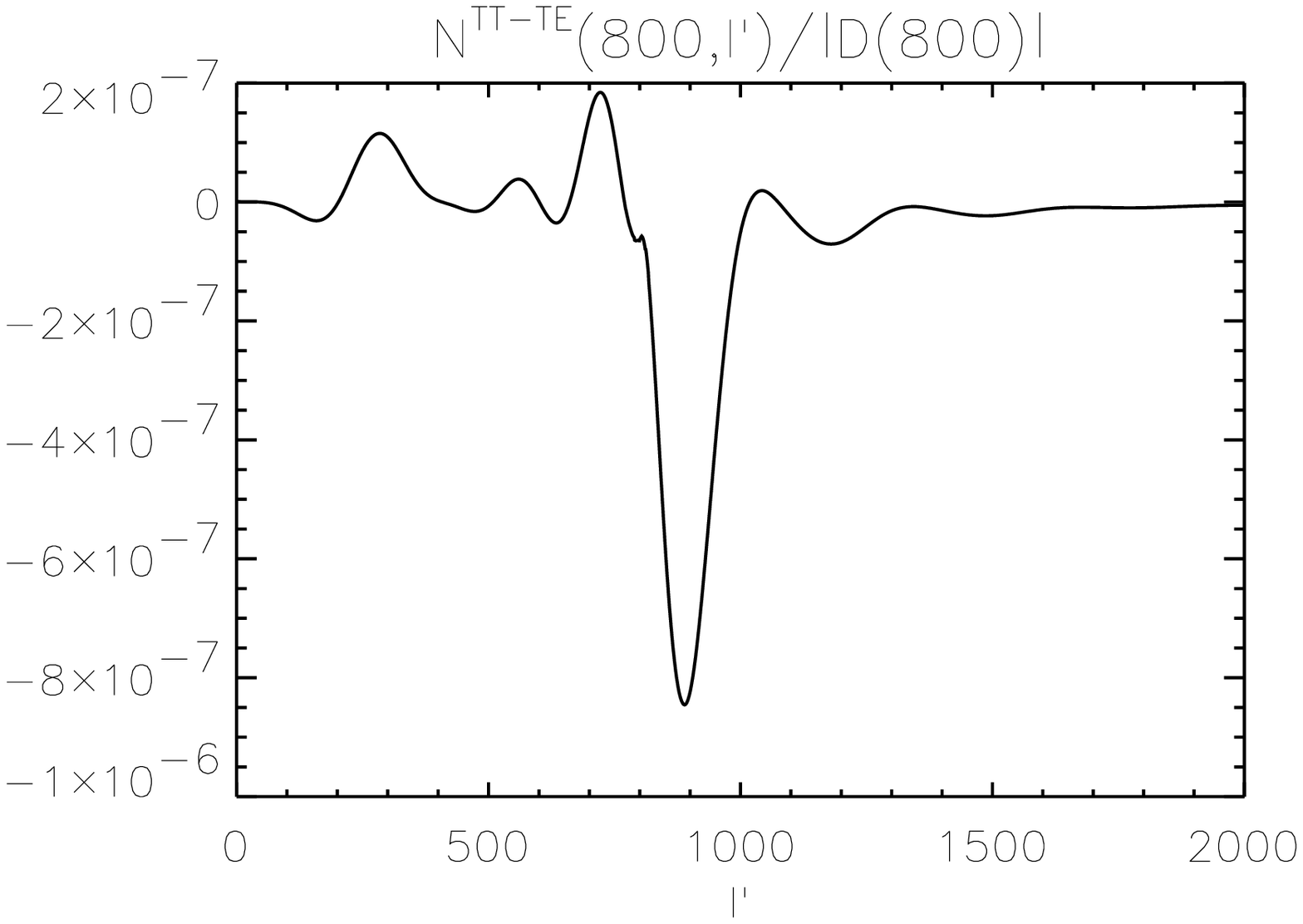}
\includegraphics[scale=0.3]{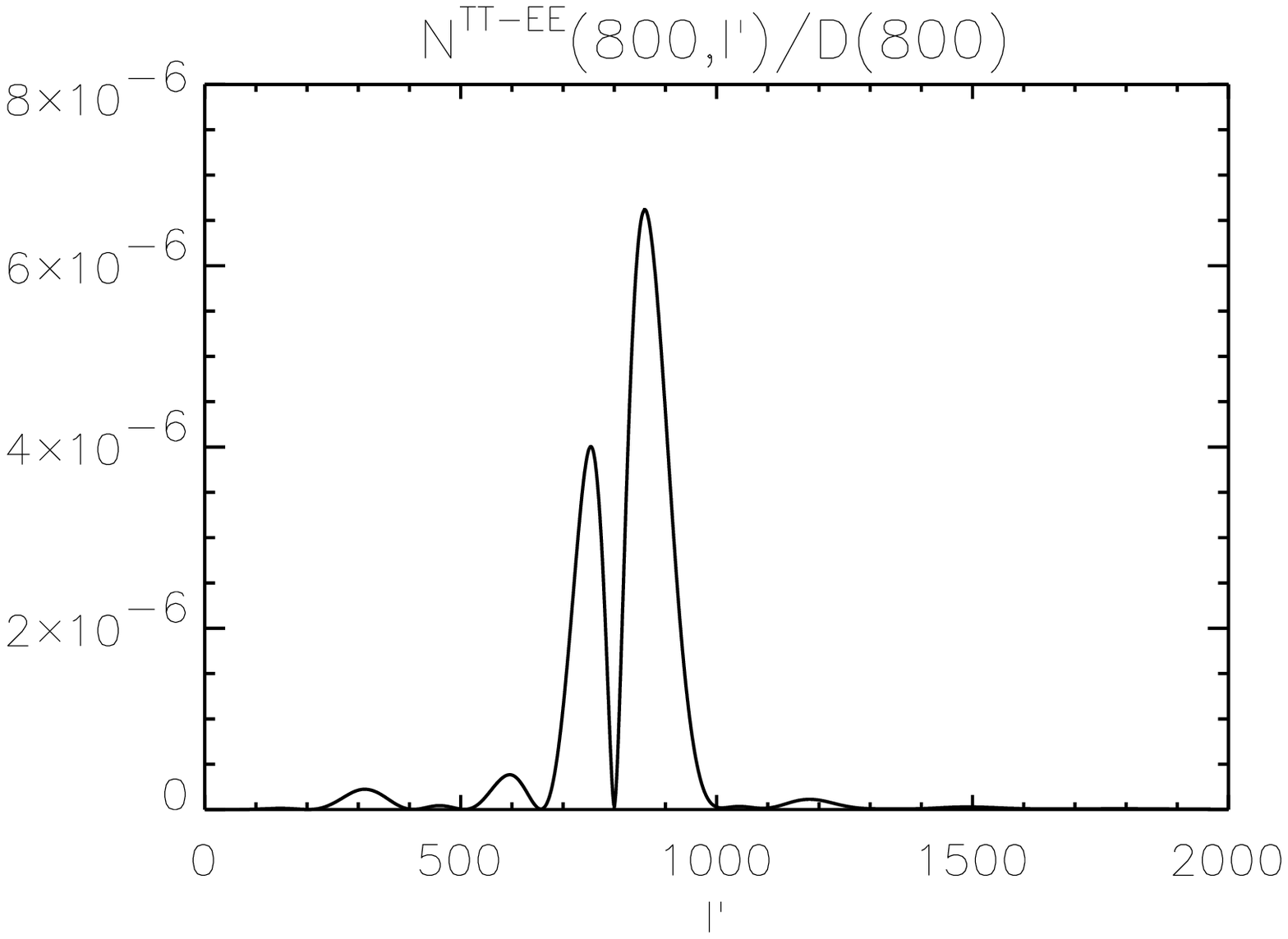}
\includegraphics[scale=0.3]{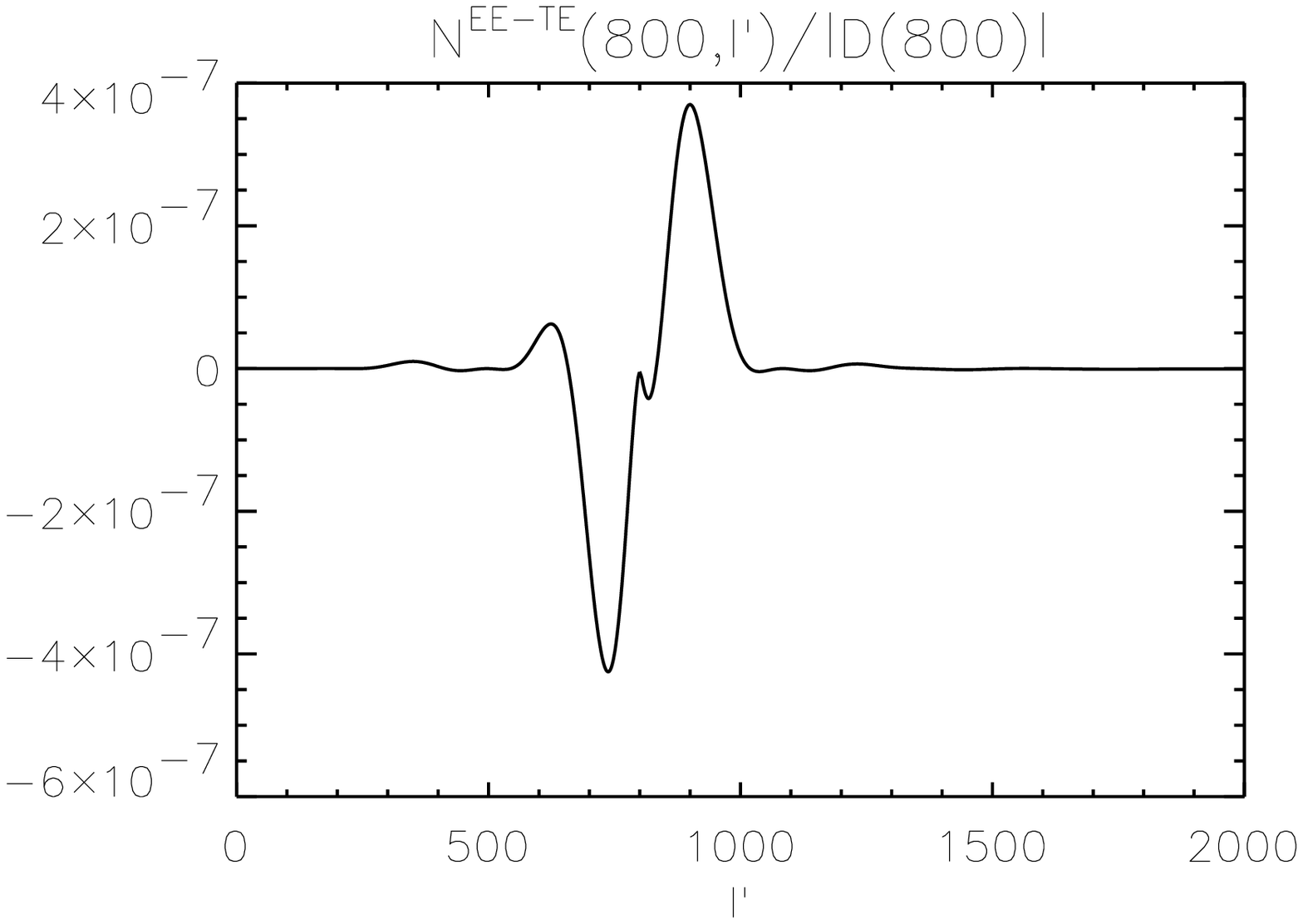}
\includegraphics[scale=0.3]{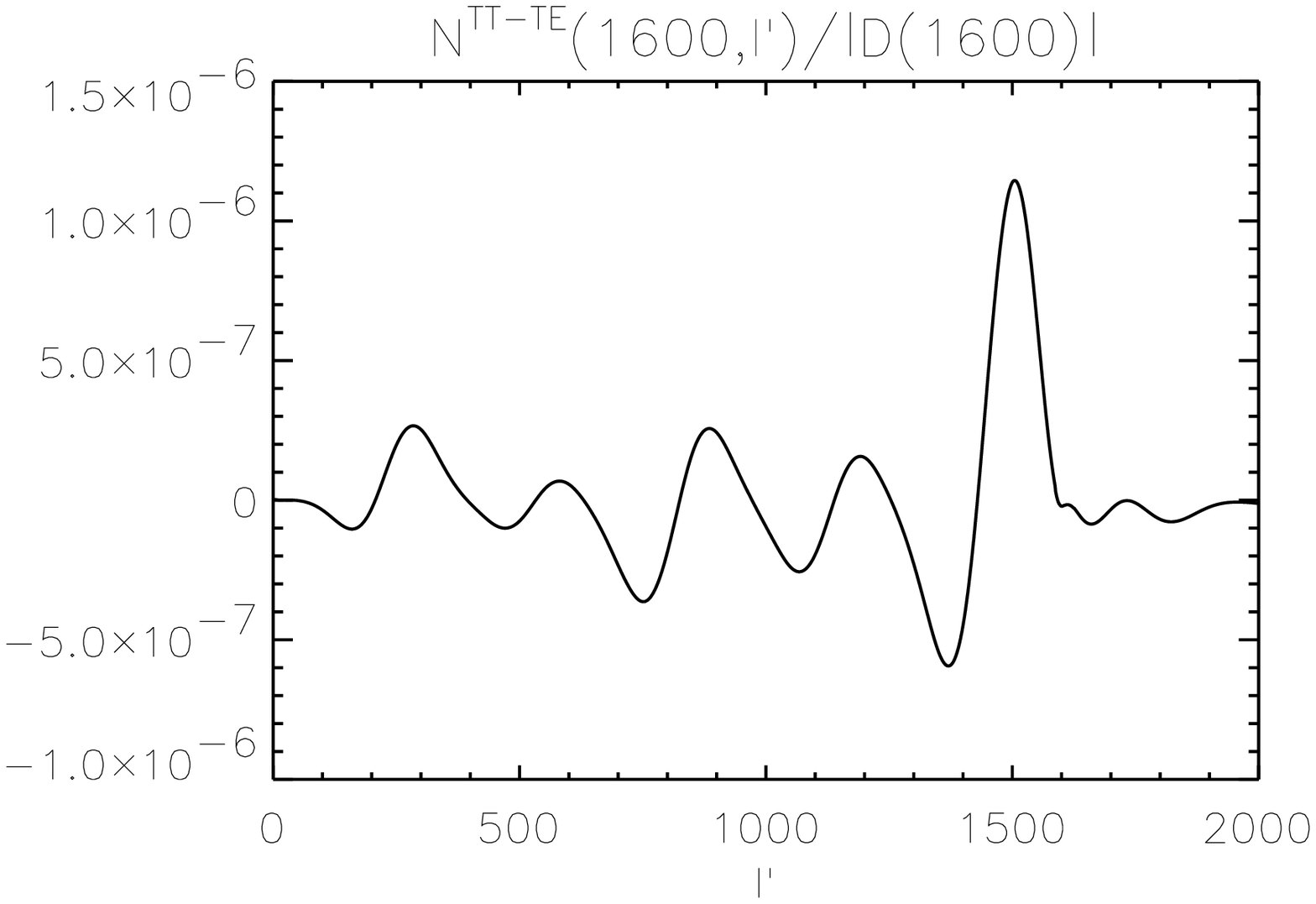}
\includegraphics[scale=0.3]{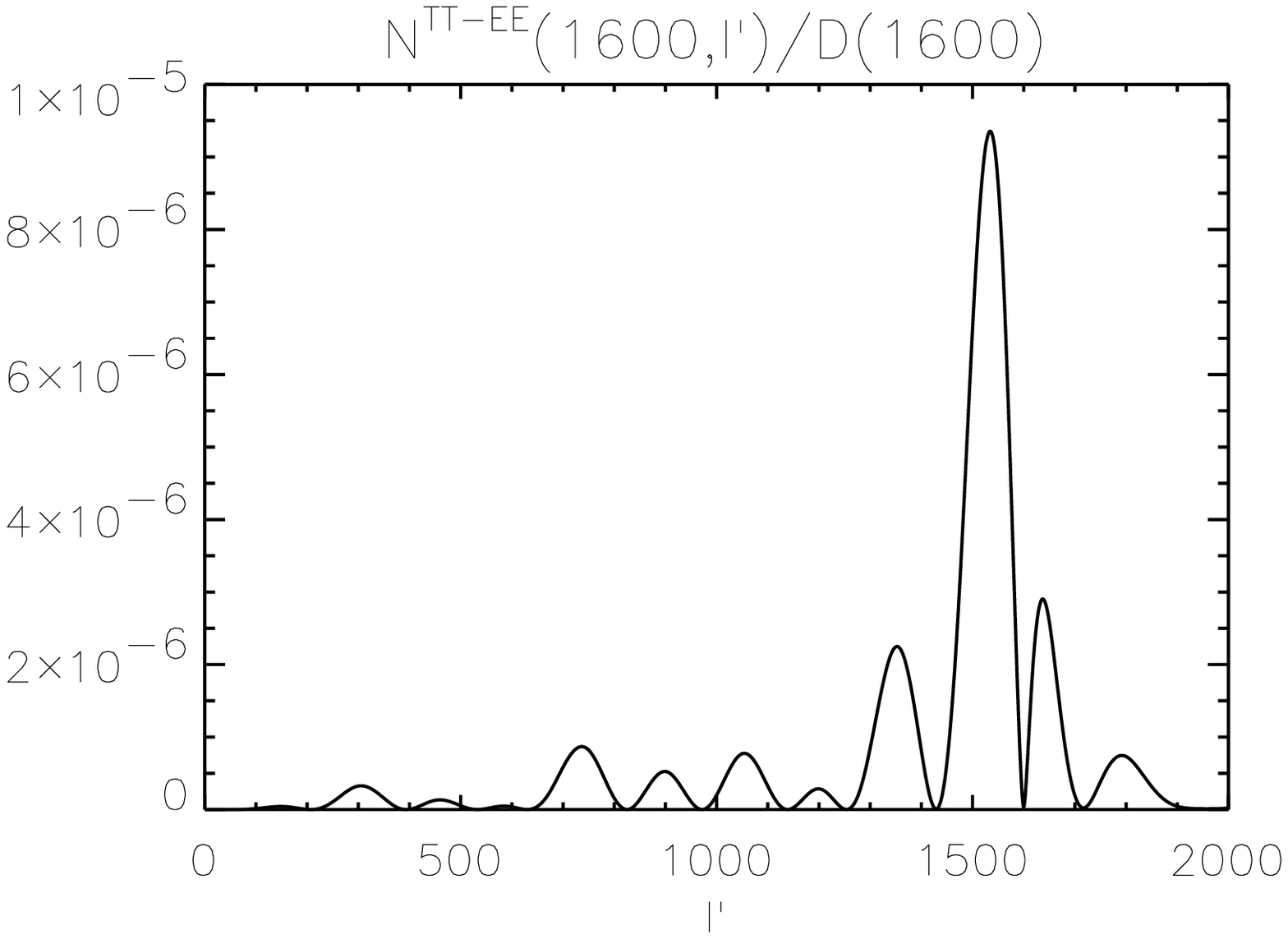}
\includegraphics[scale=0.3]{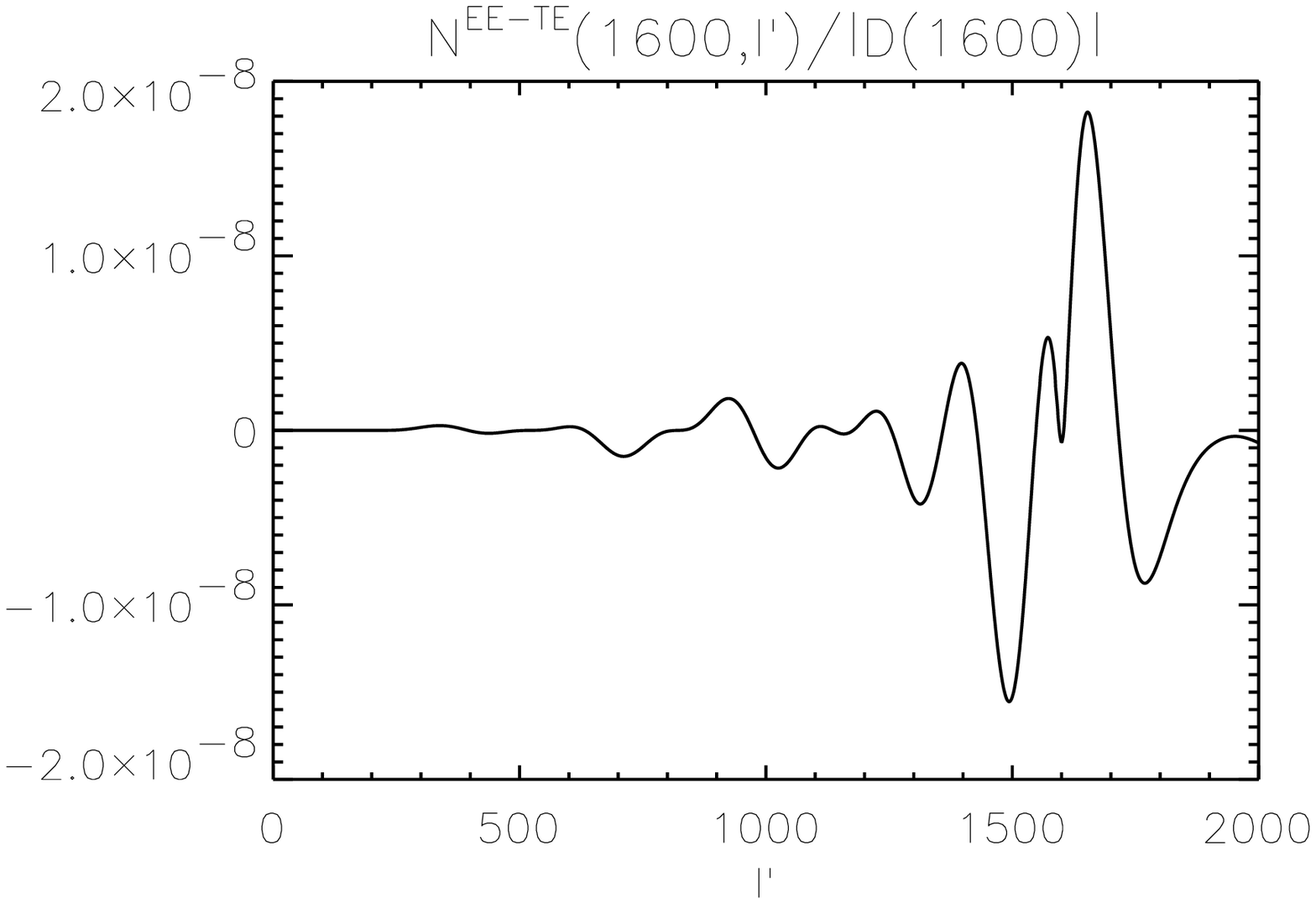}

\caption{Non-diagonal ($l\neq l'$) contribution to the covariance
matrix $\mathcal{N}_{l,l'}^{UV-XY}$ for several polarization and
various values of $l$. From left to right, $UV-XY=TT-EE$, $TT-TE$,
and $EE-TE$, and from top to bottom, $l=200$, $400$, $800$, and
$1600$. The amplitude is normalized to the diagonal value
$\mathcal{D}_{\ell}^{UV-XY}$ to have the relative contribution. Above the
middle column we represent the cross correlated spectrum
$C_{l}^{TE}$, since it is the only spectrum that generates the
term $\mathcal{N}_{l,l'}^{TT-EE}$. The cosmological parameters are given by
WMAP 3 \cite{2006astro.ph..3449S} and we assume a tensor/scalar ratio $r=1$.}
\label{Fig:Covar2}
\end{figure*}

\begin{figure*}
\begin{center}
\includegraphics[scale=.28]{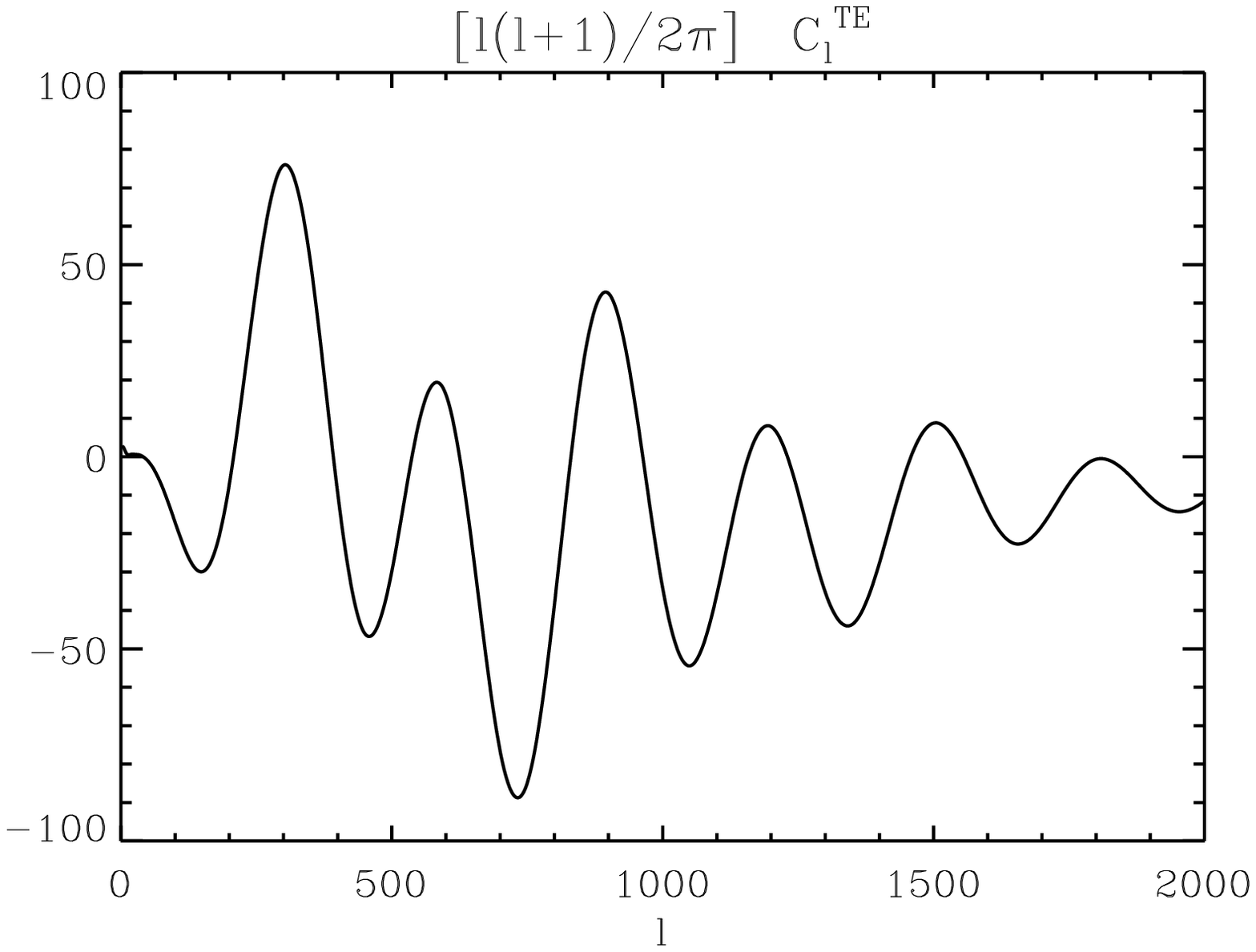}
\includegraphics[scale=.28]{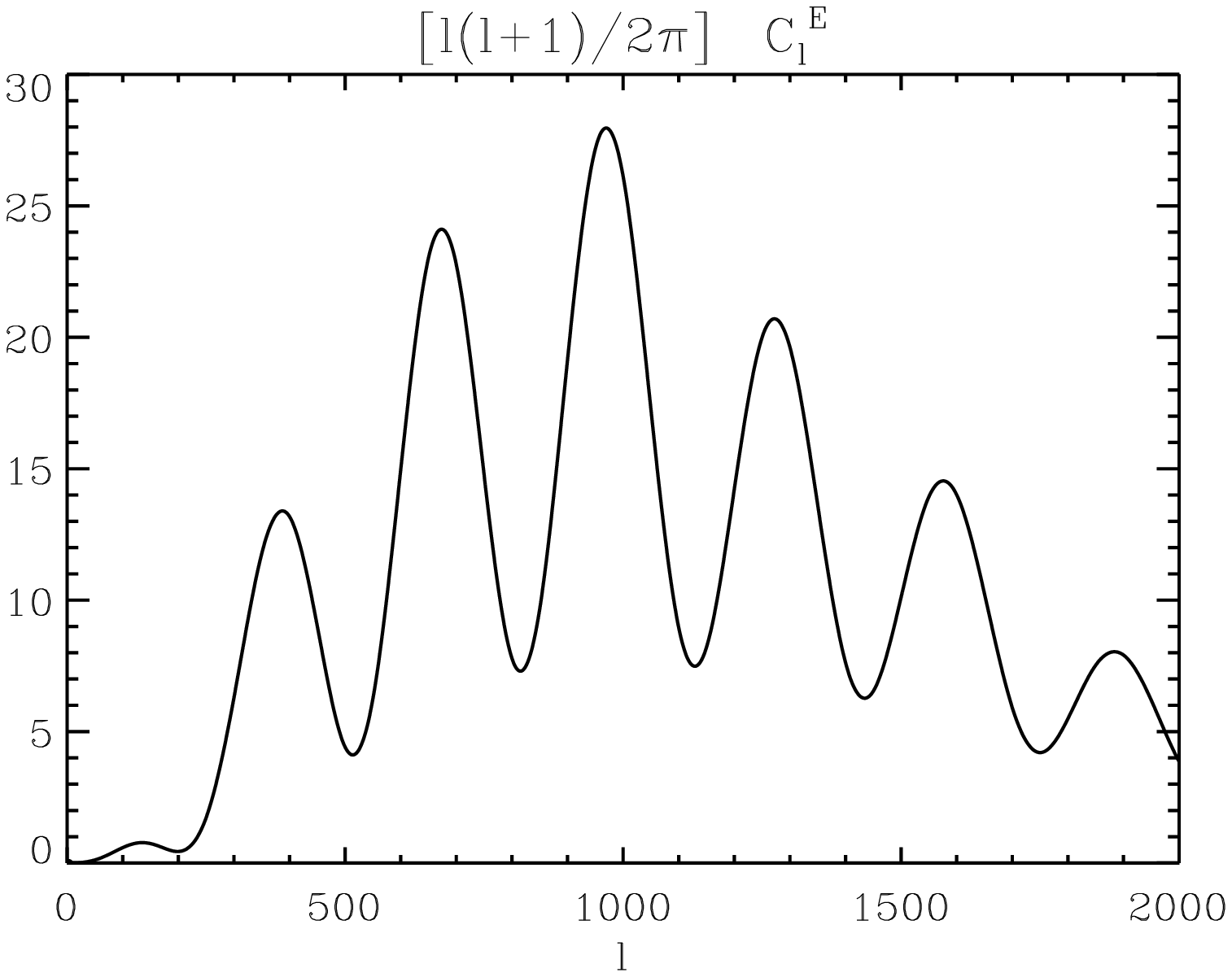}
\includegraphics[scale=.28]{specInitTE.ps}

\vspace*{1cm}

\includegraphics[scale=.28]{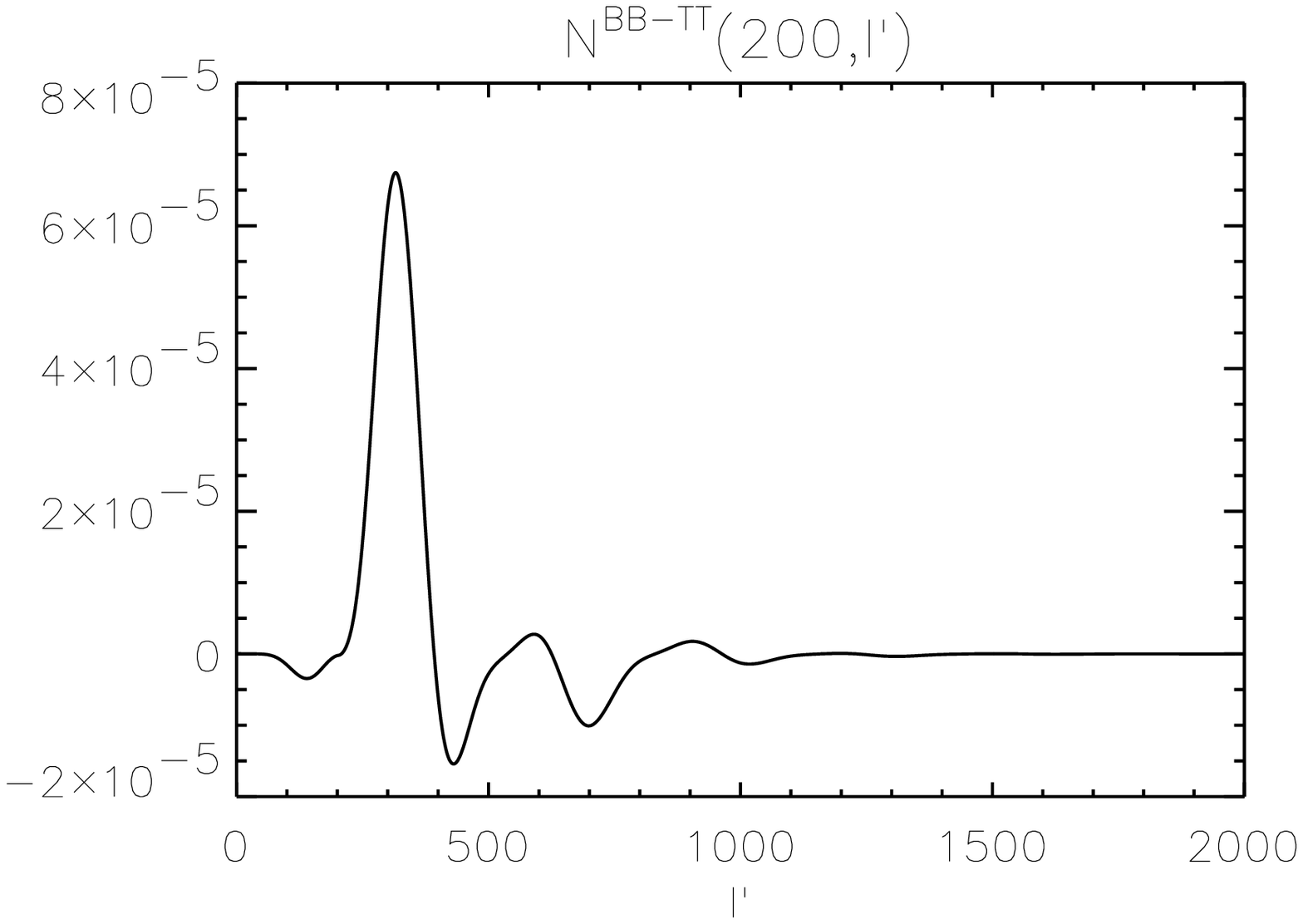}
\includegraphics[scale=.28]{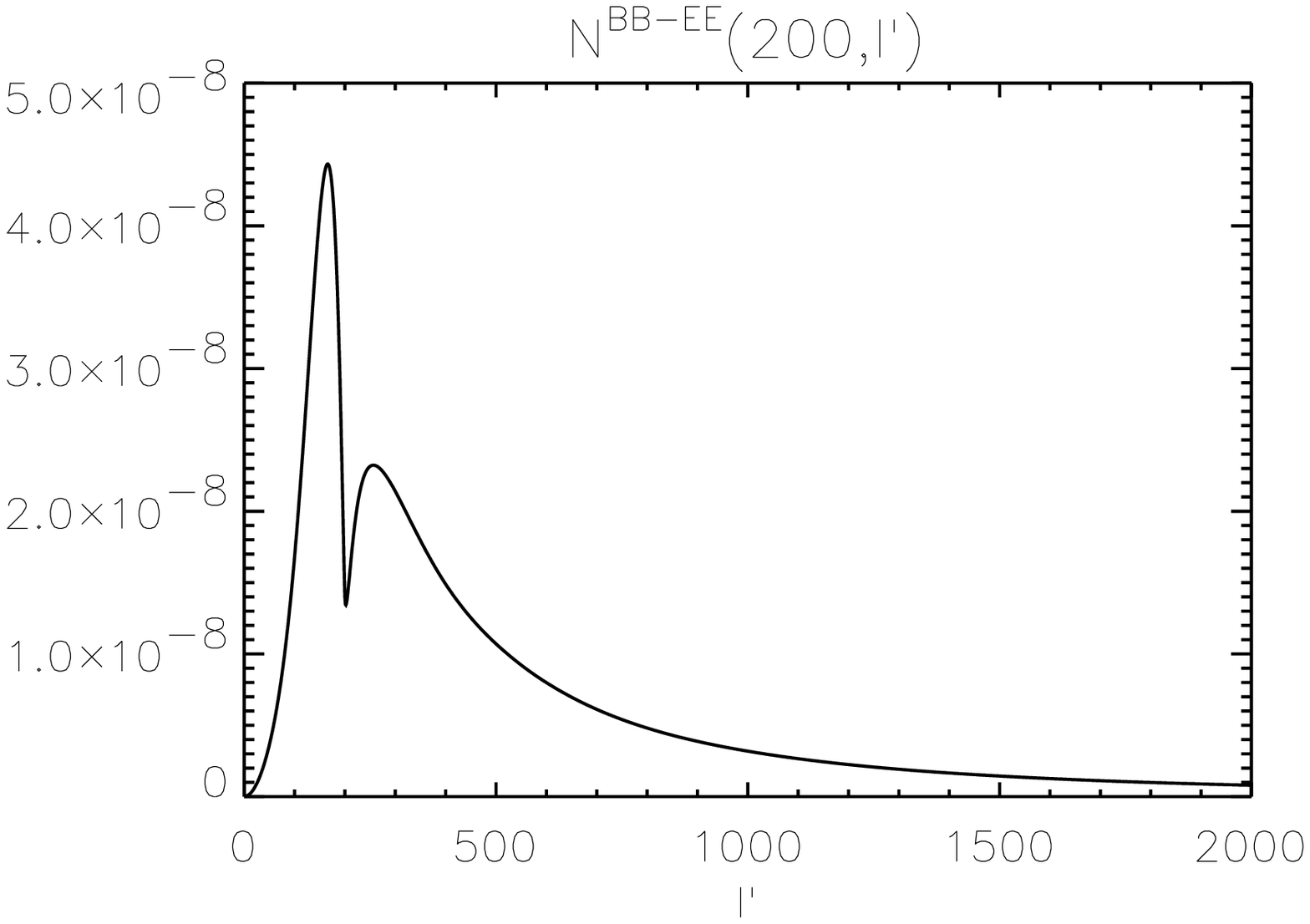}
\includegraphics[scale=.28]{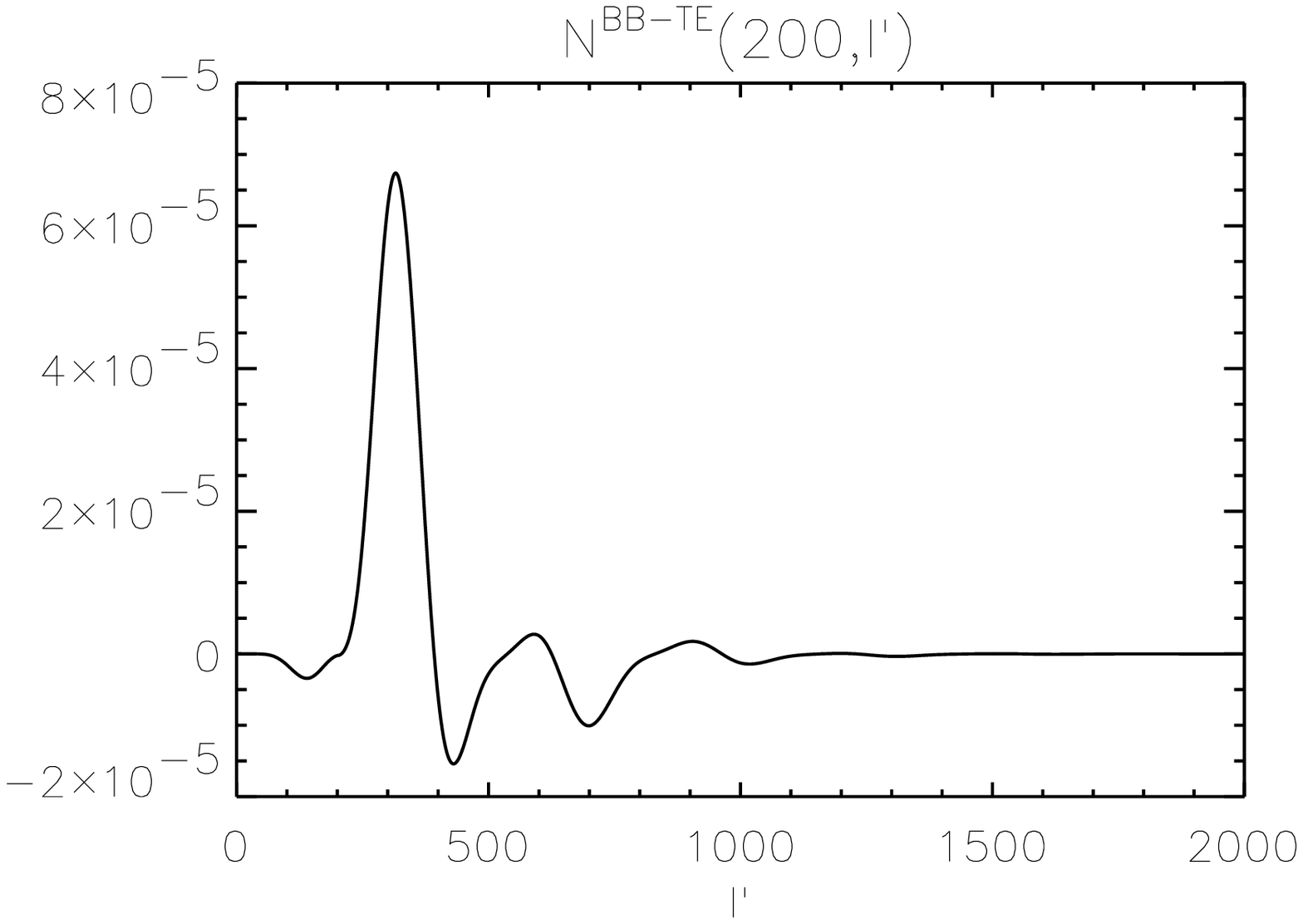}
\includegraphics[scale=.28]{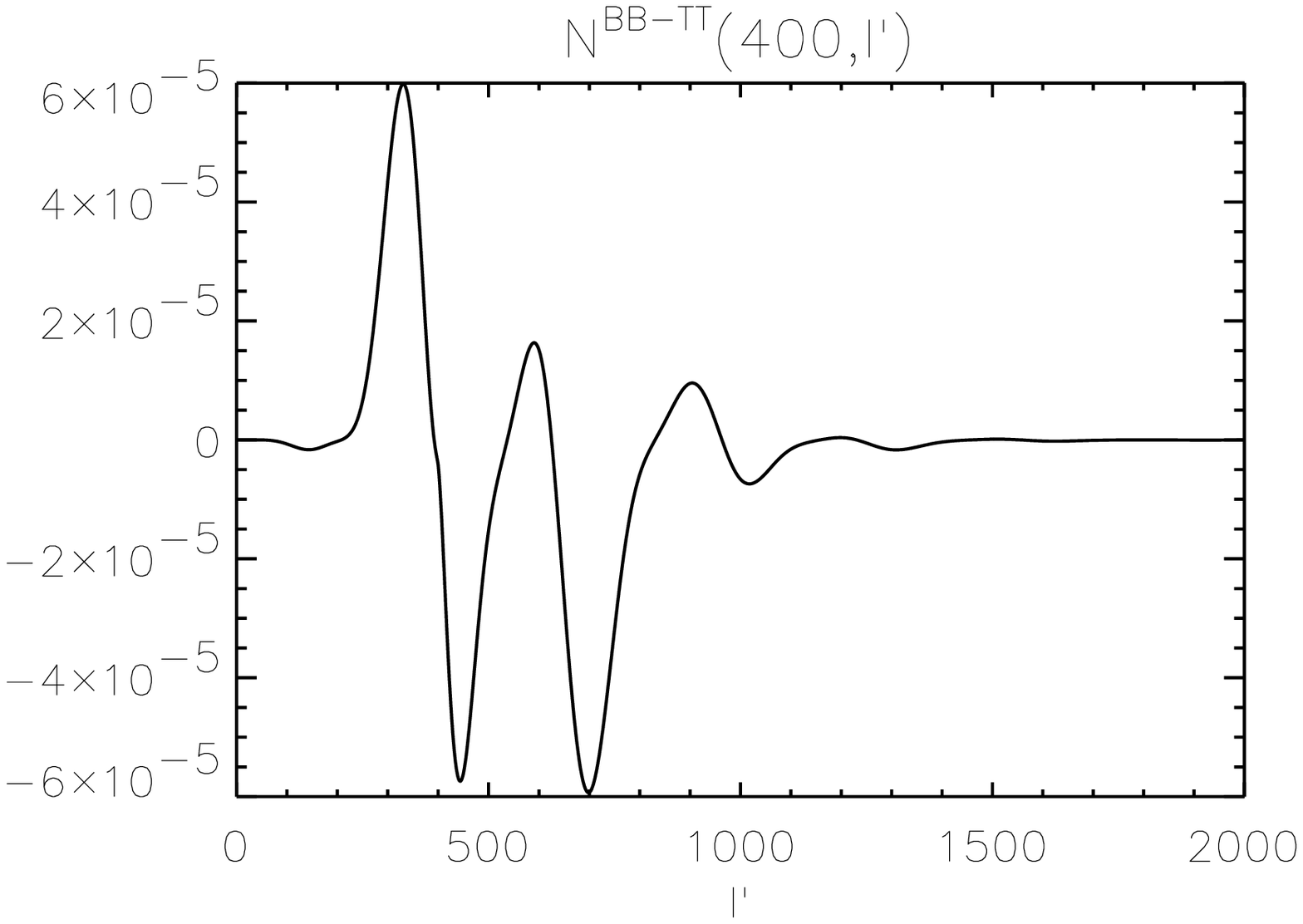}
\includegraphics[scale=.28]{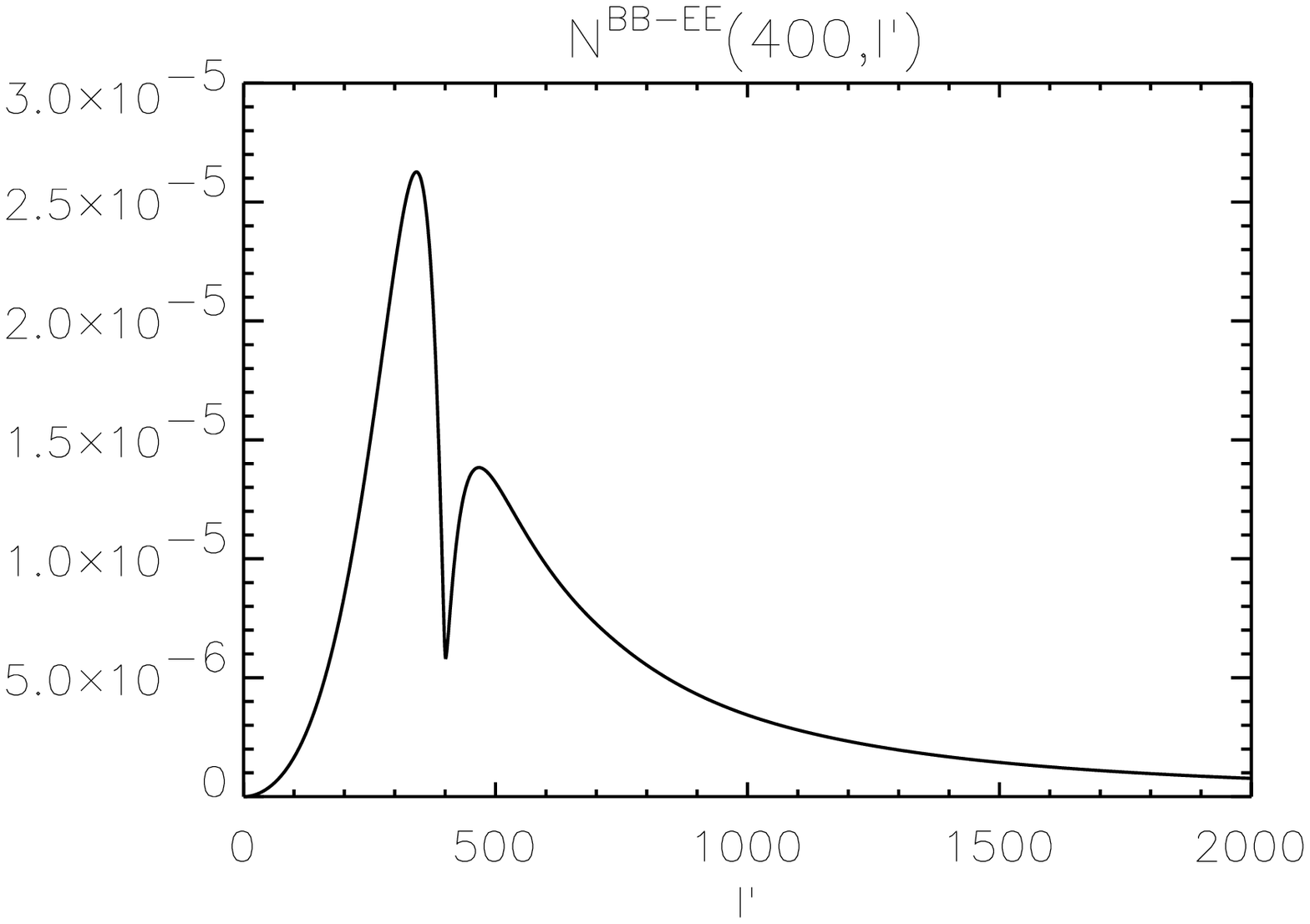}
\includegraphics[scale=.28]{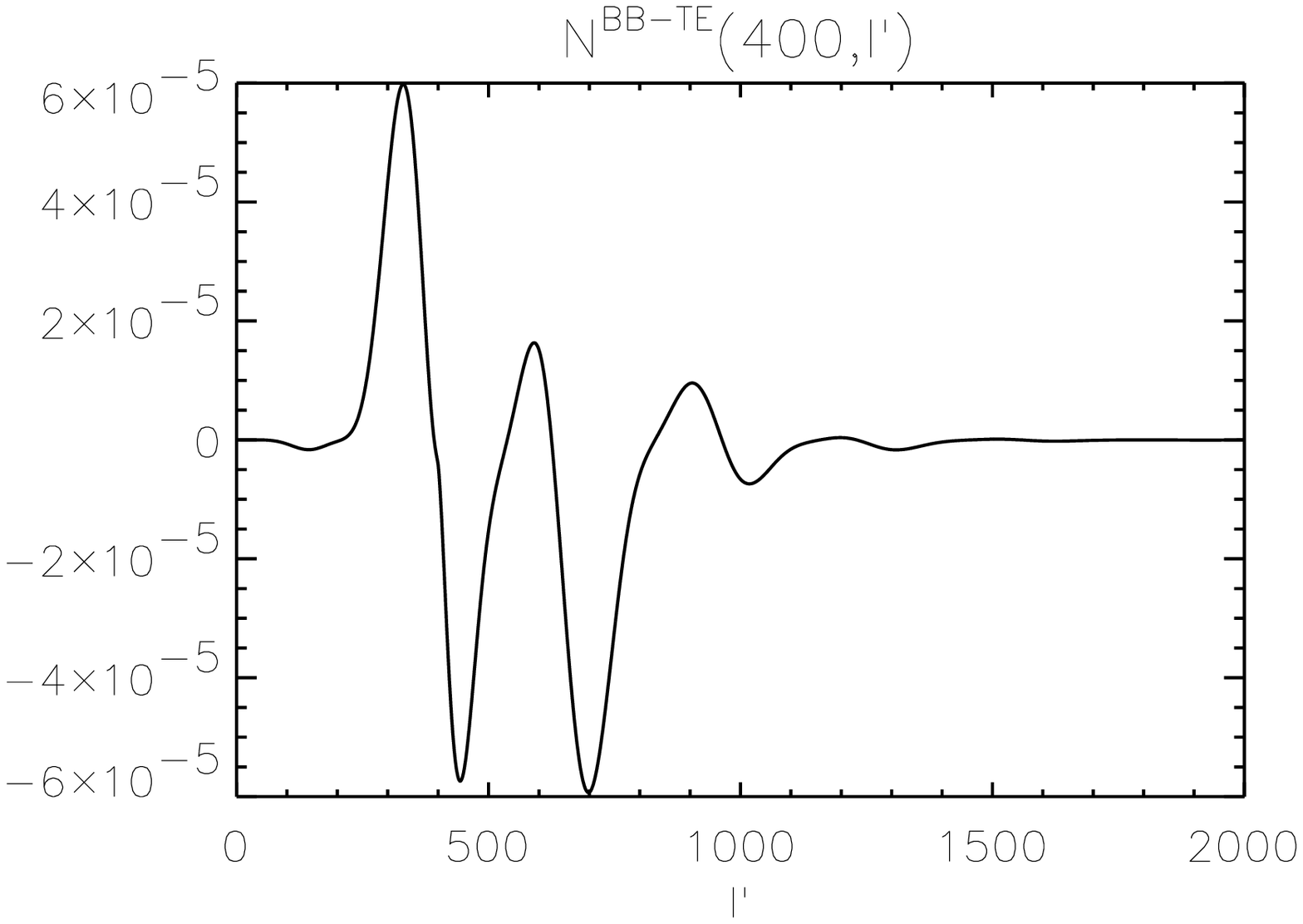}
\includegraphics[scale=.28]{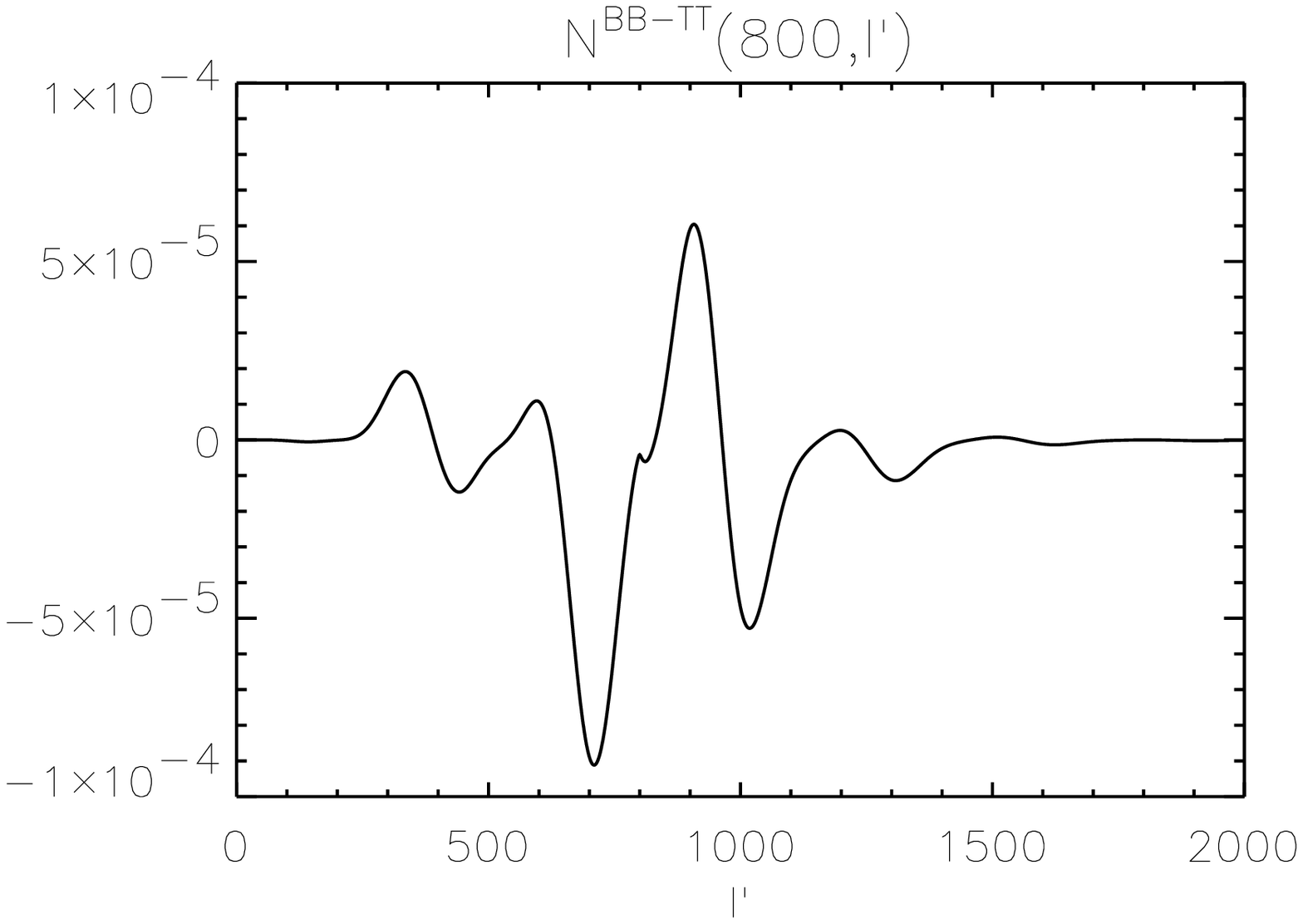}
\includegraphics[scale=.28]{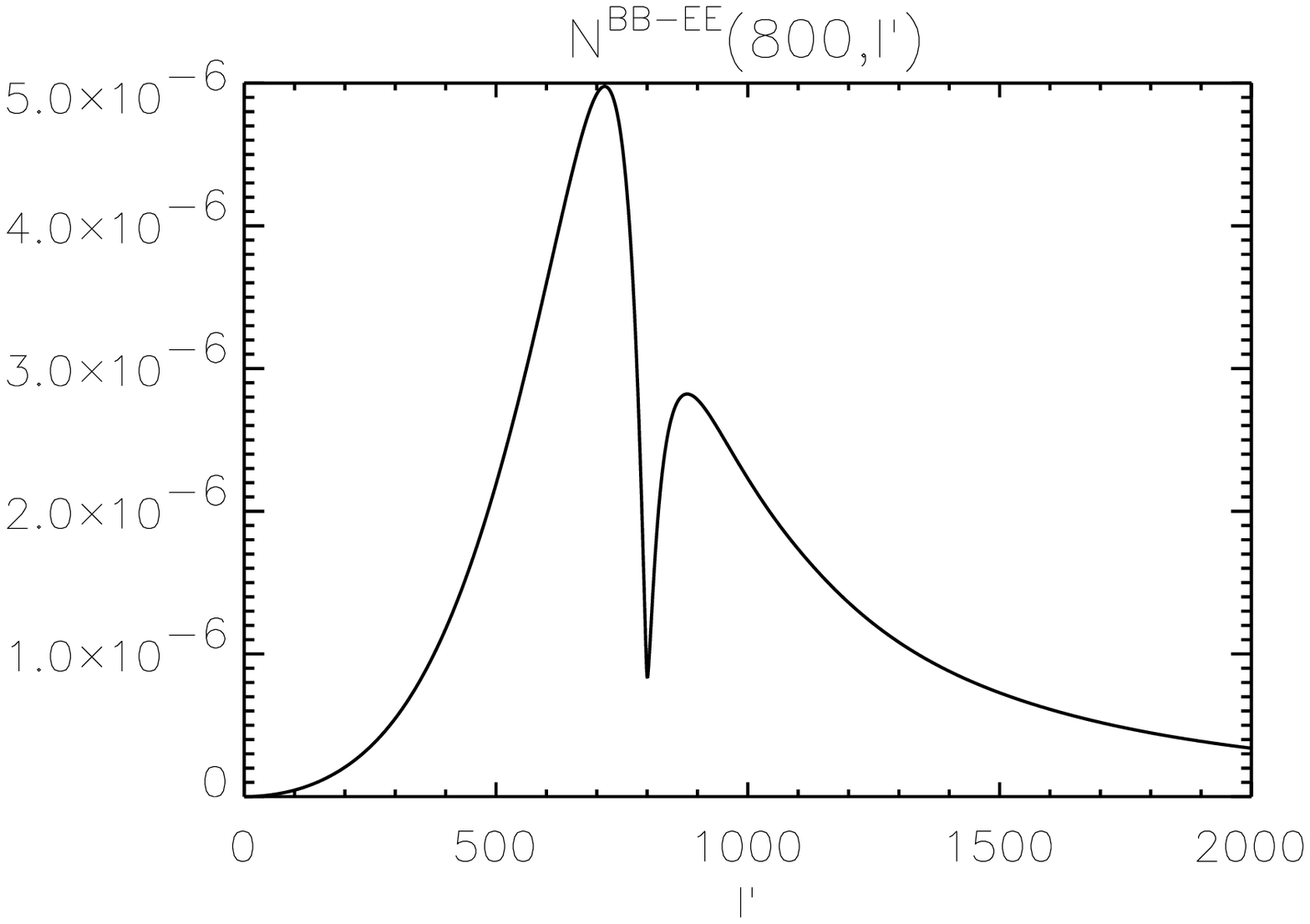}
\includegraphics[scale=.28]{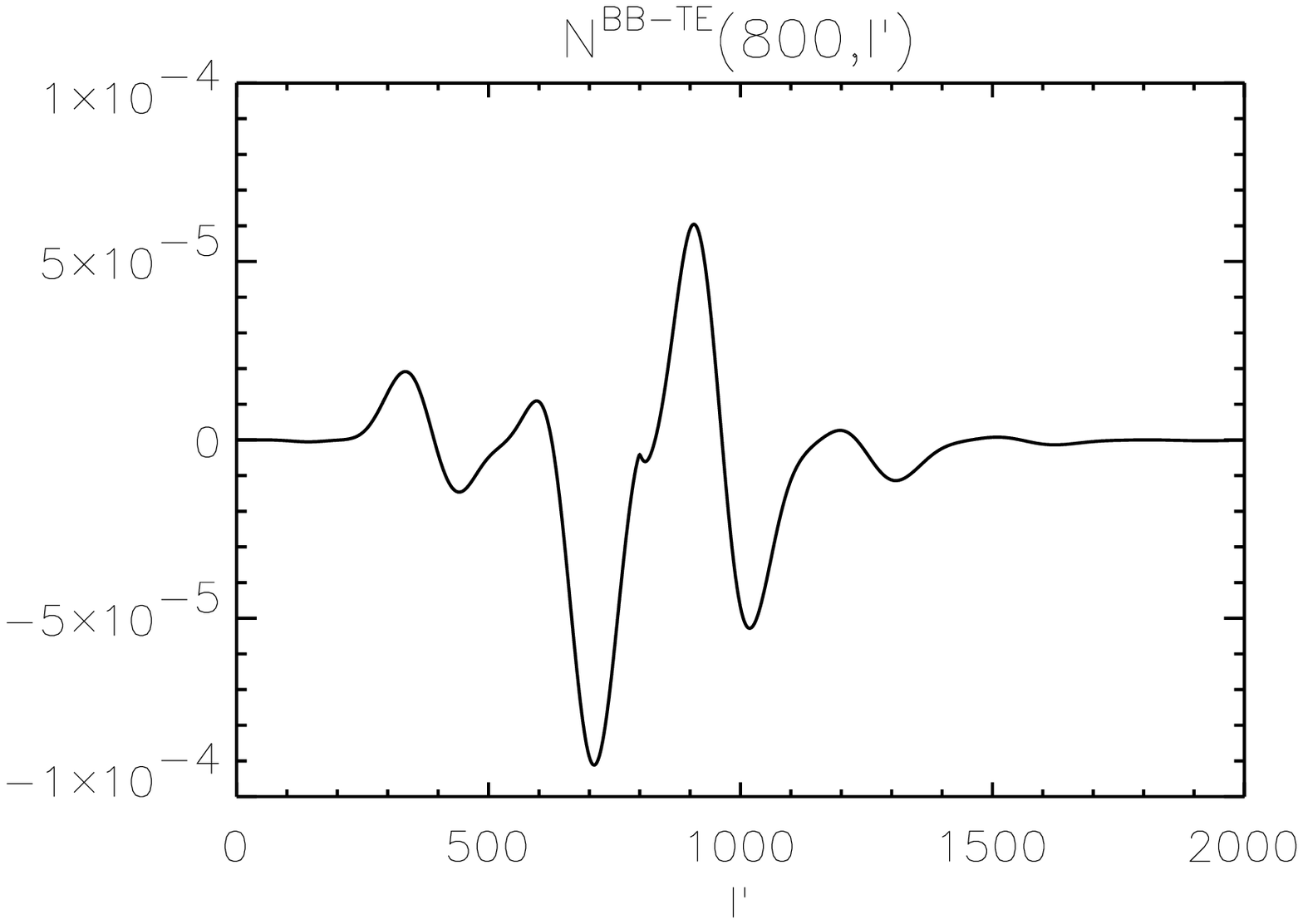}
\includegraphics[scale=.28]{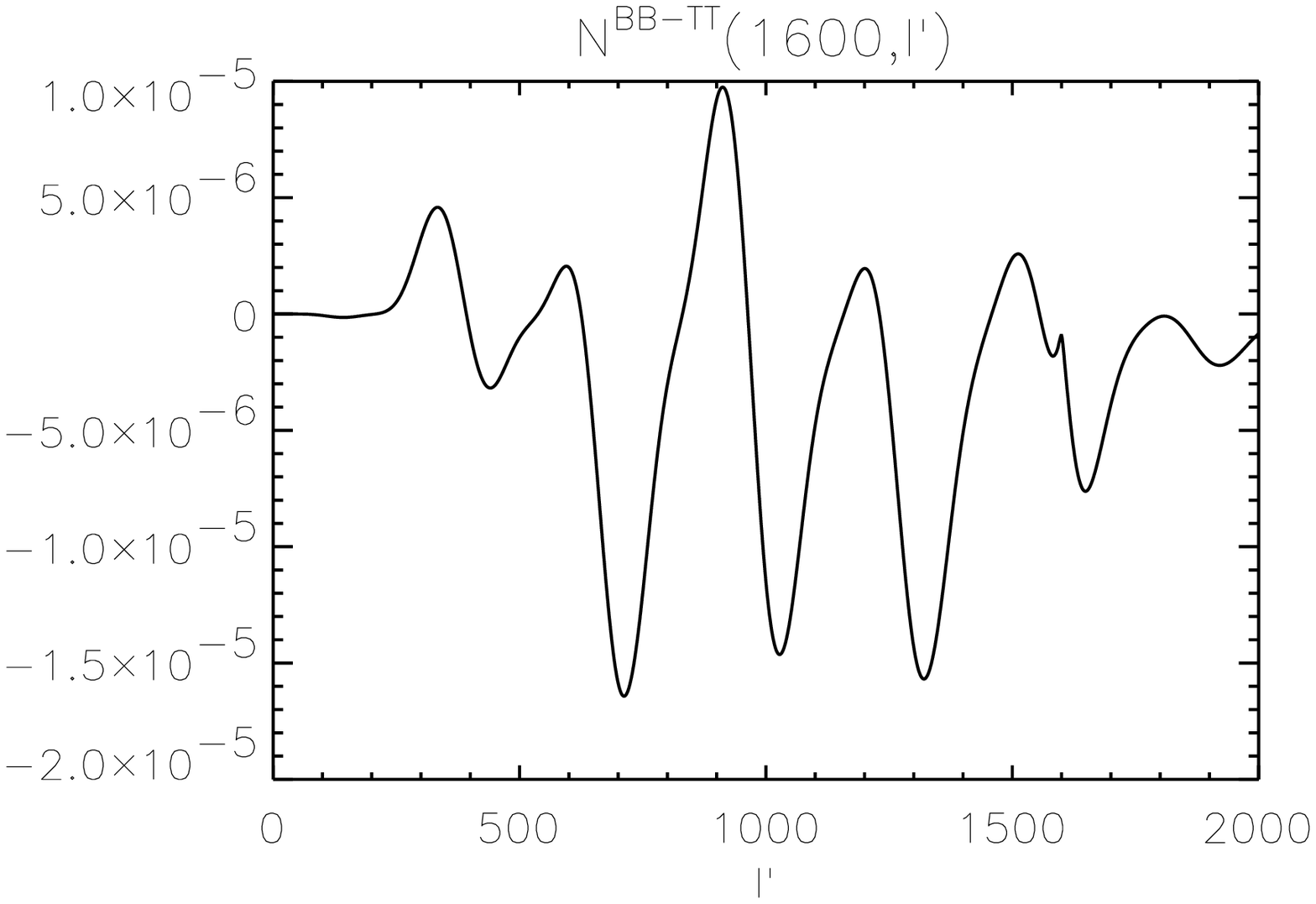}
\includegraphics[scale=.28]{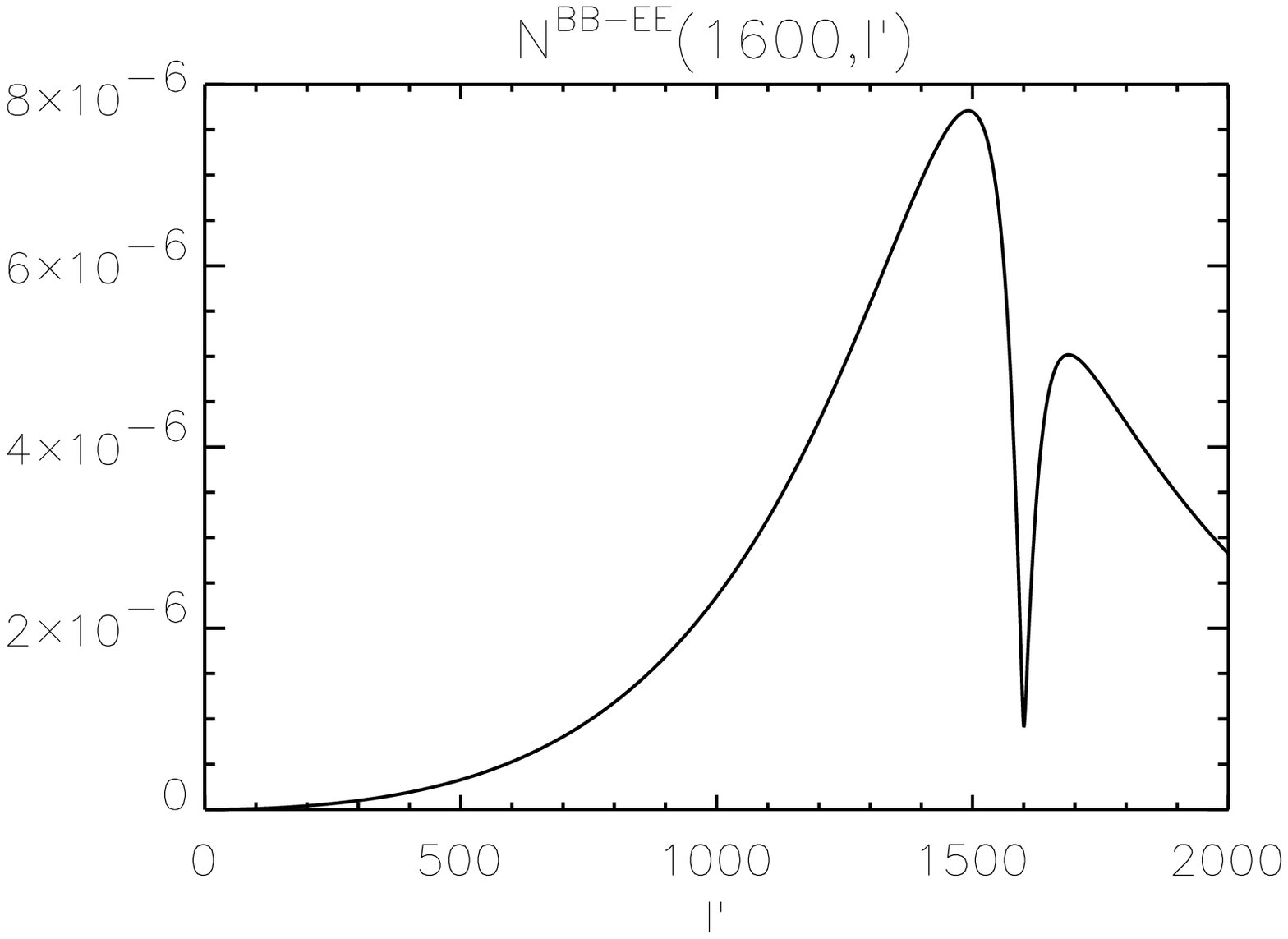}
\includegraphics[scale=.28]{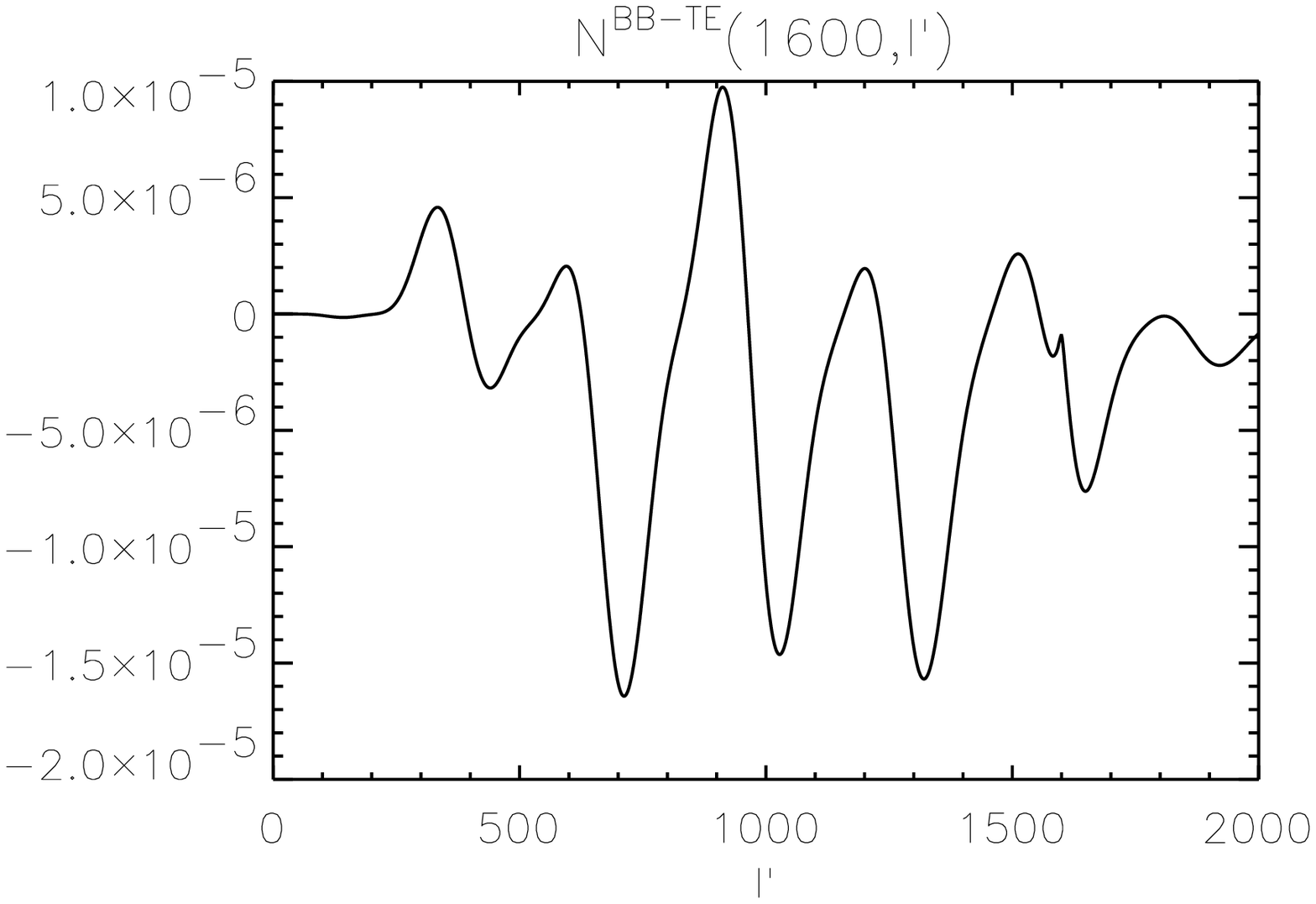}

\caption{Non-diagonal ($l\neq l'$) contribution to the covariance
matrix $\mathcal{N}_{l,l'}^{UV-XY}$ for several polarization and
various values of $l$. From left to right, $UV-XY = BB-TT$,
$BB-EE$, and $BB-TE$, and from top to bottom, $l=200$, $400$,
$800$, and $1600$. The amplitude is not renormalized by any
Gaussian value since these terms are vanishing in the Gaussian
limit. On the first line are represented the $C_l$s that play the
most important role in the non-diagonal contribution below. The
cosmological parameters are given by WMAP 3 \cite{2006astro.ph..3449S}
and we assume a tensor/scalar ratio $r=1$.}
\label{Fig:Covar4}
\end{center}
\end{figure*}

\section*{References}
\bibliographystyle{unsrt}
\bibliography{biblio}

\begin{thebibliography}{10}

\bibitem{2006astro.ph..3449S}
Spergel D~N {\it et al}.
\newblock {Wilkinson Microwave Anisotropy Probe (WMAP) Three Year Results:
  Implications for Cosmology}.
\newblock {\em ArXiv Astrophysics e-prints}, March 2006.

\bibitem{2006astro.ph..4101B}
{Bock}~J {\it et al}.
\newblock {Task Force on Cosmic Microwave Background Research}.
\newblock {\em ArXiv Astrophysics e-prints}, April 2006.

\bibitem{Bartelmann:1999yn}
Bartelmann M and Schneider P.
\newblock Weak gravitational lensing.
\newblock {\em Phys. Rept.}, 340:291--472, 2001.

\bibitem{1998PhRvD..58b3003Z}
Zaldarriaga M and {Seljak} U.
\newblock {Gravitational lensing effect on cosmic microwave background
  polarization}.
\newblock {\em \prd}, 58(2):023003--+, July 1998.

\bibitem{Kinney:1998md}
Kinney~W H.
\newblock Constraining inflation with cosmic microwave background polarization.
\newblock {\em Phys. Rev.}, D58:123506, 1998.

\bibitem{2006astro.ph..4143S}
Seljak U and Slosar A.
\newblock B polarization of cosmic microwave background as a tracer of strings.
\newblock {\em Phys. Rev.}, D74:063523, 2006.

\bibitem{2006astro.ph..4141P}
Wasserman~I Pogosian~L and Wyman M.
\newblock {On vector mode contribution to CMB temperature and polarization from
  local strings}.
\newblock {\em ArXiv Astrophysics e-prints}, April 2006.

\bibitem{Lesgourgues:2005yv}
Pastor~S Lesgourgues~J, Perotto~L and Piat M.
\newblock Probing neutrino masses with cmb lensing extraction.
\newblock {\em Phys. Rev.}, D73:045021, 2006.

\bibitem{Lesgourgues:2006nd}
Lesgourgues J and Pastor S.
\newblock Massive neutrinos and cosmology.
\newblock {\em Phys. Rept.}, 429:307--379, 2006.

\bibitem{Lewis:2006fu}
Lewis A and Challinor A.
\newblock Weak gravitational lensing of the cmb.
\newblock {\em Phys. Rept.}, 429:1--65, 2006.

\bibitem{2003PhRvD..68h3002H}
{Hirata}~C M and {Seljak} U.
\newblock {Reconstruction of lensing from the cosmic microwave background
  polarization}.
\newblock {\em \prd}, 68(8):083002--+, October 2003.

\bibitem{2003PhRvD..67h3002O}
{Okamoto} T and {Hu} W.
\newblock {Cosmic microwave background lensing reconstruction on the full sky}.
\newblock {\em \prd}, 67(8):083002--+, April 2003.

\bibitem{2001PhRvD..63d3501B}
{Bernardeau}~F {Benabed}~K and {van Waerbeke} L.
\newblock {CMB B polarization to map the large-scale structures of the
  universe}.
\newblock {\em \prd}, 63(4):043501--+, February 2001.

\bibitem{2000ApJ...540...14V}
{Bernardeau}~F {Van Waerbeke}~L and {Benabed} K.
\newblock {Lensing Effect on the Relative Orientation between the Cosmic
  Microwave Background Ellipticities and the Distant Galaxies}.
\newblock {\em \apj}, 540:14--19, September 2000.

\bibitem{1999PhRvD..59l3507Z}
{Zaldarriaga} M and {Seljak} U.
\newblock {Reconstructing projected matter density power spectrum from cosmic
  microwave background}.
\newblock {\em \prd}, 59(12):123507--+, June 1999.

\bibitem{Lewis:2005tp}
Lewis A.
\newblock Lensed cmb simulation and parameter estimation.
\newblock {\em Phys. Rev.}, D71:083008, 2005.

\bibitem{2006PhRvD..74l3002S}
{Smith}~K M, {Hu} W, and {Kaplinghat} M.
\newblock {Cosmological information from lensed CMB power spectra}.
\newblock {\em \prd}, 74(12):123002--+, December 2006.

\bibitem{2006astro.ph..7494L}
{Li} C, {Smith}~T L, and {Cooray} A.
\newblock {Non-Gaussian Covariance of CMB B-modes of Polarization and Parameter
  Degradation}.
\newblock {\em \prd}, 75:083501, 2007.

\bibitem{2006JCAP...03..007S}
{Shapiro} C and {Cooray} A.
\newblock {The Born and lens lens corrections to weak gravitational lensing
  angular power spectra}.
\newblock {\em Journal of Cosmology and Astro-Particle Physics}, 3:7--+, March
  2006.

\bibitem{2000PhRvD..62d3007H}
Hu~W.
\newblock {Weak Lensing of the CMB: A Harmonic Approach}.
\newblock {\em \prd}, 62(4):043007--+, July 2000.

\bibitem{2002ARA&A..40..171H}
{Hu} W and {Dodelson} S.
\newblock {Cosmic Microwave Background Anisotropies}.
\newblock {\em \araa}, 40:171--216, 2002.

\bibitem{2001PhRvD..64h3005H}
Hu~W.
\newblock {Angular trispectrum of the cosmic microwave background}.
\newblock {\em \prd}, 64(8):083005--+, October 2001.

\bibitem{2000PhRvD..62f3510Z}
{Zaldarriaga} M.
\newblock {Lensing of the CMB: Non-Gaussian aspects}.
\newblock {\em \prd}, 62(6):063510--+, September 2000.

\bibitem{1997A&A...324...15B}
{Bernardeau} F.
\newblock {Weak lensing detection in CMB maps.}
\newblock {\em \aap}, 324:15--26, August 1997.

\bibitem{2002PhRvD..65f3512C}
{Cooray} A.
\newblock {Weak lensing of the cosmic microwave background: Power spectrum
  covariance}.
\newblock {\em \prd}, 65(6):063512--+, March 2002.

\bibitem{Challinor:2004pr}
Challinor A and Chon G.
\newblock Error analysis of quadratic power spectrum estimates for cmb
  polarization: sampling covariance.
\newblock {\em Mon. Not. Roy. Astron. Soc.}, 360:509--532, 2005.

\bibitem{Seljak:1996ti}
Seljak U.
\newblock Measuring polarization in cosmic microwave background.
\newblock {\em ArXiv Astrophysics e-prints}, 1996.

\bibitem{Zaldarriaga:1996xe}
Zaldarriaga M and Seljak U.
\newblock An all-sky analysis of polarization in the microwave background.
\newblock {\em Phys. Rev.}, D55:1830--1840, 1997.

\bibitem{Kamionkowski:1996zd}
Kosowsky~A Kamionkowski~M and Stebbins A.
\newblock A probe of primordial gravity waves and vorticity.
\newblock {\em Phys. Rev. Lett.}, 78:2058--2061, 1997.

\bibitem{2004PhRvD..70d3002S}
{Hu}~W {Smith} K~M and {Kaplinghat} M.
\newblock {Weak lensing of the CMB: Sampling errors on B modes}.
\newblock {\em \prd}, 70(4):043002--+, August 2004.

\bibitem{Smith:2005ue}
Challinor~A Smith~S and Rocha G.
\newblock What can be learned from the lensed cosmic microwave background
  b-mode polarization power spectrum?
\newblock {\em Phys. Rev.}, D73:023517, 2006.

\bibitem{Bernardeau:2001qr}
Gaztanaga~E Bernardeau~F, Colombi~S and Scoccimarro R.
\newblock Large-scale structure of the universe and cosmological perturbation
  theory.
\newblock {\em Phys. Rept.}, 367:1--248, 2002.

\end{thebibliography}

\end{document}